# Numerical Modeling of Orbit-Spin Coupling Accelerations in a Mars General Circulation Model: Implications for Global Dust Storm Activity

Michael A. Mischna and James H. Shirley

*Jet Propulsion Laboratory, California Institute of Technology, Pasadena, CA 91109*

Government Sponsorship acknowledged

**Abstract:**

We employ the MarsWRF general circulation model (GCM) to test the predictions of a new physical hypothesis: a weak coupling of the orbital and rotational angular momenta of extended bodies is predicted to give rise to cycles of intensification and relaxation of circulatory flows within atmospheres. The dynamical core of the GCM has been modified to include the orbit-spin coupling accelerations due to solar system dynamics for the years 1920-2030. The modified GCM is first subjected to extensive testing and validation. We compare forced and unforced model outcomes for large-scale zonal and meridional flows, and for near-surface wind velocities and surface wind stresses. The predicted cycles of circulatory intensification and relaxation within the modified GCM are observed. Most remarkably, the modified GCM reproduces conditions favorable for the occurrence of perihelion-season global-scale dust storms on Mars in years in which such storms were observed. A strengthening of the meridional overturning (Hadley) circulation during the dust storm season occurs in the GCM in all previously known years with perihelion-season global-scale dust storms. The increased upwelling produced in the southern hemisphere in southern summer may facilitate the transport of dust to high altitudes in the Mars atmosphere during the dust storm season, where radiative heating may further strengthen the circulation. Significantly increased surface winds and surface wind stresses are also obtained. These may locally facilitate saltation and dust lifting from the surface. The numerical simulations constitute proof of concept for the orbit-spin coupling hypothesis under evaluation.

## 1. Introduction

A recently developed orbit-spin coupling hypothesis (*Shirley*, 2016) predicts the existence of an acceleration field that may modulate large-scale atmospheric circulatory flows through constructive and destructive interference effects. In this paper we test the predictions of this physical hypothesis by incorporating the orbit-spin coupling accelerations within a state-of-the-art global circulation model of the Mars atmosphere. The Mars atmosphere is optimal for this purpose, due to the availability of an extended observational record for Mars exhibiting marked interannual variability, its short thermal time constant, and the relative simplicity of its circulation in comparison with those of the Earth and Sun.

According to the physical hypothesis detailed in *Shirley* (2016), the rate of change of planetary orbital angular momentum ($d\mathbf{L}/dt$, or $\dot{L}$) with respect to the solar system barycenter may be implicated as a forcing function for atmospheric variability. The derived orbit-spin coupling term yields a small horizontal acceleration ($\sim 10^{-5}$ ms$^{-2}$) within the Mars atmosphere that varies spatially and with time. This acceleration, which we refer to as a "coupling term acceleration" (CTA), while instantaneously small, may cumulatively yield wind velocity changes of several tens of m s$^{-1}$ on seasonal timescales. Under the physical hypothesis investigated here, the global circulation is predicted to follow an intensification/relaxation cycle that tracks the variability of $\dot{L}$. Statistical evidence that this is in fact the case for Mars is presented in the companion paper by *Shirley and Mischna* (2016, and references therein).

Given the overall lack of robust, global atmospheric circulation measurements on Mars, testing of this physical hypothesis is ideally suited to numerical simulation through the use of a general circulation model (GCM), which provides the best means of simulating global behavior of the Martian atmosphere. With the use of such a tool, we can observe and track the individual components of the wind field across the globe, and assess other changes to the overall atmospheric state.

We explore the influence of the CTA on the Martian atmosphere with a GCM that has been modified to account for this new acceleration term. We demonstrate that the CTA can have a non-negligible influence on wind speeds and global circulation patterns. The cyclic variability of the putative forcing function is largely decoupled from the annual cycle of solar irradiance that drives the normal seasonal variability on Mars (*Shirley*, 2015; *Shirley and Mischna*, 2016). The relative phasing of the two cycles introduces a considerable degree of interannual variability in the Martian atmosphere— variability which may lead to the occasional formation of global-scale dust storms (GDS) or other intermittent atmospheric phenomena.

This paper (and the investigation it describes) may be naturally subdivided into two parts. In the first part, we develop and validate the numerical model, and employ this to evaluate specific predictions of the physical hypothesis. In the second part, we employ the modified GCM to investigate the response of the atmosphere in past years that were characterized by the occurrence of global-scale dust storms. We contrast these model results with those obtained for the intervening years lacking observed GDS events.

In Section 2 we describe the GCM employed for this study. Section 3 describes the orbit-spin coupling hypothesis under investigation, and provides a detailed explanation of how the accelerations resulting from the coupling are incorporated into the model. Validation and initial calibration of our modified GCM is described in Section 4.

In Section 5 we continue to explore the properties and behaviors of the modified GCM, through an in-depth consideration of model outcomes obtained both with and without the CTA. We focus on the late northern spring ("aphelion") season for these comparisons, as the Mars atmosphere is relatively dust-free and more quiescent at these times (*Shirley et al.*, 2015).

Our attention shifts to the southern spring and summer ("dust storm") season in Section 6, where we continue to compare forced and unforced GCM simulations in order to determine the consequences for the circulation of the Mars atmosphere of the inclusion of the CTA. We employ a catalog of 21 past Mars years known to have either included global-scale dust storms or to have been free of such storms (*Shirley and Mischna*, 2016). In Section 7 we perform a statistical evaluation of model-derived global mean daytime surface wind stress values for the years with and without global-scale dust storms. A discussion of the results of Sections 5, 6 and 7 is provided in Section 8, where we also provide a GCM-based GDS forecast for Mars years 33 and 34. We summarize and conclude in Section 9.

The occurrence of global-scale dust storms in some years but not in others is a dominant feature of the interannual variability of the Mars atmosphere. An intensification of both surface wind stresses and the large scale circulation seems to be required, in order both to initiate dust lifting in multiple areas and to efficiently transport the dust to high altitudes and over wide regions (*Haberle,* 1986; *Shirley*, 2015, and references therein). Accordingly, for this preliminary investigation, we focus primarily on near-surface winds (in the lowest model layer), on surface wind stresses, and on global winds. Other key atmospheric indices and phenomena, such as temperature, pressure, and thermal tides, are not examined here, although these will be considered in later work.

## 2. The MarsWRF GCM

We use the Mars Weather Research and Forecasting (MarsWRF) GCM for this investigation. MarsWRF is a Mars-specific implementation of the PlanetWRF GCM (*Richardson et al*., 2007), which, itself, is a global model derived from the terrestrial mesoscale WRF model (*Skamarock and Klemp*, 2008). MarsWRF solves the primitive equations using a finite difference model on an Arakawa-C grid (*Arakawa and Lamb*, 1977). The prognostic equations solved by MarsWRF are based on conserved variables, and it has been shown (*Richardson et al*., 2007), that quantities like angular momentum show no long-term trend on the decadal scales we are considering here. The horizontal resolution of the model is variable and selectable at run time, and the 40-layer vertical grid (from 0-80 km) being used follows a modified-sigma (terrain-following) coordinate. In the present investigation, we have chosen a resolution of 5° x 5°, which corresponds to a grid of 72 x 36 (lon x lat). The total present-day atmospheric $CO_2$ budget is tuned to fit the Viking Lander annual pressure curves (~6.1 mb), and both surface albedo and thermal inertia are matched to MGS-TES observations (*Christensen et al*., 2001; *Putzig et al*., 2007). Water ice albedo and emissivity are fixed at 0.45 and 1.0, respectively, while corresponding values for $CO_2$ ice are independently chosen for each hemisphere (*Guo et al*., 2009).

2.1. Dust, water ice, and water vapor

Dust plays a key role in shaping the temporal variability of the Mars atmosphere, similar to that played by water in the terrestrial atmosphere, by absorbing and re-radiating solar radiation, locally heating the atmosphere, and thereby strongly influencing the atmospheric circulation on all scales. Because we wish to isolate, to the greatest extent possible, the effects of the putative orbit-spin coupling mechanism, we have chosen to exclude atmospheric dust from the MarsWRF GCM in most model runs described here.

Although the model also incorporates elements of the surface/atmosphere system such as subsurface vapor diffusion and a full atmospheric water cycle, these components are either turned off or left in their default state during this investigation, as they play no substantive role in the short-term evolution of atmospheric winds. Water vapor plays, at most, a minor role in atmospheric dynamics (*Lewis*, 2003; but also see *Kahre et al.*, 2015). Radiative transfer is applied using the scheme detailed in *Mischna et al.* (2012).

One important consequence of our choice to exclude dust is illustrated in Fig. 1. This figure compares meridional streamflow ("Hadley cell") plots for MarsWRF model runs performed both with and without dust. Panels (a) and (b) show the zonal mean zonal wind and zonal mean meridional streamfunction, respectively, for a dust-free simulation during the aphelion season from $L_s$=70°-90°. Panel (c) is a streamfunction simulation for the same season, but including radiatively active dust obtained using a dust profile from the Mars Climate Database (*Forget et al.*, 1999). In panels (b) and (c), positive values of the streamfunction (shaded blue-green) correspond to counterclockwise circulations. Closed contours follow the general trajectory of the meridional overturning (Hadley) circulation. In this season (late northern spring), panels b and c thus indicate rising air in the north and descending air in the south. By far the largest portion of the atmospheric mass transfer occurs in the lowest two scale heights, below the 50 Pa pressure level (~25 km), due to the increased density of the air in the lower atmosphere. (Note the difference in the vertical scales of panels (a) and (b/c).) The overturning circulation in the dust-free scenario is comparatively shallow in comparison with that for the simulation including dust, due to the absence of radiative heating effects of dust in the dust-free model run. The circulations are morphologically similar, particularly near the surface. Similar peak values are obtained for the counterclockwise circulations in both simulations. The simulations compare quite well to previously published depictions of mass streamfunction at this season (cf. *Forget et al*., 1999; *Richardson and Wilson*, 2002; *Lewis*, 2003).

Figure 1 demonstrates that MarsWRF simulations without dust can adequately represent the broad-scale features of the atmospheric circulation of Mars. The addition of dust to the model, addressed in Section 8, and to be further explored in subsequent investigations, will provide greater realism, particularly with respect to the vertical dimensions of the overturning circulation.

2.2. The baseline model

We first establish a multi-year MarsWRF model run without any external forcing from the coupling term acceleration, using the GCM in its nominal, dust-free configuration. All years in this, our 'baseline' simulation, represent the same notional annual cycle. Slight variations in model output reflect the effects of year-by-year 'weather' in the martian system, but every year relies upon identical model physics (insolation, orbital position, etc.). For the baseline simulation, we have run MarsWRF for 27 Mars years, spanning the full period of spacecraft exploration of Mars, from MY (Mars Year) 8 to MY 34 (~1968-2018), although we discard the first model year (MY 8) to avoid any issues to do with model spinup. Shorter, three-year control runs, spanning either side of the isolated GDS observations of MY -16 (1924), MY -8 (1939) and MY 1 (1956) (see section 6.1), have also been performed and produce similar results.

As previously noted, given our ultimate interest in understanding the origin and evolution of GDS activity on Mars, we choose to use the near-surface wind speed and wind stress as key diagnostic measures, since these quantities have previously been closely linked to the onset of GDS activity (*Basu et al*. 2006; *Newman*, 2002a, 2002b). Wind stress ($\tau$) may be defined as

$$\tau = \rho u_{drag}^2, \qquad (1)$$

where $\rho$ is the atmospheric density and $u_{drag}$ is the near-surface drag velocity. Higher values of surface stress have been, in part, linked to increased dust activity. Figure 2 shows baseline model results for six Mars years, spanning MY 25-30, with the annual insolation cycle indicated in red, and global average horizontal wind speed (top) and global average surface stress (bottom) in black. The surface wind stress levels displayed in Fig. 2 are in good agreement with values obtained from other investigations of this topic (*e.g. Ayoub et al*., 2014).

While globally averaged values are not at the root of what is likely the more localized origin of atmospheric dust on Mars, they provide a reasonable portrayal of the seasonal vigor of the near-surface circulation, and will be a valuable metric when the global-scale CTA are later considered. The interannual variability in the baseline case, as seen in the annually repeating pattern of both black curves in Fig. 2, is relatively small, reflecting the highly repeatable nature of Martian climate system as modeled by MarsWRF. In the baseline case, there are no components that explicitly provide any variability in year-to-year forcing within the model.

As an initial validation of correct baseline model behavior, we see that both surface stresses and winds peak during the perihelion season (southern summer) as expected. This more energetic period corresponds to the dust storm season on Mars (*Zurek and Martin*, 1993; *Shirley*, 2015).

2.3. Diurnal variability

Before moving forward, we must briefly review the topic of diurnal variability, as this will emerge to play a key role in Section 5 below. Mars exhibits strong diurnal cycles in a number of atmospheric parameters, including temperature, surface pressure, and wind speed (*Schofield et al.*, 1997; *Smith et al.*, 2006; *Tyler et al*, 2002; *Toigo et al.*, 2002; *Holstein-Rathlou et al.*, 2010; *Spiga and Lewis*, 2010; *Petrosyan et al.*, 2011; *Chojnacki et al.*, 2011; *Choi and Dundas*, 2011; *Read et al.*, 2015). Strong solar heating of the surface and atmosphere during the day gives rise to atmospheric turbulence and convective activity, and generally stronger surface winds, while relatively rapid cooling during the night leads to more stable atmospheric conditions, changes in wind directions, and somewhat lower wind speeds. These pronounced differences in daytime and nighttime conditions are well-represented in the MarsWRF GCM. We will return to this topic in Section 5.5 below.

3. **Incorporation of orbit-spin coupling within the MarsWRF GCM**

3.1 The coupling term acceleration

We wish to evaluate the orbit-spin coupling mechanism of *Shirley* (2016), to which the reader is referred for a detailed discussion. *Shirley* (2016) derives a mathematical expression for a weak coupling between the orbital and rotational angular momenta of extended bodies that allows an exchange of minute quantities of angular momentum between them. The atmospheres of extended bodies may play a role in this exchange; accordingly, the motions of the atmosphere under investigation must be affected. The coupling term describes an acceleration field that varies with location and with time across the surface of an extended body. The CTA is written in the following form:

$$CTA = -c(\dot{L} \times \omega_\alpha) \times r \quad (2)$$

where $\dot{L}$ (=$dL/dt$) represents the rate of change of Mars' barycentric orbital angular momentum, $\omega_\alpha$ is the angular velocity of Mars' rotation about its spin axis, $r$ is a position vector identifying a location in the Mars body-fixed coordinate system, and $c$ is a scalar coupling coefficient. The parameter $\dot{L}$ varies slowly, on a timescale in part determined by the period of the slow 'dance' made by the Sun around the solar system barycenter (*Shirley*, 2015). The appropriate value of the coefficient $c$ will be estimated in Section 4 below; it is a measure of the efficiency of the coupling between the orbital and rotational angular momentum reservoirs. The value of $c$ is constrained by solar system observations to be quite small (*Shirley*, 2016).

The cross product of $\dot{L}$ with $\omega_\alpha$ (Eq. 2) is a vector that is perpendicular to both, and which lies within the equatorial plane of Mars. Crossing this vector with $r$ yields acceleration vectors that lie approximately within the tangent plane to the surface at all locations, as they are everywhere perpendicular to the (local) planetary radius vector, *r*. A typical spatial pattern of the predicted accelerations for a moment in time is shown in Fig. 3, along with the geometry of the terms of Eq. 2 for a single point on the surface.

From Fig. 3, we observe that this global acceleration field contains two zero-points, or nodes (one is visible in the lower right of the figure), which correspond to surface locations where the cross product of $\dot{L}$ and $\omega_\alpha$ is (anti-) parallel to the radius vector, *r*. These reference points can be employed to advantage for purposes of visualization. The global acceleration field can usefully be visualized from two different perspectives: that of an inertial observer, and that of an observer who is situated at some particular location on the surface of the subject body. An inertial observer, looking down upon the body with the perspective of Fig. 3, would see the planet rotating beneath and "through" the displayed pattern of accelerations, which would remain approximately fixed from this perspective, over the course of a day.

From the perspective of an observer situated on the surface, however, the directional component of the acceleration vector at his or her location will not remain constant over short times, but will, instead, rotate in azimuth, while the magnitude changes, over the course of a day. If the observer is located on the equator, the acceleration will disappear, twice each day, as the planetary rotation carries the observer through the nodal points. For points on the surface near the poles, the diurnal change in magnitude is small, while near the equator itself, the change is more substantial, essentially oscillating between north- and south-pointing once per day. As discussed in *Shirley* (2016), in tropical latitudes, the meridional components of the acceleration are, on average, substantially larger than the zonal components. At high latitudes, however, the zonal and meridional components are roughly equal in magnitude.

In addition to the rapid diurnal evolution seen by the local observer, there is a seasonal-time-scale pulsation to the acceleration field that tracks the magnitude of $\dot{L}$, with the strongest CTA occurring when $\dot{L}$ is greatest (*Shirley and Mischna*, 2016). When $\dot{L}=0$, the CTA become zero everywhere. Further discussion of this source of variability is deferred to Section 3.3 below.

Our prior discussion of atmospheric diurnal variability of Section 2.3 carries certain implications with respect to Fig. 3. There are significant differences between the night side and the sunlit side atmosphere of Mars, with the dayside, for instance, exhibiting stronger convective activity and higher mean wind speeds. For a moment let us suppose that the source of illumination (i.e., the Sun) is shining from the lower left in Fig. 3, directly above the equator of the subject body. We may then visualize the terminator as corresponding to the great circle of longitude that passes through both the rotational poles and the nodes of the illustrated acceleration field. On the sunlit side of the body, the acceleration vectors are directed predominantly southward, while on the night side the vectors are predominantly northward. During the course of a year, as the body orbits the Sun, the illuminated hemisphere of the global acceleration field will slowly complete one revolution around the planet over the period of one Mars year. This means that the relationship between acceleration vectors of a particular orientation, and the dayside or night side is not fixed over time. As we will see, these variable configurations of the coupling term accelerations with respect to the illuminated hemisphere can have significantly different impacts on the global winds, particularly near the surface.

3.2. Incorporation of the CTA within MarsWRF

The design of MarsWRF allows us to implement the CTA code through use of a namelist file, which permits us to turn the CTA on or off at runtime, as well as to prescribe a user-defined value for the coupling coefficient, *c*. (The nature, implications, and determination of *c* are discussed in Section 4 below). Furthermore, it provides the ability to perform model simulations for specific Mars years of our choosing, by incorporating actual $\dot{L}$ values corresponding to the modeled year(s), derived from ephemeris tables.

In order to incorporate the CTA into the system, we have modified the dynamical core of MarsWRF to include augmentations to the *u*- and *v*-wind tendencies (*dU/dt* and *dV/dt*) that are contributed by the CTA. This requires knowledge of both $\dot{L}$ and $\omega_\alpha$. Both quantities are well determined; the former can be obtained from solar system ephemerides, and the latter is an observable quantity. Using formulae and methods as described in *Shirley* (2015) and *Shirley and Mischna* (2016), a table of the J2000 ecliptic Cartesian components of $\dot{L}$ with two-Julian-day resolution from 1920 to 2030 has been obtained. The tabulated values are available by request from the authors.

Certain coordinate transformations are necessary in order to represent the $\dot{L}$ vector in the native body-fixed coordinate system of the MarsWRF GCM. Procedures for these conversions are detailed in Appendix A.

Each GCM grid point (Fig. 3) is represented by a body-fixed vector (*x*,*y*,*z*) having its origin at the center of the planet and defined such that the *z*-axis is parallel to the planetary spin axis, the *x*-axis passes through (0°,0°), and the *y*-axis is defined by the right-hand rule. For each GCM timestep (nominally 180 s), the magnitude and direction of the CTA at each model gridpoint are calculated using the following approach:

1. Query the (*x*,*y*,*z*) components of the $\dot{L}$ vector for the present model time.
2. Calculate the cross product $\dot{L} \times \omega_\alpha$ and transform to the body-fixed system (Appendix A)
3. Calculate the cross product $(\dot{L} \times \omega_\alpha) \times r$, which yields the vector acceleration term, and scale using the coupling efficiency coefficient, *c*.
4. Employ the *u*- and *v*-wind acceleration components of the CTA vector to modify the corresponding wind tendencies accordingly;
5. Repeat for every model gridpoint, and
6. Repeat for all timesteps in the model run.

The CTA magnitudes and directions vary with time, due to Mars' rotation (through $\omega_\alpha$) as well as through temporal changes in $\dot{L}$; thus the acceleration field must be updated for each model timestep.

3.3. Variability with time of the forcing function ($\dot{L}$)

The output from the unforced GCM (Section 2) is 'year agnostic'—absent any year-specific forcing, all years are like every other year (see Fig. 2). According to the physical hypothesis under investigation, atmospheric variability may be introduced through the addition of the CTA. The variability of the CTA is, in turn, principally due to $\dot{L}$, which has therefore been characterized as a "forcing function" (cf. *Shirley and Mischna*, 2016). The variability of $\dot{L}$ is ultimately due to dynamical interactions between the target body (i.e. Mars) and the rest of the solar system (*Shirley*, 2015).

The rate of change of the orbital angular momentum of Mars with respect to the solar system barycenter for six recent Mars years is shown in Fig. 4. As determined in *Shirley* (2015), the mean period of the cyclic variability of Mars' barycentric orbital angular momentum is ~2.2 (Earth) yr, which is significantly longer than the (heliocentric) annual period of 1.88 (Earth) yr. The annual cycle is indicated in red in Fig. 4; peaks in the irradiance correspond to perihelia of the Mars orbit, and troughs to the aphelia. The differences in the periods of these cycles give rise to complex variations with time of the relationships between them. For example, in MY 28, a negative peak in $\dot{L}$ occurs near-simultaneously with aphelion, while in MY 25, a positive peak in $\dot{L}$ aligns with aphelion.

The time period illustrated in Fig. 4 is optimal for purposes of model validation (as performed in Section 5) due to the fortuitous availability of two positive extrema of the $\dot{L}$ waveform during aphelion seasons (MY 25 and 26), two negative extrema of the $\dot{L}$ waveform during aphelion seasons (MY 28 and MY 29), and two zero-crossing cases (MY 27 and MY 30), also occurring close to aphelia. Vertical lines in Fig. 4 illustrate the relative phasing of the $\dot{L}$ waveform with respect to aphelion.

Under the present physical hypothesis, the relative phasing of the two waveforms of Fig. 4 is central to the question of the origins of interannual variability of the Mars atmosphere. The extrema of the green curve are suspected to represent periods of circulatory intensification, while the zero crossings are times when the coupling term accelerations of Equation 2 and Fig. 3 must completely disappear. Before moving forward, we must thus consider the illustrated phase relationships in somewhat greater detail. In the process we will touch upon a number of key open questions that inform the present investigation and represent a focus of attention in later sections of this paper.

3.3.1. "Polarity" of the $\dot{L}$ waveform

As previously described in *Shirley* (2015, 2016) and *Shirley and Mischna* (2016), Mars is gaining orbital angular momentum, at the expense of other members of the solar system family, when $\dot{L}$ is positive. Mars is correspondingly yielding orbital angular momentum during intervals when $\dot{L}$ is negative in Fig. 4. For ease of reference, periods when $\dot{L}$ is positive will be termed "positive polarity" intervals, while the term "negative polarity" refers to intervals when $\dot{L}$ is below the zero line. Table 1 lists the assigned polarity values for the aphelion season dates of Fig. 4. Periods when the $\dot{L}$ waveform is near the zero line, as in MY 27 and MY 30 of Fig. 4, are labeled "transitional" intervals. These labels will later allow us to efficiently characterize and distinguish between effects produced by the CTA under these fundamentally different forcing conditions.

3.3.2. Relative phasing of the $\dot{L}$ and solar irradiance waveforms

An angular measure of the relative phasing of the insolation cycle and $\dot{L}$ waveforms is provided in the middle column of Table 1. We take the same approach as in *Shirley and Mischna* (2016), characterizing each Mars year by the phase ($\phi$) of the $\dot{L}$ curve for the seasonal interval under investigation. In this scheme, $\phi=0°$ and $\phi=180°$ correspond to the upward and downward zero-crossing (transitional) points of the $\dot{L}$ curve, respectively, and $\phi=90°$ and $\phi=270°$ correspond to the positive and negative extrema of $\dot{L}$. Thus, a phase value of 122° *at aphelion*, (for MY 25, see Table 1), describes a point on the $\dot{L}$ waveform located sometime after its positive peak has occurred (at $\phi=90°$; see Fig. 4). Later, when we address the dust storm season in more detail, the relative phasing of the $\dot{L}$ waveform and the annual irradiance cycle will be characterized using the $\dot{L}$ waveform phase at the time of *perihelion*.

The prior results of *Shirley and Mischna* (2016), employing the phase assignments and polarity designations as described here, provide strong circumstantial evidence in support the physical reality of the orbit-spin coupling mechanism envisioned. All of the nine known global dust storm years from the historic record exhibit phase values near positive and negative extrema of the $\dot{L}$ waveform during the dust storm season (cf. *Shirley and Mischna*, 2016, Fig. 6). However, an important open question remaining from the prior analysis concerns the differences in the frequency of global-scale dust storms in positive polarity cases (7 of 8 occurrences) versus negative polarity cases (2 of 6 occurrences). It is not yet obvious why GDS more regularly occur in positive polarity episodes than in negative polarity episodes, despite generally similar absolute magnitudes of $\dot{L}$. The question of how the atmospheric response may differ under positive polarity and negative polarity conditions is a key topic of Sections 5 and 6 below.

The angular phase parameter as defined above and in *Shirley and Mischna* (2016) is a relatively crude metric for characterizing the dynamical variability displayed in Fig. 4. Numerical modeling (as performed here) potentially offers far greater insight into the suspected relationships between the putative forcing function and the atmospheric response.

3.4. An explicit statement of the hypothesis under investigation

Under the physical hypothesis outlined in *Shirley* (2016), intervals of circulatory intensification are predicted to occur when the $\dot{L}$ waveform is near (positive and negative) extrema, while the zero crossings of this waveform are times when the CTA of Eq. 2 and Fig. 3 must completely disappear. Our investigation is designed to validate or disqualify the above statement, for the case of the Mars atmosphere, through the incorporation and modeling of the dynamically determined orbit-spin coupling accelerations within the MarsWRF GCM.

We must here caution against a too-literal interpretation of the term 'intensification.' The accelerations are not uniform everywhere, but instead exhibit substantial variability as a function of latitude, longitude, and time (Fig. 3, and *Shirley*, 2016, Fig. 6). While it is possible that a linear or monotonic "speeding up" of some pre-existing circulatory flow might occasionally occur in response to the CTA, it is more likely that the adjustment of the atmosphere will take the form of structural or morphological changes, including modified patterns of large scale flows. Given the complexity of atmospheric global circulation models, we expect that simple linear relationships of atmospheric observables to altered forcing levels are likely to be the exception, rather than the rule.

**4. Preliminary model validation and calibration: Constraining *c***

4.1. On *c*

The leading coefficient, *c,* of the coupling term (Eq. 2) acts as a scalar multiplier for the accelerations determined by the triple product of that term. If *c*=0, then no accelerations will arise, and the GCM will behave identically to the unforced model simulations as described in Section 2. If *c* is too large, unrealistically high winds will be produced within the model, and any possible correspondence with reality of the modified GCM output will be lost. From this perspective, *c* may be viewed as a tunable parameter that may be employed for optimizing the correspondence of model output with observations. One should be rightly skeptical of selectable threshold values or other "fudge factors" that may arbitrarily bring model outcomes into better agreement with observations, as physical realism is likely to be lost. In this case, however, we, instead, consider *c* not as an arbitrary "fudge factor", but rather as an important inherent physical property of a system, whose actual value may be iteratively constrained by means of experiment.

In *Shirley* (2016) the coupling efficiency coefficient, *c,* is compared analogously with the coefficient of friction, $\mu$, with which it shares many characteristics in common. The coefficient of friction is a fundamental property of physical systems, but its value is not easily obtained analytically; instead we mainly employ empirically determined values for engineering applications. Friction may result from multiple causes, at a variety of scales, and, with considerable effort, we may be able to disentangle and quantify the key physical interactions that lead to it. A coupling of the orbital and rotational angular momenta of extended bodies is likely to involve multiple responses and interactions of the co-rotating physical systems of which they are comprised; atmosphere, oceans, and crust may participate in different ways. It will undoubtedly require a considerable effort in order to quantify these interactions in a meaningful way. Given our present state of knowledge, it more practical and more desirable to initially characterize the total system response with the coefficient $c$.

From a quantitative standpoint, the coefficient $c$ can be interpreted as the fractional portion of the orbital angular momentum that may participate in the excitation of atmospheric motions. As noted in *Shirley* (2016), solar system observations constrain this value to be quite small; for this reason, the mechanism is characterized as a "weak" coupling.

4.2. Determination of $c$

Starting with the modified MarsWRF model, we have performed a series of iterative tests, varying the coupling coefficient from $10^{-16}$ to $10^{-9}$. We were initially concerned that the driven models might exhibit pathological behaviors, including runaway cases. There is no *a priori* guarantee that our modifications to the algorithms should necessarily provide stable solutions, given the non-linear nature of atmospheric processes and interactions in a GCM. However, in general, the parameterized (frictional) damping mechanisms already included within the MarsWRF GCM were found to be sufficient to maintain the system within reasonable limits. We nonetheless continue to remain alert to any unanticipated consequences of the addition of CTA to the model.

Figure 5 shows, in green, the globally averaged near surface wind speeds obtained for three coupling coefficients: $10^{-13}$, $5 \times 10^{-13}$ and $10^{-12}$, for the MY 25-30 period. Overlaid in black are the globally averaged winds for the baseline case, as previously illustrated in Fig. 2. When setting $c=10^{-13}$ we see only very small differences between the black and green curves—for this value of $c$, the coupling term accelerations are evidently too small to have an appreciable impact on the circulation. Similar results (not shown) are obtained for all values of $c$ smaller than this. Conversely, for $c=10^{-12}$ we find significantly higher values of globally averaged near-surface winds. The winds are in some cases nearly twice as strong as in the baseline model, which is inconsistent with observations. Thus we conclude that $c$ must necessarily be less than $10^{-12}$ in order for our model output to be consistent with observations.

The center panel of Fig. 5 shows what appears to be a 'Goldilocks solution', with $c=5 \times 10^{-13}$ It shows a modest influence of the CTA, and does not produce unrealistically high wind speeds. We now further evaluate the model solutions employing this value of $c$ through a comparison of forced and unforced surface wind stress values.

In Fig. 6 we plot MarsWRF model-derived global mean daytime surface wind stress under the influence of the CTA, in blue, comparing it to the baseline case in black. The difference between the two is shown in orange. The instantaneous magnitude of $\dot{L}$ (arbitrary units, but properly spanning the zero line), representing the driving function for the CTA, is shown in green. Also, for comparison, the annual irradiance cycle (arbitrary units) is shown in red. In agreement with expectations, the stress differences are small near the zero-crossing times of the $\dot{L}$ waveform. Elsewhere, surface stresses are consistently larger with the CTA than without, and the magnitude of the difference (in orange) correlates reasonably well with the magnitude of $\dot{L}$, regardless of its sign. In the most dramatic example, for the perihelion season of MY 25, the magnitude of $\dot{L}$ is comparatively large relative to the other years, and it is at this time that we see the greatest increase in surface stress. During the remaining years, having smaller peak values of $\dot{L}$, there is less difference between the blue and black curves, but the pattern of the CTA influence remains consistent.

Table 2 compares the global mean daytime wind stress magnitudes at aphelion for MY 25-30, obtained with $c=5\times10^{-13}$, with the baseline model stress values for the same years. The largest difference obtained amounts to a little over 20%. (As previously noted, we preferentially employ aphelion season values for the tests of this Section and in Section 5, due to the relative stability and repeatability of the atmospheric circulation at this season). We conclude, on the basis of the comparisons of Fig. 6 and Table 2, that a value of the coupling efficiency coefficient $c$ of $5\times10^{-13}$ yields an appropriately detectable (but not overwhelming) level of added atmospheric acceleration, as desired for purposes of the present investigation. This value of $c$ has accordingly been employed for all subsequent model runs including the CTA.

At this point it is appropriate to briefly address the question of the actual levels of CTA accelerations applied within the GCM. The largest positive value of $\dot{L}$ attained during the interval 1920-2030 occurred in the year 1982 (MY 15, a global dust storm year; *Shirley and Mischna*, 2016). Using the adopted value of $c=5\times10^{-13}$ we obtain a maximum value of the coupling term acceleration of $2.227\times10^{-4}$ ms$^{-2}$ for this episode. Peak acceleration values for most other years are considerably smaller.

Finally, as cautioned previously, the value of $c$ as determined here is preliminary only, and specific only to the present implementation of the present model for the present planet. We should expect to revise this value as more sophisticated model simulations, notably those including dust, are performed.

## 5. Further model validation and initial results for the aphelion season

With the CTA included in the model, and a suitable value of the coupling coefficient identified, we may begin to evaluate the hypothesis set out in Section 3.4. Our hypothesis states that temporal variability of the CTA induces a periodic intensification and relaxation of atmospheric circulation. Specifically, we expect that during the positive and negative extrema of the $\dot{L}$ cycle, the global circulation will intensify, while during periods when $\dot{L}$ approaches zero, relaxation towards the 'baseline' state will occur. Such behavior may be expressed within our global circulation model in a number of ways. We will compare forced and unforced simulations of large-scale wind fields and time-averaged surface stresses.

To this point, we have largely assessed the behavior of the model over the course of the entire year to identify general trends, and have done a preliminary survey of model behavior during the aphelion period. We will continue to focus on this particular period within the annual dust cycle to evaluate performance under different atmospheric forcing conditions. Mars reaches aphelion in the northern spring season, at $\sim L_s=71°$. Within this section, we will be time averaging over the interval from $L_s=70°$-$90°$ to capture behavior across this season. By first turning our attention to the aphelion season, we can minimize the influence of several factors in the climate system, and isolate the effects of the CTA without being overly concerned with potential complications due to behaviors such as atmospheric dust feedback.

We continue to consider the six-Mars-year period between MY 25 and MY 30 (Figs. 4-6) because it exhibits striking differences in the phasing of $\dot{L}$ with the insolation curve, and hence with respect to the aphelion season. Referring back to Figure 4, in MYs 25 and 26, aphelion lines up reasonably closely with the positive polarity peak in $\dot{L}$, which occurs slightly before aphelion in MY 25, and slightly after in MY 26. In MY 27, aphelion corresponds closely to the transitional period of the $\dot{L}$ curve. Stepping forward through time, a similar pattern is repeated, but with opposite phasing of $\dot{L}$. Our hypothesis indicates that we should expect to see the greatest enhancement to aphelion circulation when the $\dot{L}$ waveform is near extrema, i.e., in MY 25, 26, 28, and 29, as compared to the transitional periods in MY 27 and MY 30. We also wish to determine whether the circulatory intensification in positive polarity conditions (MY 25 and 26) is similar to, or different from, that in negative polarity conditions (MY 28 and 29). We will consider, in turn, the transitional cases, the positive polarity cases, and the negative polarity cases.

5.1. 'Transitional' episodes (MY 27 and 30)

Figure 7 shows global zonal and meridional wind fields and time-averaged daytime surface stresses for the aphelion season of MY 27. Baseline model results ($c=0$) are displayed in the left hand column, while results from model runs including the CTA are shown in the center column of the figure. Differences between the forced and unforced model outcomes (forced minus baseline) are shown in the right column. We only show results for MY 27 in this section, as very similar plots are obtained for the MY 30 case. (The MY 30 version of Fig. 7 is provided in the online Supplementary Data, Figure S1).

Zonal winds during the aphelion season are dominated by the strong westerly flow in the southern hemisphere, which is strengthening as the winter season approaches (Fig. 7a). Much weaker seasonal westerly flow is also seen in the northern hemisphere at lower altitudes in Fig. 7a, while the equatorial regions exhibit weak easterly flow near the surface, which strengthens at higher altitudes (blue colors). Notably, the large-scale morphology of the zonal flows appears to be virtually identical for the forced and unforced model results (center and left column, respectively). The minor zonal wind speed differences revealed in Fig. 7c range between $\pm 6$ ms$^{-1}$.

The middle row of Fig. 7 illustrates the nature of the aphelion season circulation via the mass streamfunction. The meridional streamflow plots of the forced and unforced cases depict nearly identical large-scale flows. These are dominated by a counterclockwise circulation near the surface (in green). Air rises in the north, in the spring hemisphere, and sinks in the south. The light brown features in the right hand, 'difference', panel (Fig. 7f, at right) show only very minor differences in the clockwise circulations present over the northern tropics and northern mid-latitudes.

The bottom three panels of Fig. 7 (Figs. 7g-i) are contour plots of daytime surface wind stresses measured in N m$^{-2}$. As with the illustrations above, the differences between the forced and unforced model results are quite small (as shown in the difference map of Fig. 7i). As noted earlier in Section 2, there are substantial differences in atmospheric conditions (including wind speeds and directions) between the dayside and night side. We have chosen to illustrate only the dayside values of surface wind stress, as we believe these to be more representative of the time-of-day conditions leading to dust lifting and mixing. Partitioning between day and night is done by assessing the solar zenith angle at every model grid point. Values <90° indicate the Sun is above the horizon, which is classified as 'day'. Values >90° are classified as 'night'. In Figure 7g-i, and all subsequent figures of this style, polar latitudes having homogeneous shading correspond to regions in polar night for which no daytime surface stresses (or differences) can be calculated (e.g. poleward of ~60° S in Figure 7g-i).

These results are fully consistent with our expectations, as, during a transitional period of the $\dot{L}$ curve, the CTA approach zero, and the system relaxes towards the baseline case. Irrespective of the choice of $c$, the small value of $\dot{L}$ around zero crossing results in small CTA during the aphelion season of MY 27.

The general agreement of the forced and unforced model results of Fig. 7 documents an important feature of the atmospheric response to the coupling term accelerations. This agreement indicates that there is very little retention or "carryover" of the momentum added to (or subtracted from) the atmosphere across the transitional intervals in our model. The atmospheric "memory" of the excitation is short-lived; an efficient damping of the modified wind velocity components is implied. In other words, the MY 27 aphelion season retains no memory of the potential strengthening of the circulation from the MY 26 positive polarity season, which we address presently.

5.2. Positive polarity episodes (MY 25 and 26)

Figure 8 shows the same model fields as Fig. 7, but for the aphelion season of MY 25. During this year, aphelion occurs near a positive peak in the $\dot{L}$ curve. The differences obtained between the baseline and CTA cases must be solely due to the CTA. While the overall morphology of the zonal mean zonal winds in the baseline and forced model panels is quite similar, the difference panel (Fig. 8c) reveals key differences. The addition of the CTA clearly impacts the zonal circulation, weakening the equatorward edge of the winter jet in the southern hemisphere, and strengthening higher level winds in the northern hemisphere. The coherence in the pattern displayed in the difference panel suggests this is not an arbitrary change to zonal circulation, nor is it negligible, with changes to the zonal wind in excess of 10 ms$^{-1}$.

Meridional flows for the aphelion season of MY 25 are illustrated in the center row of Fig. 8. Recall (from Fig. 3) that the CTA are oriented largely northward/southward in the mid-latitudes. It is thus reasonable to expect that the overturning meridional circulation will be modified with respect to the baseline case due to the inclusion of the CTA and, indeed, this is what is observed.

Figure 8f shows the difference in streamfunction in MY 25 with and without the CTA. The CTA here appear to reduce the strength of the counter-clockwise circulation in the tropics, as represented by the brown region in the lower latitudes extending to the surface. This opposes the orientation of the unforced model bulk flow (in green, in Fig. 8d). In the mid-latitudes of each hemisphere of Fig. 8f, two smaller blue-green regions are present, indicating weak counterclockwise flow enhancements. In the south, this appears to intensify the pre-existing counterclockwise motion at higher altitudes.

Furthermore, by mapping the forced and unforced model surface wind stresses (Figure 8g-i), it is clear that surface stresses are enhanced when the CTA are non-zero. The enhanced wind stress with the inclusion of the CTA points to stronger winds as the mechanism increasing the stress (there is a negligible change in near-surface density). Locations of peak enhancement of the surface stress, shown in the warmest colors of Figure 8i, tend to coincide with the latitudes of greatest streamfunction enhancement in Figure 8f. So, although the instantaneous magnitude of the CTA is extremely small, it nonetheless evidently contributes to changes in wind speeds and, therefore, circulation.

Illustrations of wind fields and stress maps for MY 26 are relegated to the Supplementary Data, Figure S2, as the plots are similar in all important respects to those of Fig. 8.

5.3. Negative polarity episodes (MY 28 and MY 29)

Large-scale wind fields and surface stress maps for the negative polarity aphelion season of MY 28 are shown in Fig. 9 (and MY 29 in Fig. S3), which shares an identical format with Figs. 7 and 8. The difference plots, in the right hand column of Fig. 9, exhibit a number of interesting features in comparison with the corresponding plots of the positive polarity episode (Fig. 8). The zonal wind speed difference plots (Figs. 8c and 9c) are strikingly different. In MY 28 the prominent westerly flow in southern mid- to high latitudes is weakened, relative to the baseline case (note the cool colors in Fig. 9c for these locations), whereas in MY 25 (Fig. 8c) this circulation is strengthened. Oppositely directed changes in wind velocities are also observed at most altitudes for latitudes between 45° S and the equator, and for high altitudes from 45° S to 45° N.

The mass streamfunction difference for the aphelion season in MY 28 (Fig. 9f) mainly indicates an *enhancement* of the normal seasonal overturning circulation in the tropics, with (green) counterclockwise contours seen in Fig. 9f at nearly the same locations as in Fig. 9e. This situation is quite different from that depicted for the positive polarity case in Fig. 8f.

We additionally see quite interesting changes in the plot of surface stress differences in Fig. 9i. Overall, a greater fraction of the lower and mid-latitudes experience enhanced surface stresses as compared to the positive polarity case (Fig. 8i). Furthermore, there appear more localized areas of strong surface stress enhancement in the higher southern latitudes in the negative polarity case, which are absent in the positive polarity case.

5.4. Summary of MarsWRF model validation activities

Figures 7, 8, and 9 illustrate clear differences in the response of the global atmospheric circulation for the transitional forcing episodes (Fig. 7), the positive polarity forcing episodes (Fig. 8), and the negative polarity forcing episodes (Fig. 9). With reference to the physical hypothesis under investigation, we may conclude on the basis of these comparisons that the predicted intensification of circulatory flows at times of positive and negative extrema of the forcing function (i.e., the $\dot{L}$ waveform) is confirmed by these experiments. We can likewise confirm that a relaxation of circulatory flows, converging toward the unforced baseline model results, is replicated within the GCM outcomes, for the transitional (zero-crossing) episodes (Fig. 7).

The introduction of the CTA within the GCM does not appear to give rise to pathological consequences that would suggest fundamental problems with this implementation. The adopted value of the coupling efficiency coefficient ($c=5\times10^{-13}$) gives rise to moderate increases in the "intensity" of the large scale circulation and in the peak values of the surface wind velocities and surface wind stresses (Table 2). The differences in model outcomes for the transitional, positive polarity, and negative polarity conditions introduce a level of atmospheric interannual variability within the modified GCM that is not seen in the baseline model.

However, while we have observed clear differences in the modeled outcomes for the positive and negative polarity episodes during the aphelion season, the fundamental causes of these differences in the large-scale circulation remain somewhat obscure. Significantly different outcomes for dust storm seasons with positive and negative polarities were also noted in *Shirley and Mischna* (2016). In the following subsection we address the question of the origins of these differences in somewhat greater detail.

5.5. Differences in GCM outcomes linked with polarity differences

In connection with our description of the global pattern of the CTA displayed in Fig. 3, we noted that, in the daytime hemisphere, the global acceleration field would slowly change its orientation, completing one cycle over the period of one Mars year. We, here, explore the possibility that this source of variability, in conjunction with the differences of the dayside and night side circulations as noted in Section 2.3, may together explain some portion of the differences noted between positive polarity and negative polarity episodes. As previously noted, the daytime circulation is relatively vigorous, while the night side circulation is less energetic and generally stably stratified.

The insights gained from Figs. 7-9 suggest the following hypothesis, which may easily be tested with the data in hand. We now ask: Could the CTA interfere, constructively or destructively, with the normal seasonal patterns of near-surface winds, in such a way as to effect a modulation of the large-scale meridional flows? In MY 25 (Fig. 8f), we observed a reduction in the overturning circulation during a positive polarity episode, while intensified meridional flows were indicated for the negative polarity episode of MY 28 (Fig. 9f).

All of our prior illustrations of the meridional overturning circulation (Figs. 1 and 7-9d-f) have been obtained from full-day averaging over intervals of 20° of $L_s$. On the other hand, to this point, we have employed only the *dayside* surface wind speeds and surface wind stresses for our investigations (Figs. 7-9g-i). There are two reasons for this. First, as noted previously in Section 2.3, the more vigorous dayside winds are more likely to be directly relevant to processes that lead to dust lifting. A second reason becomes apparent from an inspection of Fig. 3. This indicates that the accelerations applied at antipodally positioned locations are opposed in direction. Diurnal averages could thus include substantial cancellation, and thus be much less suitable for assessing near-surface wind speeds relevant to the dust lifting problem. To adequately address the current question, it is clear that we must expand our focus to consider the night side wind values as well as the dayside contributions.

This separation is accomplished and displayed in Fig. 10. Partitioning the full Martian sol into day and night halves shows differing trends of near-surface wind speeds between the two. The partitioning between day and night is done as described in Section 5.1. In Figures 10-13, polar locations with zero wind speed or difference correspond to regions either in perpetual daytime (or nighttime) for which no calculations were made, depending on the exclusively dayside or night side identity of the specific panel.

The top row of Fig. 10 shows mean near-surface winds during daytime hours at aphelion at each grid point for the positive polarity case (MY 25). As before, panels on the left represent the baseline cases, the center panels illustrate the cases with the CTA, and the panels on the right illustrate the differences. The wind speed differences between the baseline and CTA cases must be strictly due to the presence of the CTA. The illustrations in the lower row (Figs. 10d-f) show the identical fields, but averaged over the nighttime hours. To more easily see the spatial distribution of the differences in the north/south component (only) of the wind, color contours have been superimposed. Contiguous sets of grid points with northward-directed differences are shaded in green, while areas with southward flow enhancements are shown with brown tones. This color scheme corresponds to that adopted in Figs. 7-9, where the normal (northward) near-surface seasonal meridional flow at aphelion was identified with the counterclockwise circulation, and colored green.

Examination of the difference panels of Fig. 10 reveals that the meridional components seen in the difference plot have a predominantly southward trend during the daytime (areas shaded in brown), and a northward trend at night. These differences are consistent with the progression of the CTA vectors over the course of the day.

For Fig. 11 we have zonally averaged the meridional components (only) of the difference plot vectors of Fig. 10 for all grid points. Figure 11 demonstrates that the daytime contribution of the CTA as combined with the normal seasonal meridional flow "outweighs" the opposing nighttime contribution, leading to a net diurnal mean southward flow (heavy solid line of Fig. 11). This is what likely leads to the overall retardation of the "normal" seasonal meridional circulation as seen in the southern mid-latitudes of Fig. 8f.

For MY 28, having an opposite polarity to MY 25, the behavior follows very nearly the same overall pattern, but with a complete opposite phasing to the day and night

CTA contributions, as seen in Fig. 12 (which may be compared with Fig. 10). As before, the combination of normal seasonal flow with the daytime CTA contribution is stronger than the resultant nighttime combination (Fig. 13), leading, in MY 28, to a diurnal mean contribution of the CTA of northward flow. This is what likely enhances the normal seasonal meridional circulation in MY 28 (Fig. 9f) compared to the reduced circulation found in MY 25 (Fig. 8f).

The CTA clearly act to "reinforce" the normal seasonal daytime near-surface circulatory flow pattern in MY 28. We find an enhancement in the large-scale meridional circulation (Fig. 9f) and in near-surface winds and surface stress (Figure 9i). At night, the resultant effect of the acceleration field is to oppose the mean seasonal overturning circulation, but the winds are generally weaker, and do not fully compensate for the daytime enhancement, yielding a net diurnal enhancement to the overturning circulation. In contrast, for MY 25, a positive polarity episode, the daytime CTA are oriented in a manner opposed to the normal near-surface meridional circulation for the aphelion season, giving rise to a small retardation in the dayside near-surface winds. The 'boost' (i.e., constructive interference with the normal seasonal circulation pattern) occurs at nighttime, instead, when winds are weaker. This asymmetry in the day/night influence of the CTA on near-surface winds leads to a smaller day/night range of global average surface stress in MY 25 than in MY 28, while the diurnal average remains approximately the same (Table 3).

The circulatory intensification predicted under our physical hypothesis may lead to quite different large-scale outcomes for the positive and negative polarity cases, depending both upon the waveform polarity and on the phasing of the annual and diurnal patterns of solar illumination of Mars. The results presented in Figs. 10-13 indicate that the different model outcomes obtained for the positive and negative polarity examples of Figs. 8 and 9 most likely result from a complex interplay between the diurnal illumination cycle and the polarity of the $\dot{L}$ waveform (which controls the orientation of the acceleration field) for the specific period under investigation.

## 6. Perihelion-season global circulation modeling and comparisons with observations

We now turn our attention to the perihelion season. Mars' perihelion occurs at $L_s$ ~251°, during the southern spring season on Mars. The dust storm season on Mars is roughly centered on the time of perihelion (*Zurek and Martin*, 1993; *Shirley*, 2015), and at least four global-scale dust storms (MY 1, 9, 21, 28) have occurred in close temporal proximity to perihelia.

6.1. Observations

Of the 24 past Mars years that have elapsed since the start of spacecraft observations in MY 9, we have information concerning the occurrence or non-occurrence of global-scale dust storms in 18 (Table 4). The gaps in our knowledge during the remaining 6 Mars years (MY 13, 14, 16, 19, 20, and 22; *Zurek and Martin*, 1993; *Shirley*, 2015) are due to either a lack of spacecraft observations, or poor telescopic viewing conditions from Earth. The record is continuous beginning in MY 23 (1998).

The historic record of global-scale dust storms, although incomplete, dates as far back as 1924 (MY -16; *Zurek and Martin*, 1993; *Shirley*, 2015). We have performed multi-year MarsWRF simulations covering all of the Mars years listed in Table 4.

The inception dates for the 10 known historic GDS of Table 4 are provided in the third column of Table 4. The 1977 storm year (MY 12) is unusual, in that two separate GDS were observed. It is not known whether changes in the atmosphere due to the first storm played a role in the genesis of the second; for this reason, we consider only the initial storm in the analyses below.

We are interested in the atmospheric conditions prevailing during the times immediately prior to the inception dates of the GDS. As in Section 5, the time intervals examined will span a range of 20° of $L_s$; however, for this portion of the investigation, the intervals selected for analysis must accordingly differ from storm to storm. The selected "pre-storm intervals" are identified in the last column of Table 4. As in Section 5, we will compare forced and unforced model outcomes for identical intervals.

The remaining 12 global-storm-free years of Table 4 are analyzed and compared using a "standard" interval consisting of the time period from $L_s$ 250°-270°, extending approximately from Mars' perihelion to the southern summer solstice. The solar heating of the Mars atmosphere is maximized during this period.

For brevity, we will not display graphical model outcomes individually for all of the 21 Mars years under investigation. Instead we have selected representative examples from each polarity category for discussion purposes. In the following, we will make reference to "positive polarity years", "negative polarity years," and "transitional polarity years," as identified in column 5 of Table 4. As indicated in the Table, all such identifications now refer to the state of polarity of the $\dot{L}$ waveform at the time of Mars' *perihelion*. A noteworthy feature of Table 4 is that *all* of the perihelion-season GDS (indicated by the orange shading) occurred under conditions of positive $\dot{L}$ waveform polarity during the second half of the Mars year, i.e., during the dust storm season. The equinox-season GDS (MY 12 and 25, shaded in blue in Table 4) instead exhibit negative $\dot{L}$ waveform polarity during the dust storm season.

6.2. Positive Polarity Years

Mars year 28 was chosen as an 'archetypal' example of a positive polarity year. The phase value at perihelion (Table 4) is 82.4°, indicating that a positive peak (90°) value of the forcing function occurs very soon after perihelion (cf. Figs. 4 and 6). A global-scale dust storm occurred in this year; the storm was initiated at $L_s$=262°. As we are concerned with the atmospheric conditions for times immediately prior to the time of GDS initiation, our "pre-storm" time interval of interest for MY 28 spans the period $L_s$=240-260°. Model output for this case is displayed in Fig. 14, which follows the same layout as Figs. 7-9. Plots for the other positive polarity years of Table 4 are generally similar to those of Fig. 14; these are available for inspection in the online Supplementary Data, Figures S4-S10.

The zonal winds and streamfunction plots of Fig. 14 are quite different from those of the aphelion season simulations of Figs. 7-9. The strongest westerly zonal winds (Fig. 14a and 14b, in red) are now found in the northern (late fall) hemisphere. The streamfunction plots (Figs. 14d and 14e) are now dominated by a strong *clockwise* overturning meridional (Hadley) circulation, in brown colors, with air rising in the southern hemisphere and sinking in the north. The overall wind field is more vigorous, as expected, at perihelion; the scale bars displayed cover a somewhat wider range of values than was the case for the aphelion season panels.

Regarding the zonal wind differences, examination of the CTA influence on the zonal mean zonal wind shows relatively small changes below about 1 Pa. The easterly winds above much of the southern hemisphere (in cool colors) have weakened (showing warmer colors in Fig. 14c), while at the same time the strong westerly flow in the northern hemisphere has diminished slightly (note the color values bracketing the zero value in the color bar supplied).

Notable differences are found between the baseline and forced model streamfunction plots (Figs. 14d and 14e). We see a strong enhancement of the flow resulting from the addition of the CTA, as marked by the two brown regions in the mid-latitudes of Fig. 14f which have the same orientation and locations as the brown clockwise cell seen in the large-scale flow. An enhanced southward surface flow between the equator and ~45° S here feeds the uplifting branch of the Hadley circulation. Both of these factors could conceivably facilitate dust lifting from the surface, and the entrainment and transport of dust from lower to higher altitudes in the atmosphere of Mars.

The daytime surface wind stress difference map (Fig. 14i) is also quite interesting. Figure 14i shows a broad swath of the Martian tropics and subtropics experiencing increased surface wind stresses. On the basis of the discussion in Section 5.5, this distribution can plausibly be attributed to an increase in near-surface wind speeds in these latitudes due to the CTA augmentation. The streamflow enhancement indicated in Fig. 14f indicates that an enhancement of southward winds may be responsible for the higher surface wind stresses seen in Fig. 14i. While it is premature to attempt a detailed comparison with observations, given the limitations of our dust-free simulations, it is still of interest to note that areas with increased surface wind stresses in Fig. 14i include Chryse, Margaritifer Terra, and Noachis, all of which were active sites of dust lifting prior to and during the earliest stages of the MY 28 GDS (*Wang and Richardson*, 2015).

We quantify the difference in daytime surface wind stresses between the baseline case and the forced model case using the global mean daytime values for this parameter. The baseline global mean daytime surface wind stress value (Fig. 14g) is 0.00643 N m$^{-2}$. We obtain a value of 0.00771 N m$^{-2}$ for the wind stresses of Fig. 14h, which is ~20% larger. We will tabulate and compare model-derived daytime surface wind stress values for all 21 Mars years of Table 5 in Section 7 below.

6.2.1. Meridional (streamflow) differences for positive polarity years

Changes in streamfunction due to CTA influences, similar to those of Fig. 14f, are seen in the other seven 'positive phase' years as well (Fig. 15). The regions of strongest enhancement occur in the southern hemisphere, at the latitudes of the vertical branches of the circulation, and in these years, that enhancement continues along near the surface, through the subtropics and tropics. Slight weakening of the circulation (flow changes of a counterclockwise sense, shown in green tones) is seen just north of the equator, centered at ~10 km above the surface, in all years.

In seven of these eight years, a GDS was observed (no GDS was seen in MY 27). *All* of the perihelion-season GDS of Table 4 are included here. It is notable that the Hadley cell intensification in the one exceptional case (MY 27) is among the weakest of all positive polarity cases. Further discussion of the distinctive features of MY 27 is deferred to Section 8.4.

Figure 15 reveals a remarkable consistency in modeled atmospheric behavior across all the positive polarity years. The present investigation has uncovered a common factor linking Mars' perihelion-season global-scale dust storms one with another. This is a key finding—the net effect of the CTA is evidently to enhance, or intensify, the global meridional overturning circulation in the southern spring and summer seasons of these positive polarity years.

6.3 Transitional Polarity Years

The 'transitional' category of Mars years of Table 4 corresponds to years in which the $\dot{L}$ and irradiance curves are approximately in quadrature at perihelion, and consequently these are years in which the magnitude of the CTA approach zero at some point during the dust storm season. As discussed in *Shirley and Mischna* (2016), there are seven such years in the historical record (MY 17, 18, 23, 26, 29, 30, and 32). In these years, the circulation is minimally changed between the baseline and CTA-included cases. In none of these years was a GDS observed. For these reasons we have not included a representative composite plot such as that of Fig. 14 for these years (such plots are however included in the online Supplementary Data, Figures S11-S17). Certain features of the streamflow difference plots for these years are instructive, however, and so we include these in Fig. 16 below.

In four of the seven examples of Fig. 16 (MY 18, 23, 30, and 32) we see virtually no difference between the CTA model outcomes and the baseline model runs. The phase parameter of Table 4 is found within < 20° of the "null CTA" (0° or 180°) phase values for these years. The streamflow differences of Fig. 16 for the other years (MY 17, 26, and 29) serve to illustrate gradational transitions of circulatory intensification (between zero-crossing and positive polarity cases, and between zero-crossing and negative polarity cases). The streamflow difference plots for MY 17 and 29 (with phases of 39° and 38° respectively) show emerging features similar to those of the positive polarity years of Fig. 15. As we will see, the remaining case (MY 26, with φ=213°) has features in common with the negative polarity years to be discussed next.

As with the aphelion season examples of Fig. 7, the relaxation of the atmospheric circulation to unforced (baseline) conditions in the MarsWRF GCM in transitional polarity years is consistent with expectations based on the physical hypothesis introduced in *Shirley* (2016). Further discussion is provided below in Section 8.

6.4. Negative Polarity Years

Four of the six negative polarity Mars years of Table 4 lack GDS (MY -8, 11, 24, and 31), while the two others (MY 12 and 25) are equinox-season global-scale storm years (*Shirley and Mischna*, 2016). We separately consider these two subsets in the following sections. Additional negative polarity illustrations are found in the online Supplementary Data, Figures S18-S20.

6.4.1 Global-storm-free negative polarity years

Mars year 24 is representative of this type of Mars year. MarsWRF model output for the perihelion season ($L_s$=250-270°) of this year is illustrated in Fig. 17. We will mainly compare this figure with that for MY 28 (Fig. 14).

The model-derived zonal mean zonal wind behavior in MY 24 is not greatly dissimilar to that of MY 28 (Fig. 14). In the southern mid latitudes of Fig. 17c. the CTA are associated with a weakening of the (easterly) zonal winds between about 0.1-5 Pa, as indicated by the warmer colors in the difference plot. This weakening is more pronounced than in years with positive polarity (compare Fig. 14c). In addition, the winter jet at high altitudes in the northern hemisphere is more extensively weakened.

The modification of the streamfunction due to the coupling term accelerations during this negative polarity 'archetype' year is quite different from that of Fig. 14. Whereas in MY 28, the equatorial latitudes were dominated by a region of meridional circulation enhancement (Fig. 14f), here we find a deep central region of weakening of the normal seasonal flow (in green). Bracketing this are two regions of enhanced clockwise circulation (in brown) that are shifted away from the southern tropical latitudes. This pattern is found consistently (at perihelion) in each of the storm-free negative polarity years, to varying extent, as illustrated in Fig. 18.

A comparison of the surface stress difference plots of Figs. 14i and 17i shows the regions of greatest surface stress enhancement pushed to higher northern latitudes in the negative polarity years as compared to the positive polarity years. In the positive polarity MY 28, there is a belt of increased surface stresses largely localized around the equator, whereas in the negative polarity MY 24, the strength of the enhancement in the tropics is much diminished relative to areas at higher northern latitudes. It should be emphasized that areas of greatest surface stress increase may not strictly correspond to areas of greatest surface stress proper, and therefore regions of greatest dust lifting may not ultimately map precisely to the distribution found in the surface stress lifting panel (panel i). There is a clear distinction between the two polarity types—not necessarily in their global effect on surface stress but, rather, where that effect is expressed. We previously noted that the global mean daytime surface wind stress value for the positive polarity 'archetype' was 0.00771 N m$^{-2}$. The corresponding global mean daytime value for the stresses of Fig. 17h is not markedly different, at 0.00757 N m$^{-2}$. This is well above the perihelion season baseline mean daytime value of 0.00615 N m$^{-2}$.

Mars year 24 was not a GDS year, but it was nonetheless a year characterized by significant regional dust storm activity (*Wang and Richardson*, 2015). In particular, multiple widely separated dust lifting centers were observed to be active simultaneously, between $L_s$=220-230° (cf. *Wang and Richardson*, 2015, Fig. 8). Why did this activity not lead to GDS conditions? The most significant difference we have been able to resolve through the above comparisons of Fig. 14 and Fig. 17 lies in the *enhancement* of the overturning meridional flow in the southern tropics of MY 28 (Fig. 14f), versus the *diminishment* of this same large-scale flow in MY 24 (Fig. 17f).

6.4.2. Negative polarity global-scale dust storm years (MY 12 and MY 25)

The occurrence of equinox-season global-scale dust storms in MY 12 and MY 25 has long represented a major puzzle from an atmospheric modeling standpoint (cf. *Basu et al.*, 2006), due to the relatively low solar energy input to the Mars atmosphere at their respective inception times (*Shirley*, 2015, Fig. 2). The modeling performed here does not yet provide a more satisfactory explanation. Here we will first examine the model outcomes for the MY 25 storm (Fig. 19), as it was the earliest to develop (at $L_s$=185°), turning thereafter to the first MY 12 ("1977a") storm, initiating at $L_s$=204° (Fig. 20). For brevity we will discuss the common features of Figs. 19 and 20 in parallel.

Baseline simulation zonal winds for the intervals prior to the MY 25 storm (Fig. 19a) and the MY 12 storm (Fig. 20a) are strikingly different from those displayed earlier for the aphelion season (Figs. 7-9) and for the perihelion season (Figs. 14, 17). Westerly jets are present in both hemispheres, as expected for the equinox season, with weaker easterly flows occurring above the equatorial regions. The westerly flow in the northern hemisphere is somewhat stronger in Fig. 20, as the northern fall season is more advanced during the $L_s$ range shown.

Although not visible with the scaling used here (designed to maintain consistency across all figures in this section), a two-celled Hadley circulation typical of the equinox season is in fact present; however, in both Figs 19d-e and 20d-e, the observed, clockwise cell dominates its companion.

The baseline surface wind stress maps (Figs. 19g and 20g) differ only in minor details. Baseline global mean daytime surface wind stresss for these two cases are 0.00485 N m$^{-2}$ (MY 25) and 0.00541 N m$^{-2}$ (MY 12). The corresponding values for the forced model simulations (Figs. 19h and 20h) are 0.00540 N m$^{-2}$ and 0.00557 N m$^{-2}$ respectively. While the CTA-forced model stresses are somewhat higher (by 10% and 3% respectively), these are considerably lower in absolute terms than those found for the perihelion season simulations with positive and negative polarity above. It appears that we cannot thus appeal solely to CTA-enhanced surface wind stresses as a primary factor leading to dust lifting prior to these storms.

Turning now to a consideration of MY 25 only, we immediately note that the streamflow difference plot in Fig. 19f implies a weak *retardation* of the main clockwise meridional overturning flow over the equatorial zone. This is similar to the pattern found for the negative polarity GDS-free streamflow plots of Fig. 18.

Turning next to the negative polarity MY 12 simulation of Fig. 20, we note a strong similarity of all the displayed plots to those of Fig. 19, with one exception. The streamflow differences plot (Fig. 20f) here shows only negligible differences from the baseline simulation.

The MarsWRF GCM solutions of Figs. 19 and 20 clearly do not yet provide straightforward explanations for the occurrence of equinoctial global-scale dust storms in MY 12 and MY 25. It may be that the limitations of our model, particularly the lack of atmospheric dust, may contribute to these outcomes. Further work with more sophisticated and more complete models is needed.

The large-scale circulation of the Mars atmosphere undergoes a significant reorganization around the times of the equinoxes. The meridional circulation in particular transitions from a dominantly single-celled morphology, with predominantly northward surface flows in northern summer, to a more Earth-like twin cell configuration near the vernal equinox (at $L_s$ ~180°), and thereafter to a southern summer single-cell configuration with surface winds directed predominantly southward. During this transition, we speculate that the variability in the direction of prevailing winds may expose surface dust "reservoirs" that may be unavailable at other seasons. This factor, together with the previously described interactions between the CTA and the diurnal heating cycle, may lead to conditions favorable for dust lifting that are outside the scope of the simple conceptual model for storm initiation that has emerged here. Southern hemisphere polar cap edge storms may have played an important role in the initiation of the MY 25 event (*Cantor*, 2007); if so, then this suggests that mesoscale modeling may be required in order to better understand the potential for significant dust lifting during the equinox season. In any case, as with all other hypotheses for GDS initiation on Mars, we must always recognize and acknowledge that the Mars atmosphere may have more than one circulatory "mode" capable of initiating global-scale dust storms.

6.5. Summary of dust storm season simulations results

We have been successful in reproducing atmospheric conditions (in terms of atmospheric large scale flows and surface wind stresses) that may adequately account for the observations in 19 of the 21 Mars years of Table 4 with only simple interpretation required. A modeled intensification of seasonal meridional flows is a common factor found prior to *all* of the perihelion-season GDS (Figs. 14-15). Conversely, no GDS have been recorded in transitional polarity years, when the putative forcing disappears (Fig. 16). Lastly, a destructive interference with the seasonal Hadley circulation emerges within the GCM near perihelion in negative polarity years (Fig. 18), which may account for the non-occurrence of GDS under those conditions. In common with all prior GCM investigations of global-scale dust storms, we have been unable to adequately account for the occurrence of equinox season GDS in Mars years 12 and 25. Further discussion of one or two interesting or exceptional cases (such as MY 27 and 31) is found below in Section 8.4. We now turn to a consideration of the surface wind stress differences obtained from the MarsWRF model for the 21 Mars years included in Table 4.

## 7. Statistics of global mean daytime surface wind stresses

Figure 21 displays the time variability of global mean daytime surface wind stresses obtained from both the baseline simulations (black) and the modified GCM (blue) for a period of 26 Mars years. The figure employs the identical format to Fig. 6, but now covers the period from MY 9, at the start of the period of spacecraft observations of Mars, out to MY 34, with perihelion occurring in the year 2018. Stress differences (forced minus baseline) are in gold, while the $\dot{L}$ waveform is shown in green and the insolation pattern is in red (the latter two values are arbitrarily scaled, although the $\dot{L}$ waveform is properly scaled with respect to the zero line). The considerable variability of the amplitude and period of the putative forcing function is well displayed; the origins of this variability are detailed in *Shirley* (2015). Positive, negative and transitional polarity episodes are easily distinguished in this format. In this Section we will compare numerical values of the baseline and CTA-modified stress levels.

Maps of daytime surface wind stresses (and wind stress differences) have been presented above in Figs. 7-9, 14, 17, 19, and 20. We have, in addition, obtained *global mean* daytime values of the surface stress for all of the tested intervals for each Mars year listed in Table 4. The calculated global mean daytime wind stress values, in N m$^{-2}$, are provided in Table 5.

There are a number of questions that may be addressed using the data presented in Table 5. For instance, it may be of interest to know whether the sample of model-derived surface stress levels for positive polarity years differ significantly from the sample of surface stress levels for transitional years, or from that of the negative polarity years. We require a statistical test that can address such questions. The Mann-Whitney test is appropriate for this purpose.

The Mann-Whitney test (MWT) is a non-parametric test of the equality of the means of two samples that may be employed to determine whether two samples of values are likely to have come from the same underlying distribution. (The null hypothesis for all of the following comparisons is that the paired samples of surface wind stress values evaluated have all been drawn from the same underlying population). To perform this test, the two samples (e.g., positive polarity and negative polarity) are first combined, and then ordered (or ranked). Summation of the ranks for each of the original samples yields a test statistic (cf. *Davis*, 1986). Critical values of the test statistic (corresponding to the 5%, 1%, and 0.1% levels of statistical significance) may be attained or exceeded if one or the other sample contains a disproportionate number of the highest (or lowest) ranked values. In our application, we employ *z*-values for estimating significance levels (where $z=1$ corresponds to one standard deviation of a normal distribution). The 5% significance level (probability $p \leq 0.05$) is thus attained when $z \geq 1.65$.

The physical hypothesis under evaluation here suggests that a 'relaxation' of circulatory flows should occur in the transitional years, in comparison with conditions in the positive and negative polarity years (Section 3; *Shirley,* 2016; *Shirley and Mischna*, 2016). We suspect that this condition is likely to be accompanied by relatively lower values of surface wind stresses. For our first set of tests we can thus compare the perihelion-season surface wind stress values of the transitional years of Table 5 (*n*=7) with the corresponding series of values for the positive polarity years ($n=8$) and negative polarity ($n=6$) years. When this is done we obtain $z=2.14$, $p=0.02$ for the positive polarity years comparison, and $z=1.79$, $p=0.04$ for the negative polarity years comparison. A comparison of the stress values for the 7 transitional years versus the combined sample of 14 positive and negative polarity years yields $z=2.35$, $p=0.01$. Thus, *the differences between the model-derived stresses obtained for the negative and positive polarity years as compared with the transitional years are statistically significant*. These results are listed in Table 6, along with the results for three additional comparisons, as described below.

We can employ the data of Table 5 to compare values for the perihelion season global-mean daytime surface wind stress, obtained through numerical modeling of the Mars atmosphere, for the GDS years sample ($n=9$) and the GDS-free years sample ($n=12$). This comparison yields $p=0.005$.

A comparison of the forced- and unforced model surface wind stress values obtained for the pre-GDS intervals is also possible. These values are supplied in the $6^{th}$ and $7^{th}$ columns of Table 5. We obtain a probability $p$ of 0.02 for this comparison. That is, the model with CTA forcing produces significantly higher values of the global mean daytime surface wind stress than are obtained using the baseline model, during the intervals leading up to the inception dates of past GDS on Mars.

This brings us to one final comparison, between the stress magnitudes at perihelion obtained for the negative polarity years ($n=6$) versus the positive polarity years ($n=8$). The MWT returns $p=0.37$ for this comparison, indicating that the surface wind stress levels are not significantly different. Comparing the sample means, we obtain

0.00774 ± 0.00129 (1 σ) N m$^{-2}$ for the positive polarity set, and 0.00733 ± 0.00070 (1 σ) N m$^{-2}$ for the negative polarity set. (For reference, the perihelion-season global mean daytime surface wind stress obtained in the multi-year control run was 0.00615 N m$^{-2}$).

We have found that the global mean daytime surface wind stress values for transitional years are significantly smaller than those for the negative and positive polarity years. This finding is consistent with the physical hypothesis of driven cycles of intensification and relaxation of atmospheric flows arising from orbit-spin coupling. We have further found that the daytime surface wind stresses for the perihelion seasons of the years with GDS differ significantly from those for years without GDS. While statistical significance has previously been obtained in comparisons of solar system dynamical quantities for these sets of years (*Shirley*, 2015; *Shirley and Mischna*, 2016), this marks the first time that clear interannual differences (relating directly to historic data) have emerged from numerical modeling of a planetary atmosphere with a global circulation model.

**8. Discussion**

To this point we have: 1) Successfully incorporated the coupling term accelerations of *Shirley* (2016) into the MarsWRF global circulation model; 2) tested and validated the model response to the accelerations, through modeling performed for the aphelion season; 3) verified a key prediction of the physical hypothesis, concerning the existence of driven cycles of circulatory intensification and relaxation, within the modified GCM; and 4) made considerable progress with the important open question of the origins of differences in the atmospheric response for positive polarity and negative polarity episodes at perihelion (*Shirley and Mischna*, 2016). We have found that an intensification of the atmospheric circulation occurs during both positive and negative polarity intervals; however, both surface wind stresses and the large-scale circulation are altered in different ways. We have seen that, for positive polarity years, a broad belt of the tropics experiences enhanced surface wind stresses, while for negative polarity years, the increased stresses are found in higher latitude regions, predominantly in the north.

8.1. Caveats and limitations

It is important to take note of the following caveats and limitations, which must be kept in mind with reference to the above findings and to others detailed below. In order to better isolate the effects of the coupling term accelerations, we decided to "turn off" and exclude the effects of atmospheric dust from our simulations. While we believe this tactic was successful, we must recognize that the realism of our simulations is necessarily impacted. To assess the robustness of this investigation to more realistic scenarios that contain atmospheric dust, we performed additional simulations, identical in structure to those in our dust-free cases, but with a simplified, prescribed dust loading obtained from the Mars Climate Database (*Forget et al.*, 1999). The prescribed dust is radiatively active and is a function of time and location (horizontal and vertical), but is neither lifted from the surface nor transported through the atmosphere. Results for these dusty atmospheres replicating Figures 15, 16 and 18 are found in the Supplementary Data, Figures S21-S23, respectively, and show qualitatively similar patterns to our dust-free scenario. So, while we expect that our present findings will be robust under dusty conditions, we also recognize that more sophisticated modeling is likely to reveal details that may potentially require revisions to some of the interpretations and scenarios described here.

We have sought to limit the possibility of negative consequences of this sort in a number of ways, first by employing the relatively dust-free aphelion season for our validation activities, and secondly by focusing on intervals *prior to* the initiation of GDS (as contrasted to intervals within the active GDS period). Injection of significant quantities of dust into the atmosphere of Mars must significantly alter the energy balance, and the resulting thermally driven atmospheric motions are likely to greatly dominate over the influence of the small CTA accelerations when this occurs.

We also recognize that the present exploratory investigation has barely scratched the surface. We have not attempted to study the consequences of the coupling term accelerations with respect to pressure and temperature variability or with respect to atmospheric tides or topographic influences. The specific mechanisms of dust lifting have not been addressed (cf. *Read et al*., 2015, for a recent review). We have employed mainly zonal averages, allowing us to ignore any effects or differences that may be a function of longitude and, hence, zonal topography. Mesoscale modeling, not attempted here, is likely to yield additional insights.

Another significant aspect we have neglected to discuss is the distribution of surface dust available for lifting. Briefly, it is thought that even under certain circumstances where surface stresses appear sufficient to raise dust, if a sufficient supply of readily mobile dust is not available to lift, a dust storm cannot initiate (*Haberle*, 1986; *Basu et al*., 2006; *Pankine and Ingersoll*, 2004; *Fenton et al*., 2006; *Cantor, 2007*; *Mullholland et al*., 2013). Over time, dust is transported from location to location on the surface (e.g., *Szwast et al.*, 2006), which may serve as an essential recharge mechanism for the dust lifting process. Dust availability and redistribution processes may thus serve as an additional regulator of the dust cycle which imprints on our model, and which we do not yet understand.

While the choice of the coupling efficiency coefficient, *c*, employed for this investigation may be criticized as arbitrary, it is relevant to point out that it is *not* the value of *c* that determines the agreement with observations we have obtained. Rather, it is the phasing of the dynamical forcing function with respect to the annual cycle of solar irradiance that gives rise to the systematic relationships observed. No value of *c* could be expected to improve the results, if the phasing of these two cycles was in fact unrelated to the problem of the intermittent occurrence of GDS on Mars.

8.2. Implications

The modified MarsWRF GCM clearly resolves conditions favorable for the occurrence of global-scale dust storms on Mars, in at least 7 of the 9 years in which such storms were observed. Less glamorously, the GCM with coupling term accelerations also successfully "predicts" the non-occurrence of GDS in all seven transitional polarity years. Statistically significant differences in global mean daytime surface wind stress are found in comparisons of transitional years versus global-scale dust storm years, and in comparisons of transitional years versus positive polarity and negative polarity years, both separately and in combination. Thus, despite the caveats and limitations of the previous section, we can nonetheless conclude that *the MarsWRF GCM with coupling term accelerations successfully retrodicts historic seasonal-time-scale weather anomalies* *actually observed on Mars.* We are unaware of any other numerical model, either for Mars or for the Earth, which can make a similar claim.

8.3. GDS forecasts for MY 33 and 34

What about future Mars years? As noted in *Shirley and Mischna* (2016), MY 33 has a positive polarity phase of $\varphi=104°$. With reference to Table 4, this falls closest to the phases obtained for the GDS years MY 15 (with $\varphi$ ~99°), and MY -16 and MY 9 (each with $\varphi$ ~93°). The GCM simulation of MY 33 shows a similar magnitude meridional circulation increase to these years (Fig. 22, left), and has a global daytime average perihelion-season surface stress of 0.00720 N m$^{-2}$, which is elevated above the transitional (and baseline) cases, but still at the lower end of the range for years with an observed GDS (Table 5).

Mars year 34 (Fig. 22, right) looks to be an even stronger candidate GDS year than MY 33, from the GCM results. The circulation follows the same trend as in other positive phase years, and the perihelion season global mean daytime surface stress is exceptionally high at 0.00859 N m$^{-2}$, greater than any other year in the observed record except for MY 15, although the phasing ($\phi=65°$) is perhaps not quite as favorable as that for MY 33. This phasing falls between those for MY 10 ($\phi=44°$) and MY 21 ($\phi=70°$), both of which were GDS years. Based on these findings, and comparison to the historical record, it seems there is a strong chance of a mid-season GDS during MY 33, a stronger chance in MY 34, and an extremely high likelihood of a mid-season (perihelion) GDS in one or both of these two years. As in the prior case of two successive positive polarity years, in MY 9 and 10 (Table 4), global-scale dust storms in both years is a distinct possibility. This forecast supersedes those of *Shirley* (2015) and *Shirley and Mischna* (2016), which were obtained without benefit of numerical modeling.

8.4. Future work

Work is now underway to characterize the behaviors of the Mars atmosphere (as represented in the MarsWRF GCM) under varied conditions of interactive (and self-consistent) dust loading, going beyond the 'passive' dust scenarios discussed in Section 8.1. A process of validation similar to that performed here in Section 5 will be required in order to ensure realism of model outcomes during each successive iteration of this process. A redetermination of the optimal value of the coupling efficiency coefficient, $c$, will likewise be required in each new experiment. In that connection, we may now return to a consideration of the special characteristics of MY 27. This was the only positive polarity year within our sample (Tables 4-5) in which no global-scale dust storm occurred. As we noted in passing in Section 6.2.1, the amplitude of the forcing function waveform is relatively small in this case, as are the derived surface wind stress values (Table 5). This suggests that MY 27 may represent a year in which some critical threshold for circulatory intensification by the CTA may not have been attained. If that is the case, then we may anticipate using MY 27 in future as a test case for fine-tuning $c$; if a global-scale dust storm is reproduced in the GCM with MY 27 forcing, we may conclude that the $c$ value employed was too large. MY 31 may provide a similar test case for negative polarity year simulations; the phasing for this year is nearly identical to that of MY 25, a GDS year, but the amplitude is significantly smaller, and no GDS was initiated in MY 31.

While we plan to employ the above tuning strategy for upcoming tests, we also recognize that other explanations for the non-occurrence of a GDS in MY 27 are also possible, such as an unfavorable distribution of surface dust, as noted above in Section 8.1.

Participation by other investigators and other centers in exploring the consequences for the Mars atmosphere of the orbit-spin coupling hypothesis is encouraged. To this end, upon request we will provide the basic data and algorithms employed for this investigation.

## 9. Summary and Conclusions

The MarsWRF atmospheric global circulation model was employed to test the predictions of the orbit-spin coupling hypothesis of *Shirley* (2016). The dynamical core of the GCM was modified to include "coupling term accelerations." The modified GCM was validated through extensive modeling experiments focused on atmospheric conditions and processes occurring during the less-dusty aphelion season on Mars. The physical hypothesis predicts that cycles of intensification and relaxation of large-scale circulatory flows within planetary atmospheres will occur on seasonal timescales, in response to contemporaneous changes in the planetary orbital angular momentum. The modified MarsWRF general circulation model simulations fully confirm this prediction.

A catalog of Mars years including nine years with global-scale dust storms and 12 years without global-scale dust storms (*Shirley and Mischna*, 2016) was employed to evaluate the correspondence of the modified MarsWRF model simulations with historic observations. Conditions favorable for the occurrence of global-scale dust storms were reproduced in the model in all of the years in which perihelion-season global-scale dust storms were observed. Model-derived global mean daytime surface wind stresses during the perihelion season of Mars years with global-scale dust storms are significantly larger than those obtained in years without such storms. Meridional overturning ("Hadley Cell") circulations that may lift dust to high altitudes in the Mars atmosphere in the southern summer dust storm season are strengthened in years when perihelion season global-scale dust storms were observed, and are retarded or forced only weakly in all of the years when no such storms occurred. Conditions unfavorable for the occurrence of global-scale dust storms were recognized in our simulations in all 12 of the years lacking such storms. An unprecedented degree of correspondence between numerical model outcomes and historic observations has been achieved in this investigation, suggesting that the interannual variability of Mars' weather and climate may largely be explained as a result of orbit-spin coupling and solar system dynamics. Proof of concept for the orbit-spin coupling hypothesis of *Shirley* (2016) is thereby attained.

An important simplifying condition was imposed within the GCM for this investigation. Atmospheric dust, which is known to significantly impact the atmospheric dynamics through radiative processes, was deliberately excluded, in order to better isolate the predicted effects of the coupling term accelerations. Further simulations including dust are expected to provide improved realism. Modeling results are so far unable to explain the occurrence of two global-scale dust storms occurring near the time of the autumnal equinox, in Mars years 12 and 25. Further investigation will be required in order to better illuminate the processes responsible for global-scale dust storm initiation in these cases.

One or more global-scale dust storms are forecast to occur on Mars in the upcoming southern summer dust storm seasons of Mars years 33 (2016) and 34 (2018).

**Acknowledgements**:

Support from JPL's Research and Technology Development Program and from NASA's Solar System Workings Program is gratefully acknowledged. Comments from Jon Giorgini materially improved this presentation. We have benefitted from ongoing discussions and critical feedback from Rich Zurek, Bruce Bills, Dan McCleese, David Kass, Tim Schofield, Bruce Cantor, Armin Kleinböhl, Eric deJong, Shigeru Suzuki, Laura Kerber, Don Banfield, and the members of the Mars Climate Sounder Science Team. This work was performed at the Jet Propulsion Laboratory, California Institute of Technology, under a contract with NASA. Resources supporting this work were provided by the NASA High-End Computing (HEC) Program through the NASA Advanced Supercomputing (NAS) Division at Ames Research Center as well as the High Performance Computing facilities of the Jet Propulsion Laboratory Office of the Chief Information Officer.

**Appendix A:**

This appendix describes the procedures employed in the present investigation for introducing the coupling term accelerations within the MarsWRF GCM. The interested reader may replicate our results by applying the operations described below to the data which may be obtained from the authors. The dynamical values are referenced to the J2000 ecliptic coordinate system. This is advantageous and convenient, if the user wishes to view or display the variability with time of $\dot{L}$, which represents the underlying forcing function, with a minimum of distortion. The inclination with respect to the ecliptic plane of Mars' orbital plane is less than 2°; thus the orbital angular momentum vector for Mars is nearly orthogonal to the ecliptic. In consequence, for plotting purposes, it suffices to plot the *z*-component of the time derivative of the solar system barycentric angular momentum vector, as in Figs. 4, 6, and 21 of the accompanying paper, as the components lying within the ecliptic plane are then negligibly small. If the user is only concerned with obtaining the coupling term accelerations for use within a GCM, however, it may be more efficient to begin with data represented in an equatorial coordinate system, as by so doing one may avoid an extra coordinate transformation.

The instantaneous barycentric angular momentum of Mars is obtained as follows (*Jose*, 1965):

$$L = \left[ (y\dot{z} - z\dot{y})^2 + (z\dot{x} - x\dot{z})^2 + (x\dot{y} - y\dot{x})^2 \right]^{1/2},$$

where the required quantities are the positional coordinates (*x,y,z*) and velocities ($\dot{x}, \dot{y}, \dot{z}$) of the subject body with respect to the solar system barycenter. The mass is not explicitly included (but must be supplied later as a multiplicative factor for quantitative comparisons). The requisite positions and velocities may be obtained from JPL's online Horizons ephemeris system (*Giorgini et al.*, 1996; *Giorgini and Yeomans*, 1999). A time step of 2 days was employed for the present investigation. We obtain the time derivative ($\dot{L}$) by differencing the vector components of successive time steps, dividing the values by the time in seconds, and assigning a time coordinate intermediate between those of the bracketing values of *L*. The resulting time-tagged Cartesian vector values are recorded in a file that is accessed by the modified MarsWRF global circulation model.

In order to obtain the coupling term accelerations for any particular grid point location and time on Mars, it is necessary to perform a set of coordinate transformations that resolve the components of the $\dot{L}$ vector in the Mars body-fixed coordinate system. The latter is a rotating Cartesian system with *z*-axis coincident with the direction of the north rotation pole of Mars, and with *x* axis simultaneously lying in the equatorial plane and in the plane defining the prime meridian of longitude on Mars. Thus the surface location on Mars with latitude=0° and longitude=0° has Cartesian components in the body-fixed system of (*r*, 0, 0), where *r* is the equatorial radius of Mars.

The rotation state of Mars for any particular time in the recent past and near future may be obtained using constants and formulae from *Seidelmann* (1992, Table 15.7). Small improvements to the 1992 values are found in an IAU report (*Archinal et al.*, 2009). Our calculations were performed using the 1992 values. However, for completeness, we reproduce both sets of values in Table 1 below.

Table A1. Rotational data for Mars

| Seidelmann, 1992: | IAU, 2009: |
|---|---|
| $\alpha_0$= 317.681 - 0.108 $T$ | $\alpha_0$= 317.68143 - 0.1061 $T$ |
| $\delta_0$= 52.886 – 0.061 $T$ | $\delta_0$= 52.88650 – 0.0609 $T$ |
| $W$= 176.868 + 350.8919830 $d$ | $W$= 176.630 + 350.89198226 $d$ |

Here, $\alpha_0$ and $\delta_0$ are the celestial direction (right ascension and declination) of the north pole of Mars, and the angle $W$ identifies the location of the Prime Meridian, measured along Mars' equator in an easterly direction from the node of the planet's equator on the standard equator (the equatorial plane of a terrestrial observer). $T$ is time in Julian centuries (of 36525 Julian days), measured from the J2000 epoch, while $d$ is time measured in Julian days from the same epoch. The angles are measured in degrees. The MarsWRF GCM accepts input and performs calculations using Julian dates as a run-time option.

We began by obtaining positions and velocities of Mars referenced to the J2000 ecliptic coordinate system with origin at the solar system barycenter, and thereafter obtained the components of the $\dot{L}$ vector as described above and included in the accompanying data file. Our next step was to rotate the ecliptic system of the resulting data in a counterclockwise direction about the positive $x$ axis by the value of the Earth's obliquity (23.45°), bringing the $x$-$y$ plane into coincidence with the reference equatorial plane. (The $\alpha_0$ and $\delta_0$ angles of Table A1 are referenced to this system). This initial transformation would have been unnecessary if we had first obtained the Horizons data in the equatorial system, instead of the ecliptic system.

We next rotate the system about its $z$-axis such that the new $x$-axis coincides with the ascending node of the Mars equator on the equatorial plane. The node is located at an angular separation of 90° from $\alpha_0$; using $\alpha_0$ as given in Table A1 the node would be located at $\alpha_{node} = 90° + \sim317.681°$, requiring a rotation of ~47° about $z$ in this step.

The north polar direction of Mars has a declination, $\delta$, of ~52.886° with respect to the celestial equator. We are next required to align the z-axis of our intermediate system with the north polar direction of Mars. We accomplish this by a counterclockwise rotation about the x-axis (which lies along the line of nodes) of 90° - $\delta$.

The resulting system has z along Mars' north polar direction, with the x- and y-axes now lying within Mars' equatorial plane, as desired. The x-axis direction coincides with that of the node of Mars' equator on the reference equatorial plane. Our final step is to transform about the z-axis by the angle W as calculated using the third relationship of Table A1. The angle W is negative for dates prior to the J2000 epoch, and positive thereafter. This step aligns the final coordinate frame with the rotating body-fixed frame for Mars, which is the coordinate system employed for all global circulation modeling work.

The $\dot{L}$ vectors ingested from the dynamical quantities input file (available from the authors by request) may each be subjected to the above set of transformations before the cross product with the rotational angular velocity vector, $\omega_\alpha$, is obtained. The rectangular components of the angular velocity of the axial rotation, $\omega_\alpha$, of Mars, in Mars' body fixed frame, are: $[0.0, 0.0, 7.0882181 \times 10^{-5}]$ rad sec$^{-1}$. This corresponds to a rotation period of 88,642.6632 seconds or 1.02595675 (Earth) days. The components of the rotational angular velocity vector in ecliptic coordinates, supplied for convenience, are $[3.16245 \times 10^{-5}, 3.9279289 \times 10^{-6}, 6.33144866 \times 10^{-5}]$.

At this point, it may be convenient to multiply each vector component obtained by the selected value of the scalar coupling efficiency coefficient, c. (This step may be performed at any stage of the process). For the particular case of the present investigation, we have employed a value of $5 \times 10^{-13}$ for c (see main text). The tabulated values for $\dot{L}$ may be employed as supplied for calculating accelerations. To obtain values of $\dot{L}$ for Mars in MKS units it is also necessary to multiply the tabulated values by the mass of Mars (~$0.64191 \times 10^{24}$ kg; *Seidelmann et al.*, 1992).

**References**

Archinal, B. A., and 16 others, 2009. Report of the IAU Working Group on Cartographic Coordinates and Rotational Elements: 2009, *Cel. Mech. Dyn. Astron.*, doi:10.1007/s10569-010-9320-4.

Ayoub, F., J.-P. Avouac, C.E. Newman, M.I. Richardson, A. Lucas, S. Leprince and N.T. Bridges, 2014. Threshold for sand mobility on Mars calibrated from seasonal variations of sand flux, *Nat. Commun.* 5:5096, 10.1038/ncomms6096.

Basu, S., J. Wilson, M. Richardson, and A. Ingersoll, 2006. Simulation of spontaneous and variable global dust storms with the GFDL Mars GCM, *J. Geophys. Res.* 111, E090004, 10.1029/2005JE002660.

Cantor, B. A., 2007. MOC observations of the 2001 Mars planet-encircling dust storm, *Icarus* 186, 60-96.

Choi, D. S., and C. M. Dundas, 2011. Measurements of Martian dust devil winds with HiRISE, *Geophys. Res. Lett.* 38, L24206, 10.1029/2011GL049806.

Chojnacki, M., D. M. Burr, J. E. Moersch, and T. I. Michaels, 2011. Orbital observations of contemporary dune activity in Endeavor crater, Meridiani Planum, Mars, *J. Geophys. Res*. 116, E00F19, 10.1029/2010JE003675.

Davis, J. C., 1986. Statistics and Data Analysis in Geology, John Wiley & Sons, N. Y., pp. 93-96.

Fenton, L. K., P. E. Geissler, and R. M. Haberle, 2007. Global warming and climate forcing by recent albedo changes on Mars. *Nature* 44, 646-649, 10.1038/nature05718.

Forget, F., F. Hourdin, R. Fournier, C. Hourdin, O. Talagrand, M. Collins, S. R. Lewis, P. L. Read, and J.-P. Huot, 1999. Improved general circulation models of the Martian atmosphere from the surface to above 80 km, J. *Geophys. Res*. 104, E4, 24,155-24,175.

Giorgini, J. D., D. K. Yeomans, A. B. Chamberlin, P. W. Choudas, R. A. Jacobsen, M. S. Keesey, J. H. Lieske, S. J. Ostro, E. M. Standish, and R. N. Wimberly, 1996. JPL's on-line solar system data service, *Bull. Am. Astron. Soc*. 28, 1158.

Giorgini, J. D., and D. K. Yeomans, 1999. On-line system provides accurate ephemeris and related data, NASA Tech Briefs, NPO-20416, p. 48.

Haberle, R. M., 1986. Interannual variability of global dust storms on Mars, *Science* 234, 459-461.

Holstein-Rathlou, C., and 20 colleagues, 2010. Winds at the Phoenix landing site, *J. Geophys. Res*. 115, E00E18, 10.1029/2009JE003411.

Kahre, M. A., J. L. Hollingsworth, R. M. Haberle, and R. J. Wilson, 2014. Coupling the Mars dust and water cycles: The importance of radiative-dynamic feedbacks

during northern hemisphere summer, *Icarus* 260, 477-479, 10.1016/j.icarus.2014.07.017.

Lewis, S. R., 2003. Modeling the Martian atmosphere, *Astron. Geophys*. 44, 4.6-4.14.

Martin, L. J. and Zurek, R. W., 1993. An analysis of the history of dust activity on Mars, *J. Geophys. Res*. 98, E2, 3221-3246.

Mischna, M. A., C. Lee, and M. Richardson, 2012. Development of a fast, accurate radiative transfer model for the Martian atmosphere, past and present, *J. Geophys. Res*. 117, E10009, 10.1029/2012JE004110.

Mulholland, D. P., P. L. Read, and S. R. Lewis, 2013. Simulating the interannual variability of major dust storms on Mars using variable lifting thresholds, *Icarus* 223, 344-358, doi:10.1016/j.icarus.2012.12.003.

Newman, C. E., S. R. Lewis, P. L. Read, and F. Forget, 2002a. Modeling the Martian dust cycle 1. Representations of dust transport processes, *J. Geophys. Res*. 107, E12, doi:10.1029/2002JE001910.

Newman, C. E., S. R. Lewis, P. L. Read, and F. Forget, 2002b. Modeling the Martian dust cycle 2. Multiannual radiatively active dust transport simulations, *J. Geophys. Res*. 107, E12, doi:10.1029/2002JE001920.

Pankine, A. A., and A. P. Ingersoll, 2004. Interannual variability of Mars global dust storms: An example of self-organized criticality? *Icarus* 170, 514-518.

Petrosyan, A., B. Galperin, S. E. Larsen, S. R. Lewis, A. Määttänen, P. L. Read, N. Renno, L. P. H. T. Rogverg, H. Savijärvi, T. Silli, A. Spiga, A. Toigo, and L.

Vázquez, 2011. The Martian atmospheric boundary layer, *Rev. Geophys*. 49, RG3005.

Read, P. L., S. R. Lewis, and D. P. Mulholland, 2015. The physics of Martian weather and climate: A review, *Rep. Prog. Phys*. 78, 125901 (54 pp), 10.1088/4885/78/12/125901.

Richardson, M. A., A. D. Toigo, and C. E. Newman, 2007. PlanetWRF: A general purpose, local to global numerical model for planetary atmospheric and climate dynamics, *J. Geophys. Res*. 112, E09001, 10.1020/2006JE002825.

Schofield, J. T., J. R. Barnes, D. Crisp, R. M. Haberle, S. Larson, J. A. Magalhäes, A. Seiff, and G. Wilson, 1997. The Mars Pathfinder atmospheric structure investigation, *Science* 278, 1572-1758, 10.1126/science.278.5344.1752.

Seidelmann, P. K. (Ed.), 1992. Explanatory Supplement to the Astronomical Almanac, University Science Books, Mill Valley, Ca.

Shirley, J. H., T. H. McConnochie, D. M. Kass, A. Kleinböhl, J.T. Schofield, N. G. Heavens, D. J. McCleese, J. Benson, D. P. Hinson, and J. L. Bandfield, 2015. Temperatures and aerosol opacities of the Mars atmosphere at aphelion: Validation and inter-comparison of limb sounding profiles from MRO/MCS and MGS/TES, *Icarus* 251, 26-49, 10.1016/j.icarus.2014.05.011.

Shirley, J. H., 2016. Non-tidal coupling of orbital and rotational motions: Planetary and solar atmospheres, *in preparation*.

Shirley, J. H., 2015. Solar system dynamics and global-scale dust storms on Mars, *Icarus* 252, 128-144, 10.1016/j.icarus.2014.09.038.
66

Shirley, J. H., and M. A. Mischna, 2016. Orbit-spin coupling and the interannual variability of global-scale dust storm occurrence on Mars, *in preparation*

Smith, M. D., M. J. Wolff, N. Spanovich, A. Ghosh, D. Banfield, P. R. Christensen, G. A. Landis, and S. W. Squyres, 2006. One Martian year of atmospheric observations using MER Mini-TES, *J. Geophys. Res*. 111, E12513.

Spiga, A. and S. R. Lewis, 2010. Martian mesoscale and microscale wind variability of relevance for dust lifting. *Mars* 5, 146-158, doi:10.1555/mars.2010.0006.

Strausberg, M. J., H. Wang, M. I. Richardson, and S. P. Ewald, 2005. Observations of the initiation and evolution of the 2001 Mars global dust storm, *J. Geophys. Res*. 110, E02006, 10.1029/2004JE002361.

Szwast, M. A., M. I. Richardson, and A. R. Vasavada, 2006. Surface dust redistribution on Mars as observed by the Mars Global Surveyor and Viking Orbiters, *J. Geophys. Res*. 111, E11008, 10.1029/2005JE002485.

Toigo, A. D., M. I. Richardson, R. J. Wilson, H. Wang, and A. P. Ingersoll, 2002. A first look at dust lifting and dust storms near the south pole of Mars with a mesoscale model, *J. Geophys. Res*. 107, E7, 10.1029/2001JE001592.

Tyler, D., Jr., J. R. Barnes, and R. H. Haberle, 2002. Simulation of surface meteorology at the Pathfinder and VL1 sites using a Mars mesoscale model, *J. Geophys. Res*. 107, E4, 5018, 10.1029/2001JE001618.

Wang, H. and M. I. Richardson, 2015. The origin, evolution, and trajectory of large dust storms on Mars during Mars years 24-30 (1999-2011), *Icarus* 251, 112-127, doi:10.1016/j.icarus.2013.10.033.

Zurek, R. W., and L. J. Martin, 1993. Interannual variability of planet-encircling dust storms on Mars, *J. Geophys. Res.* 98, E2, 3247-3259.

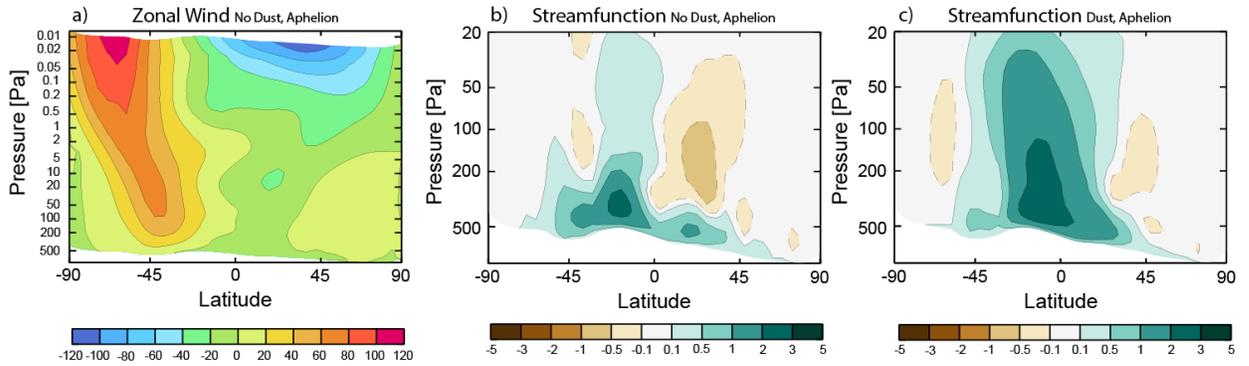

**Figure 1:** **(**a) zonal mean zonal wind for the baseline MarsWRF model without dust, in northern spring ($L_s$=70°-90°). Zonal winds during this season are dominated by the winter jet in the southern hemisphere. Much weaker westerly flow is seen in the northern hemisphere at lower altitudes, while the equatorial regions exhibit weak easterly flow near the surface, which strengthens at higher altitudes (blue colors). Wind velocities are in m s$^{-1}$. (b): Zonal mean meridional streamfunction for this same dust-free simulation. Streamflow units are $10^9$ kg s$^{-1}$. (c): Zonal mean meridional streamfunction at aphelion, with radiatively active atmospheric dust included.

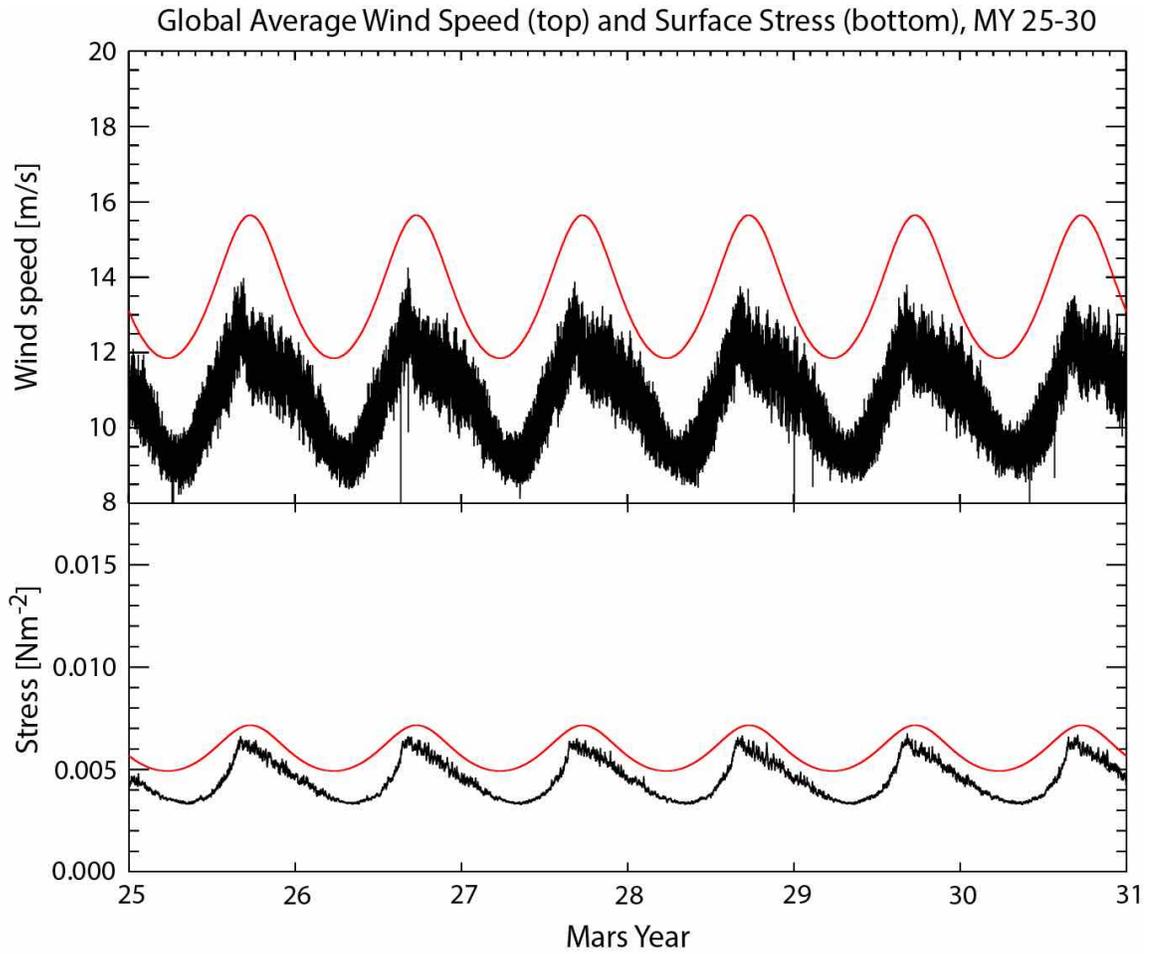

**Figure 2**: Model-generated global mean near-surface wind speeds and surface wind stresses for six Mars years, drawn from a longer MarsWRF 'baseline' simulation. (Top): global mean wind speed (in black). (Bottom): global mean surface stress (also in black). In both panels, a scaled version of the same annual insolation cycle is shown in red to illustrate the relationships of wind speed and surface wind stresses to the annual cycle.

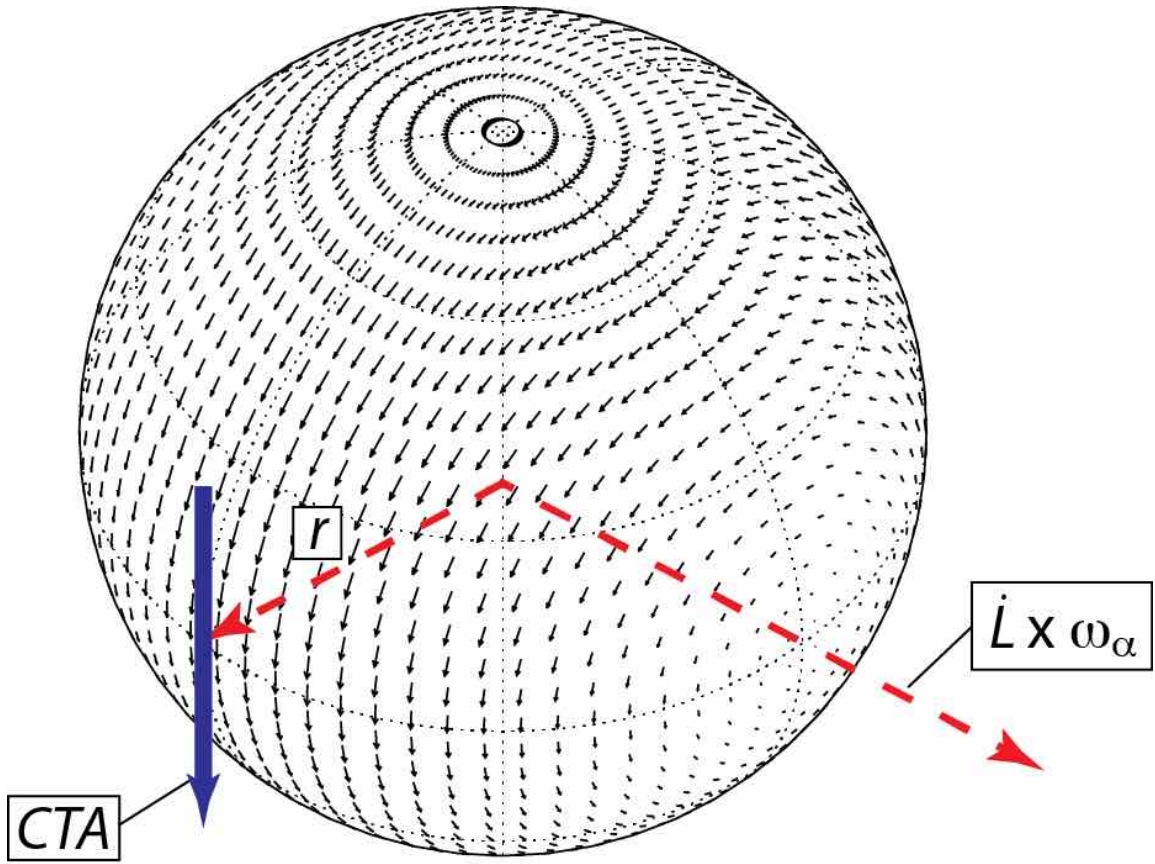

**Figure 3**: A planetary globe showing direction and magnitude of the CTA field for a single point in time at the 5° x 5° resolution of the MarsWRF general circulation model used in this study. A zero-point, or node, can be seen in the lower right of the figure, where the cross product of $\dot{L}$ and $\omega_\alpha$ is parallel to the position vector at that location. A similar node is found in a position antipodal to the visible one.

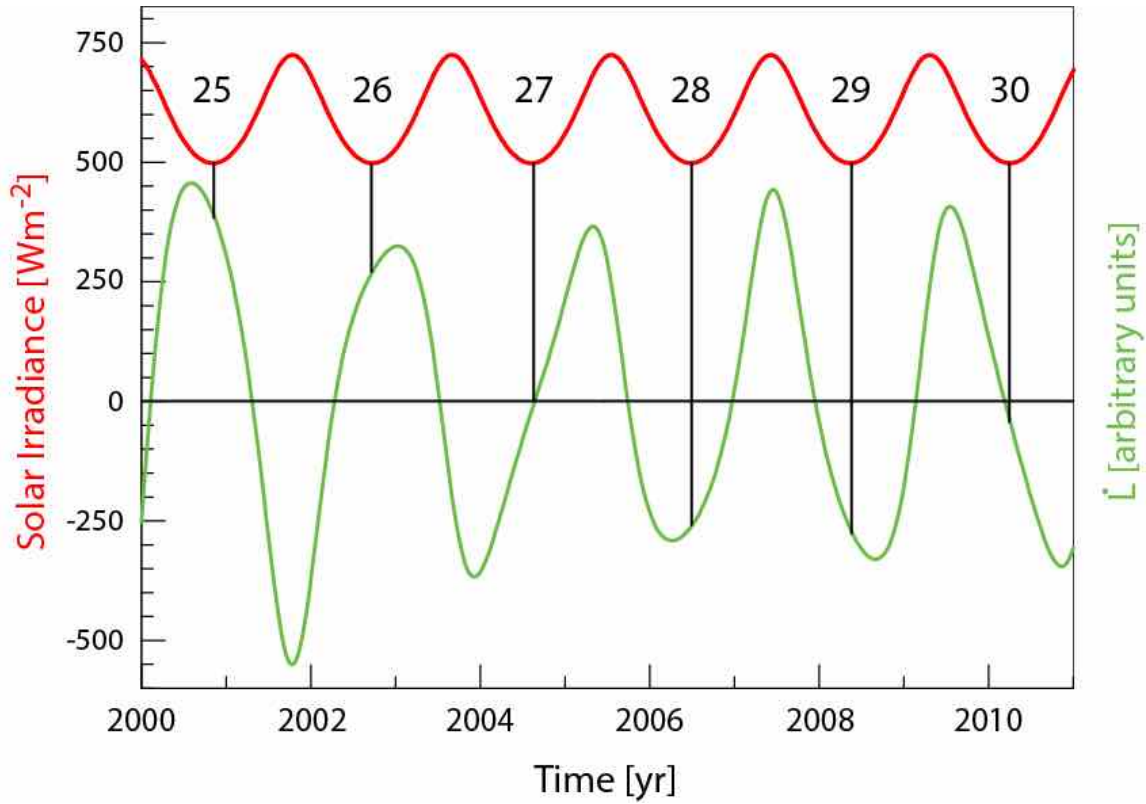

**Figure 4 (**in green**):** The signed *z* component (in J2000 ecliptic coordinates; see Appendix A) of the rate of change of the barycentric orbital angular momentum, $\dot{L}$, over six Mars years (MY 25-30), along with the annual cycle of solar irradiance (in red) for this same period. The latter repeats exactly, year after year, but is not simply phased with the $\dot{L}$ cycle. Vertical black lines are shown for reference at each aphelion to illustrate the relative phasing of the two waveforms at this season.

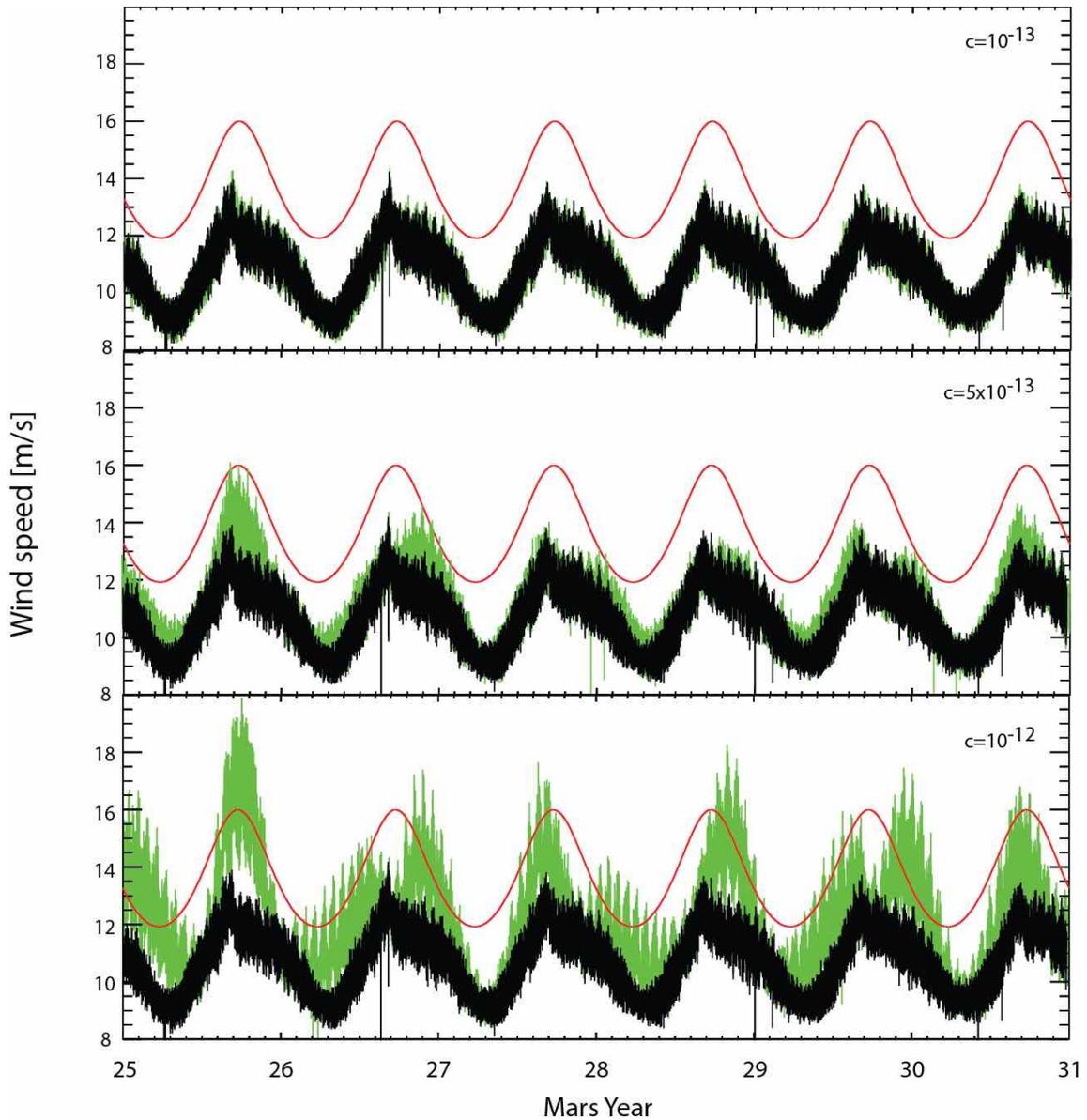

**Figure 5**: Model-derived near-surface wind speeds for three values of the coupling coefficient, *c*. The black curve, repeated in each panel, is for the baseline case (Section 2), and is the same as the corresponding curve in Fig. 2. Green curves illustrate global mean winds in the lowest model layer for the stated value of *c*. Insolation (arbitrary units) is illustrated in all panels to demonstrate phasing of seasonal behavior.

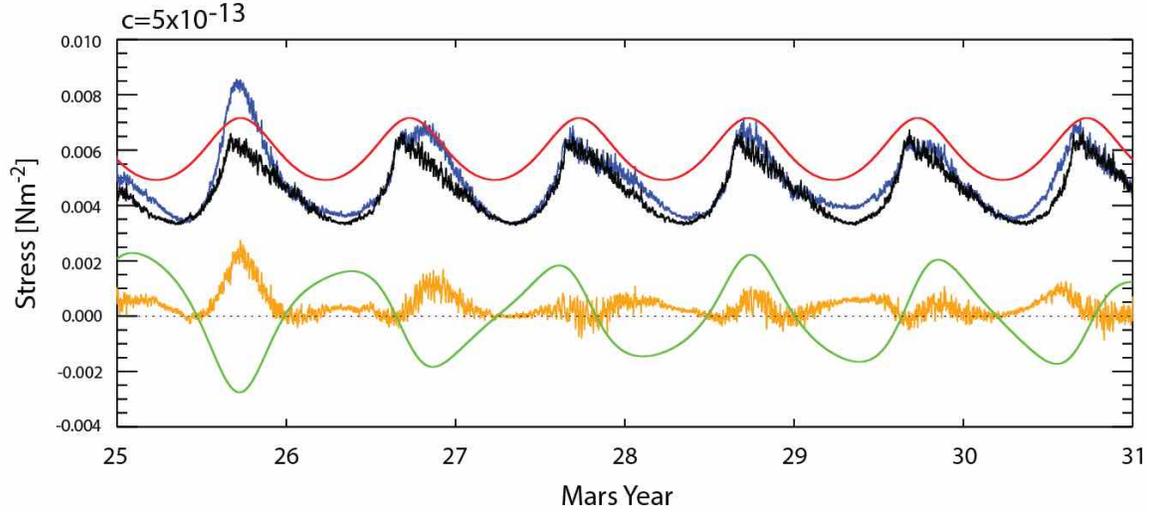

**Figure 6**: Global mean surface wind stresses (N m$^{-2}$) without CTA (black) and with the CTA (blue) for MY 25-30. The difference between the two is shown in orange. The $\dot{L}$ waveform is shown in green (arbitrary units, but scaled to the zero line), to illustrate the phasing of the differences, and the seasonal insolation cycle (also in arbitrary units) is shown in red. Stress magnitudes and stress differences are detailed in Table 2. With reference to the $\dot{L}$ waveform, in green, we note that both positive polarity intervals and negative polarity intervals are accompanied by increased surface wind stresses.

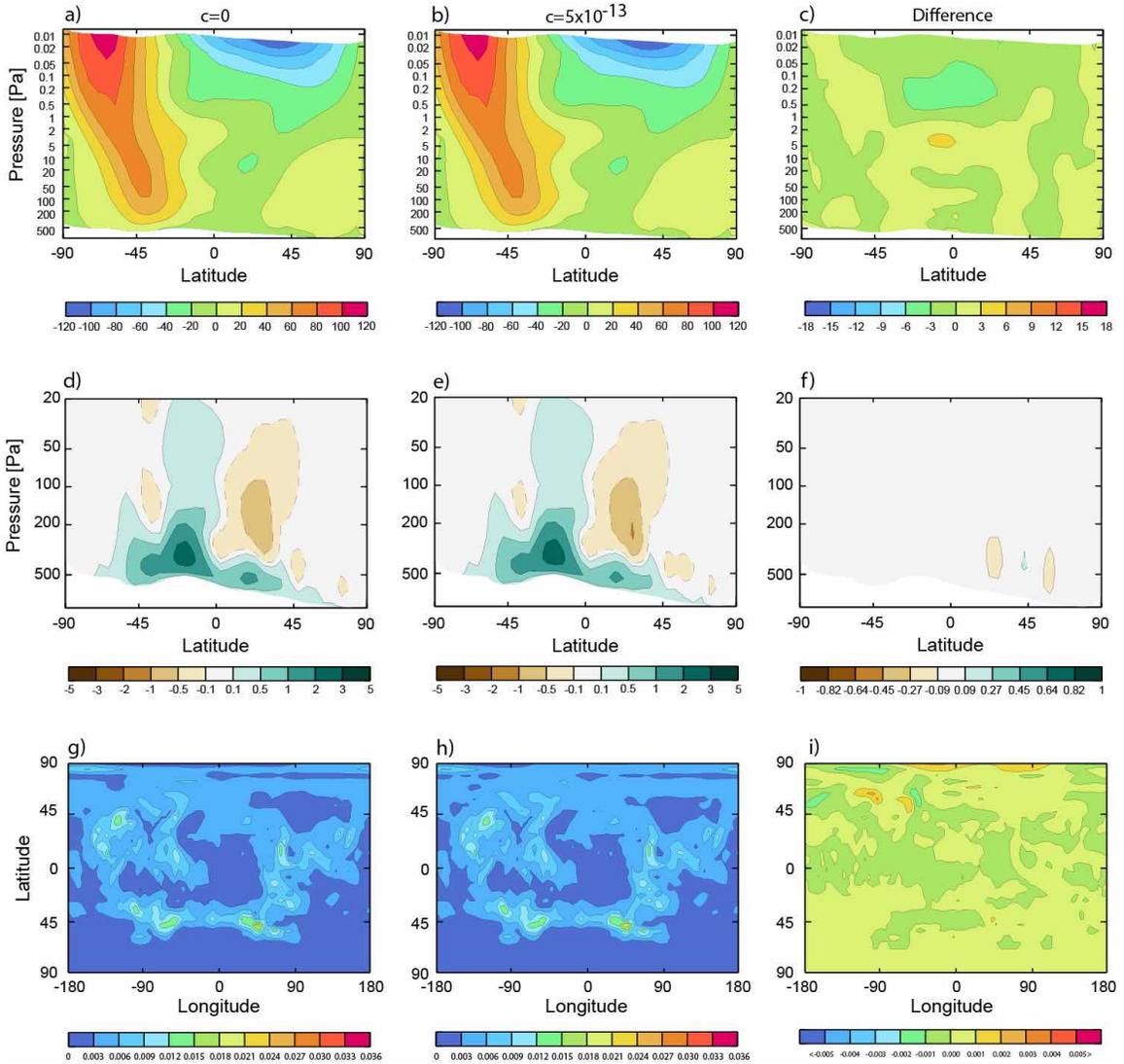

**Figure 7**. Mars Year 27 wind fields for the 'baseline' case (left column) and CTA case (middle column) along with differences between the two (right column) for (top) vertical slice of zonal-mean zonal wind [m/s], (middle) vertical slice of meridional streamfunction [$10^9$ kg/s], and (bottom) map view of daytime-averaged surface stress [N m$^{-2}$]. Note the difference in the vertical (pressure) scales of the top two rows.

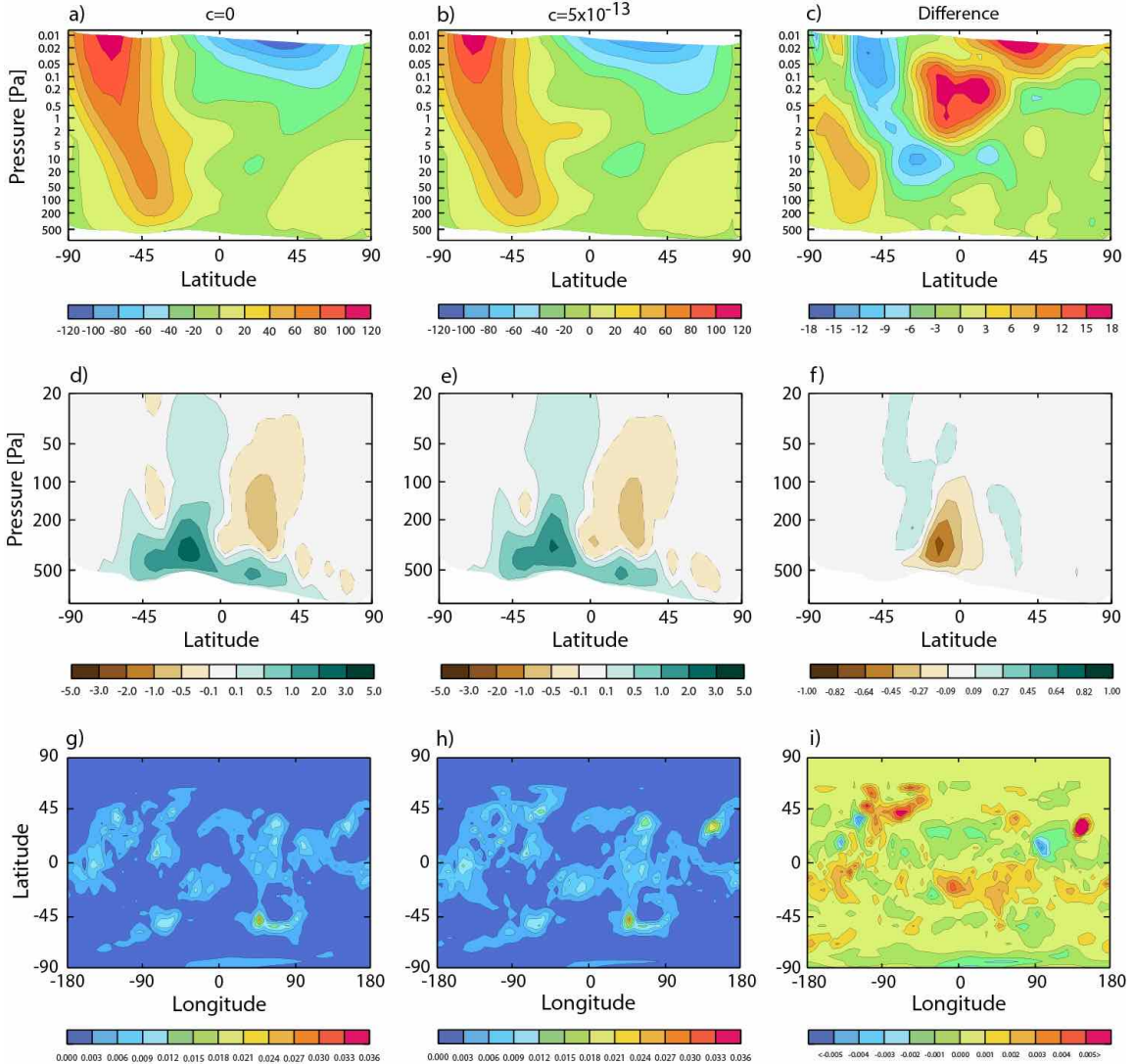

**Figure 8**: Same as Fig. 7, but for MY 25, a positive polarity case. Wind fields for the 'baseline' case (left column) and CTA case (middle column) along with differences between the two (right column) for (top) vertical slice of zonal-mean zonal wind [m/s], (middle) vertical slice of meridional streamfunction [$10^9$ kg/s], and (bottom) map view of daytime-averaged surface stress [N m$^{-2}$]. Note the difference in the vertical (pressure) scales of the top two rows.

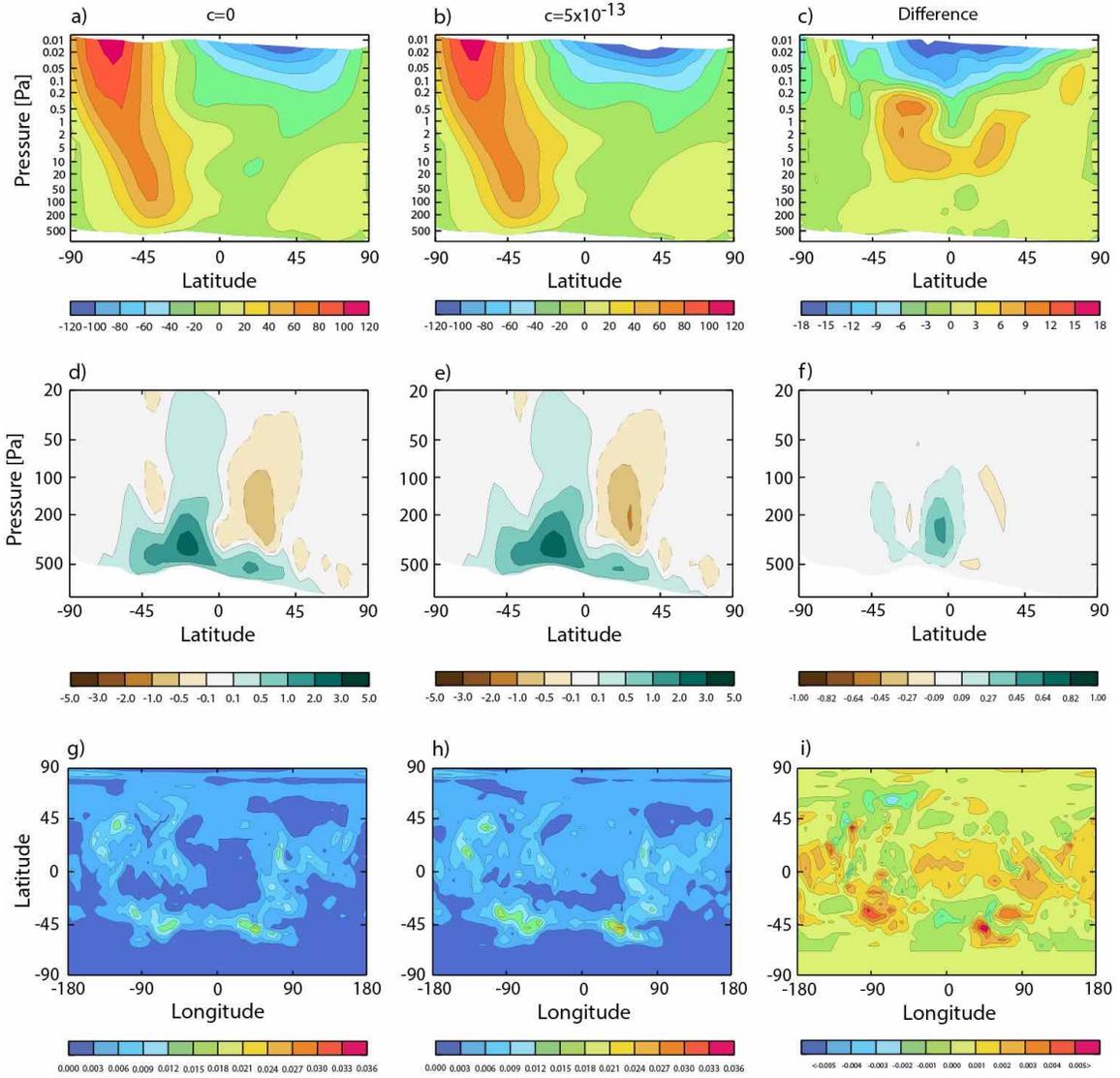

**Figure 9**: Same as Figs. 7 and 8, but for MY 28, a negative polarity case. Wind fields for the 'baseline' case (left column) and CTA case (middle column) along with differences between the two (right column) for (top) vertical slice of zonal-mean zonal wind [m/s], (middle) vertical slice of meridional streamfunction [$10^9$ kg/s], and (bottom) map view of daytime-averaged surface stress [N m$^{-2}$]. Note the difference in the vertical (pressure) scales of the top two rows.

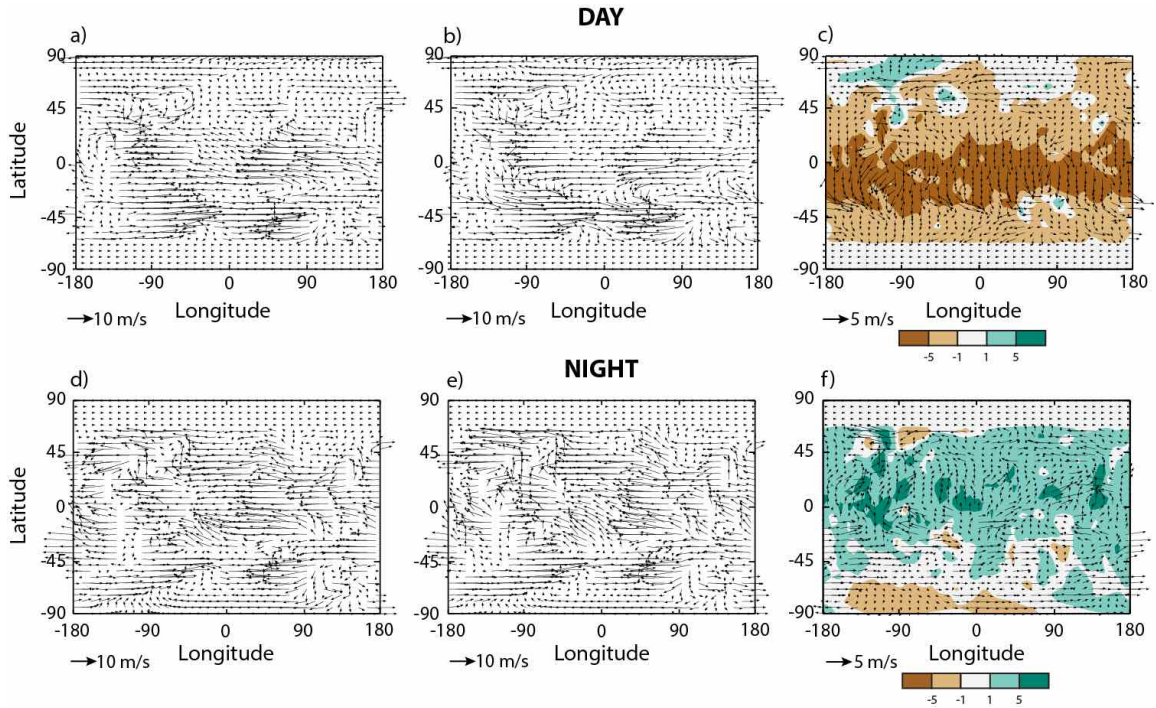

**Figure 10**: Near-surface winds during the MY 25 aphelion season averaged over (top) daytime and (bottom) nighttime hours. Left column is the baseline case ($c=0$), middle column is case with CTA. Right column shows differences between the baseline and CTA simulations. Contours in panels c and f show difference (in ms$^{-1}$) in meridional component of surface wind. Northward-directed differences are shaded in green, to be consistent with the sense of the normal seasonal surface flow of Figs. 7-9, while locations with southward-directed meridional components are highlighted with brown shading.

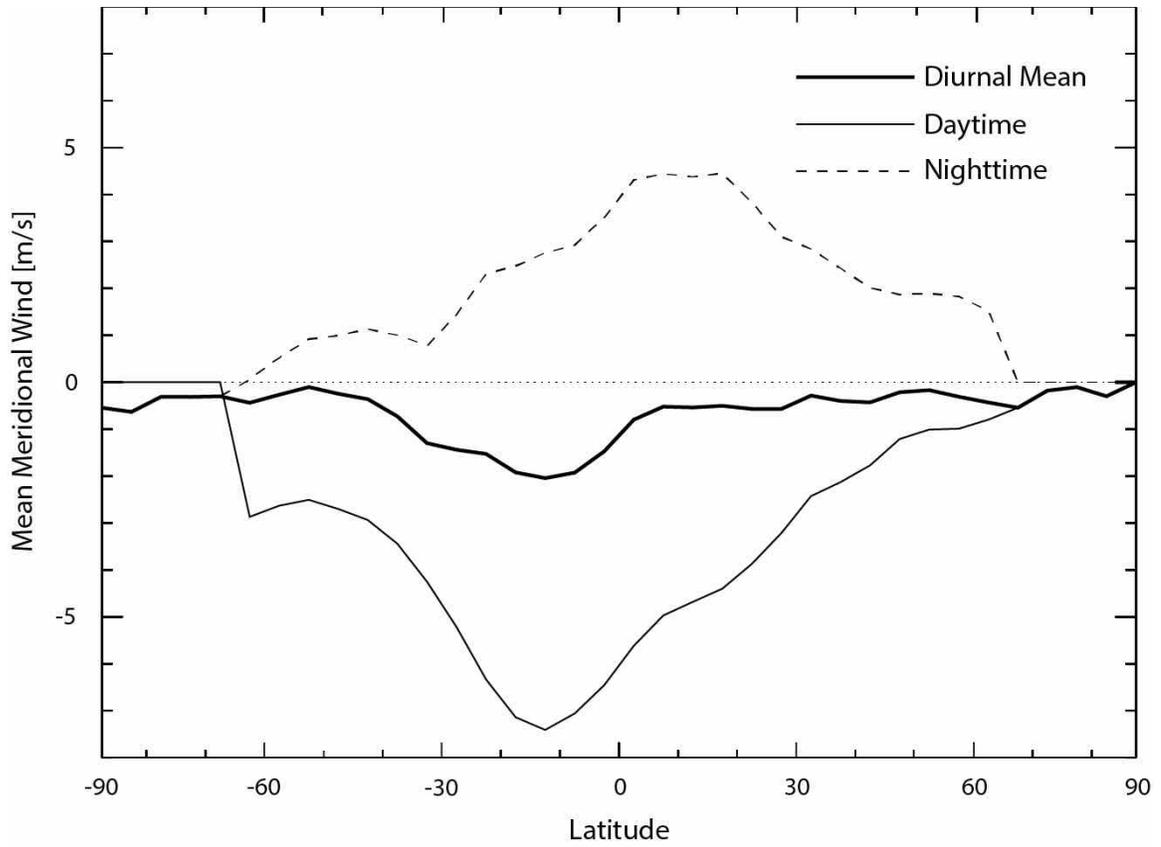

**Figure 11**: Zonal mean near-surface meridional wind difference (CTA minus baseline) for the aphelion season of MY 25, for daytime hours (thin solid line), nighttime hours (dashed line) and full day (thick solid line). Dotted line is a fiducial drawn at 0 m/s.

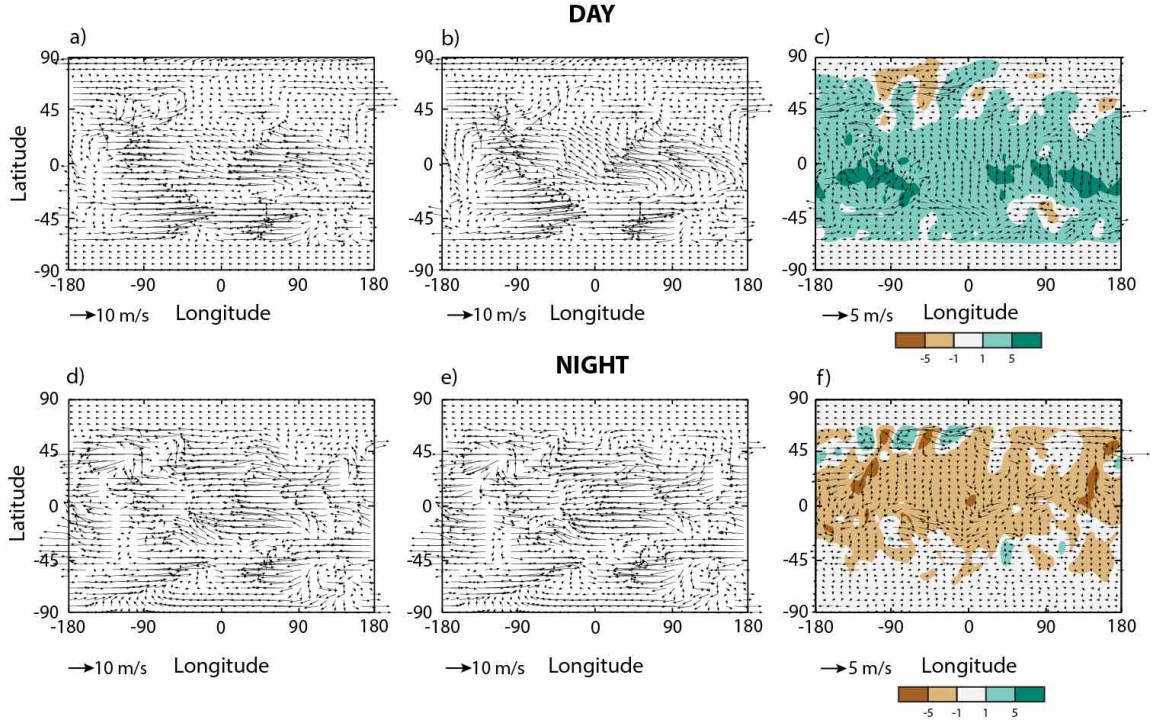

**Figure 12**: Same as Fig. 10, but for MY 28, a negative polarity episode. Near-surface winds averaged over (top) daytime and (bottom) nighttime hours. Left column is the baseline case (*c*=0), middle column is case with CTA. Right column shows differences between the baseline and CTA simulations. Contours in panels c and f show difference (in ms$^{-1}$) in meridional component of surface wind. Northward-directed differences are shaded in green, to be consistent with the sense of the normal seasonal surface flow of Figs. 7-9, while locations with southward-directed meridional components are highlighted with brown shading. Trends are opposite those seen in MY 25.

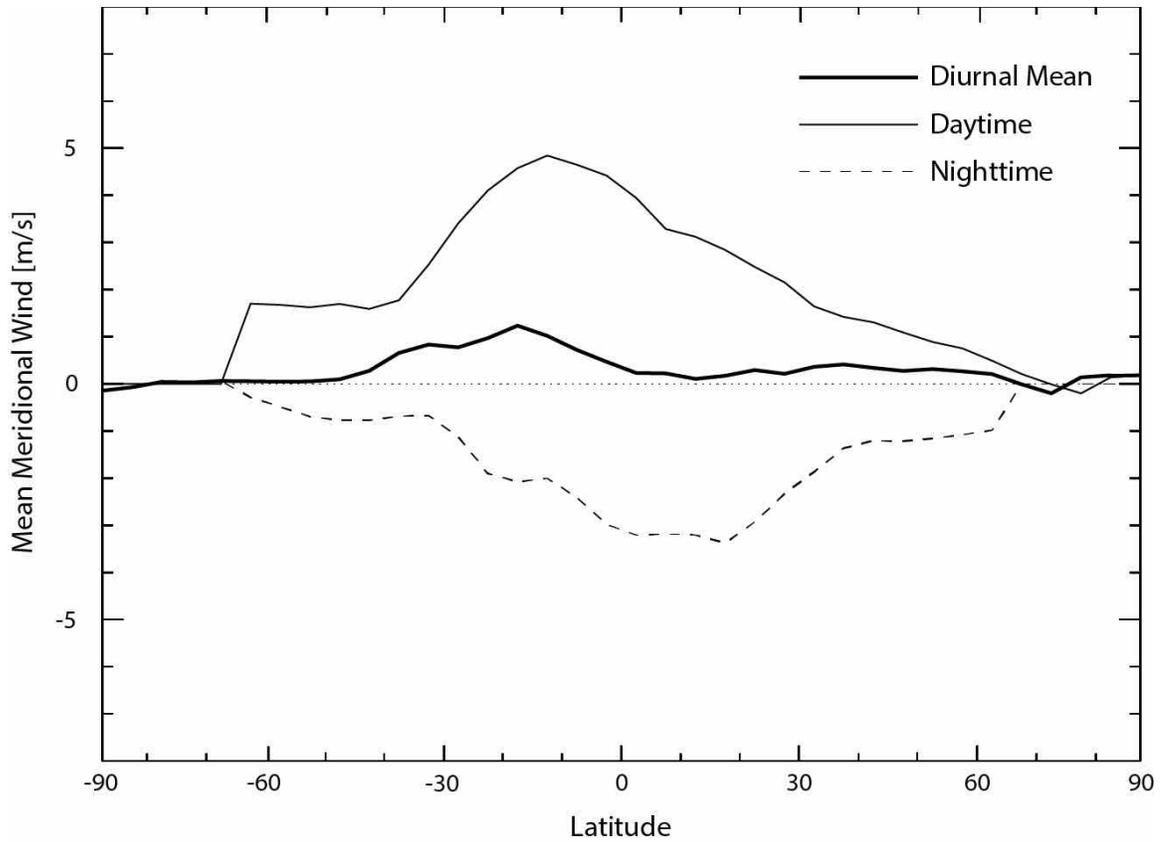

**Figure 13**: Same as Fig. 11, but for MY 28, a negative polarity case. Zonal mean near-surface meridional wind difference (CTA minus baseline) for daytime hours (thin solid line), nighttime hours (dashed line) and full day (thick solid line). Dotted line is a fiducial drawn at 0 m/s. Trends are opposite those seen in MY 25.

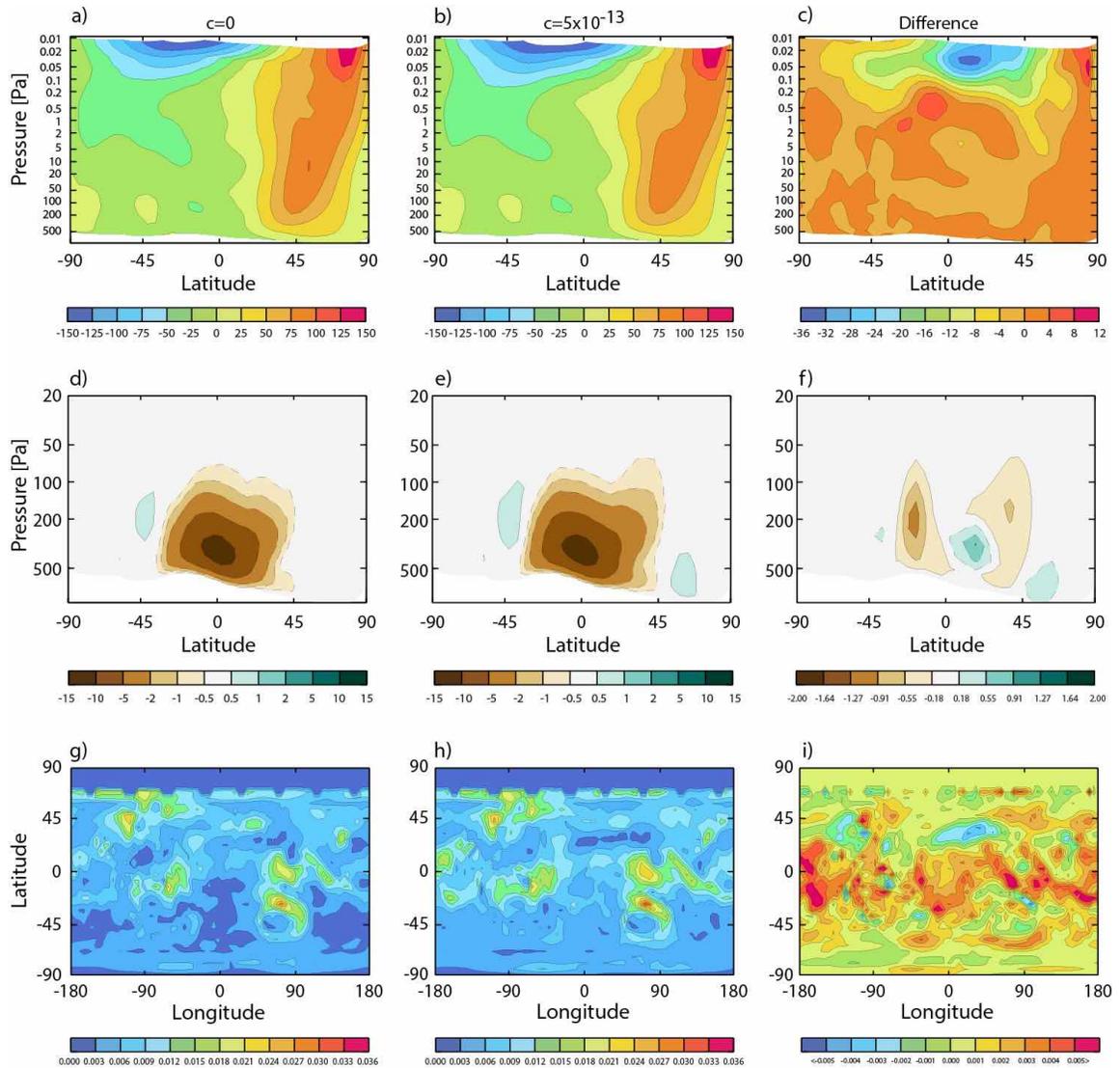

**Figure 14**: Mars Year 28 wind fields for the 'baseline' case (left column) and CTA case (middle column) along with differences between the two (right column) for (top) zonal-mean zonal wind [m/s], (middle) meridional streamfunction [$10^9$ kg/s], and (bottom) daytime-averaged surface stress [N m$^{-2}$]. Note the difference in the vertical (pressure) scales of the top two rows.

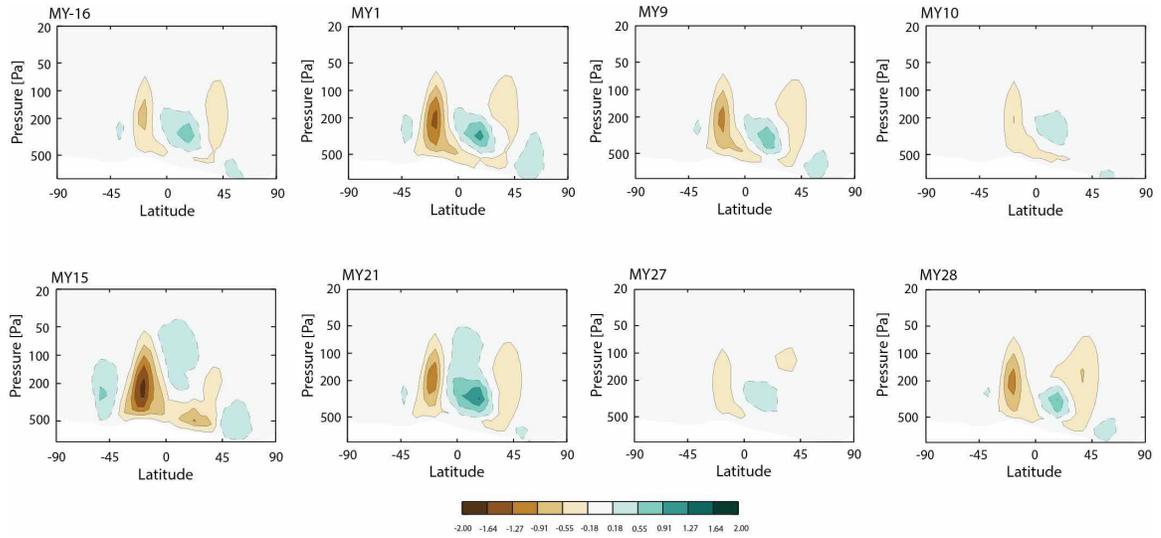

**Figure 15**: Streamfunction differences (forced minus unforced) for the positive polarity Mars years of Table 4. For years with GDS, the illustrated season corresponds to the period just prior to GDS initiation (Table 4). The Mars year 28 panel is identical to Fig. 14f. For the one non-GDS year (MY 27), the illustrated seasonal interval corresponds to the post-perihelion period ($L_s$=250-270°).

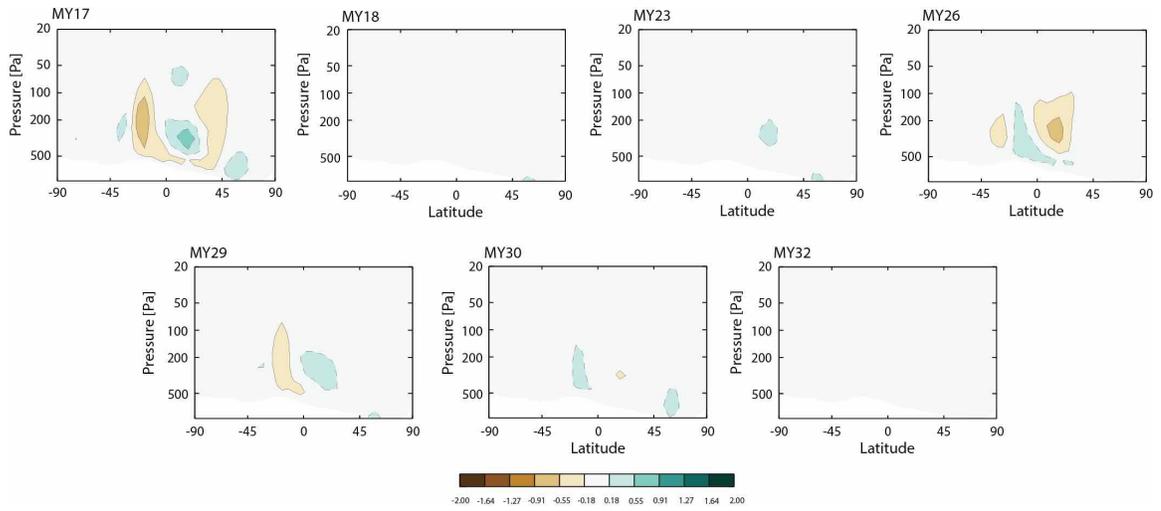

**Figure 16**: Streamfunction differences (forced minus unforced) for the transitional polarity (zero-crossing) Mars years of Table 4. In all cases, the seasonal interval modeled corresponds to the post-perihelion period ($L_s$=250-270°).

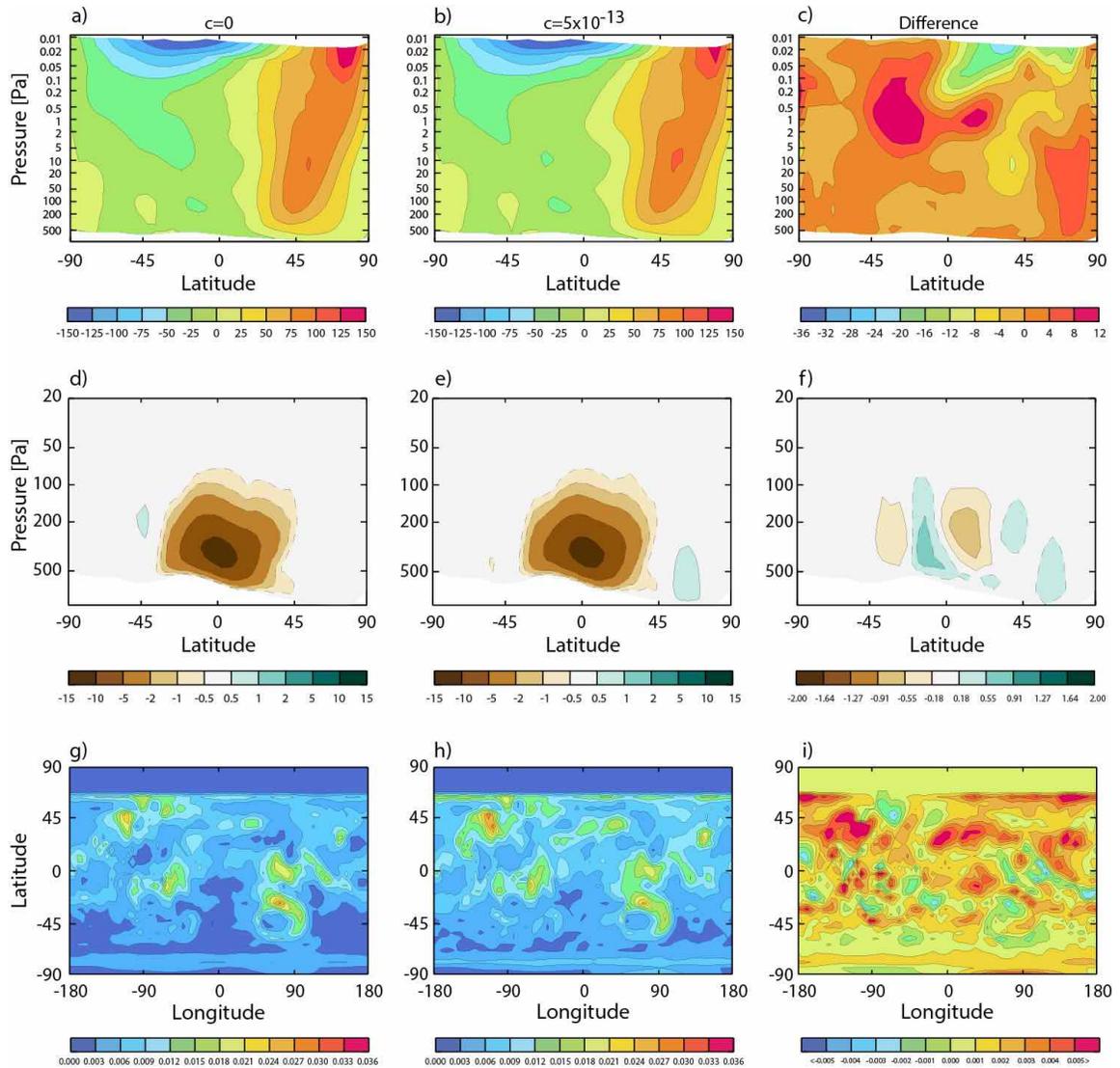

**Figure 17**: Mars Year 24 (a negative polarity, GDS-free perihelion season) wind fields for the 'baseline' case (left column) and CTA case (middle column) along with differences between the two (right column) for (top) zonal-mean zonal wind [m/s], (middle) meridional streamfunction [$10^9$ kg/s], and (bottom) daytime-averaged surface stress [N m$^{-2}$]. Note the difference in the vertical (pressure) scales of the top two rows.

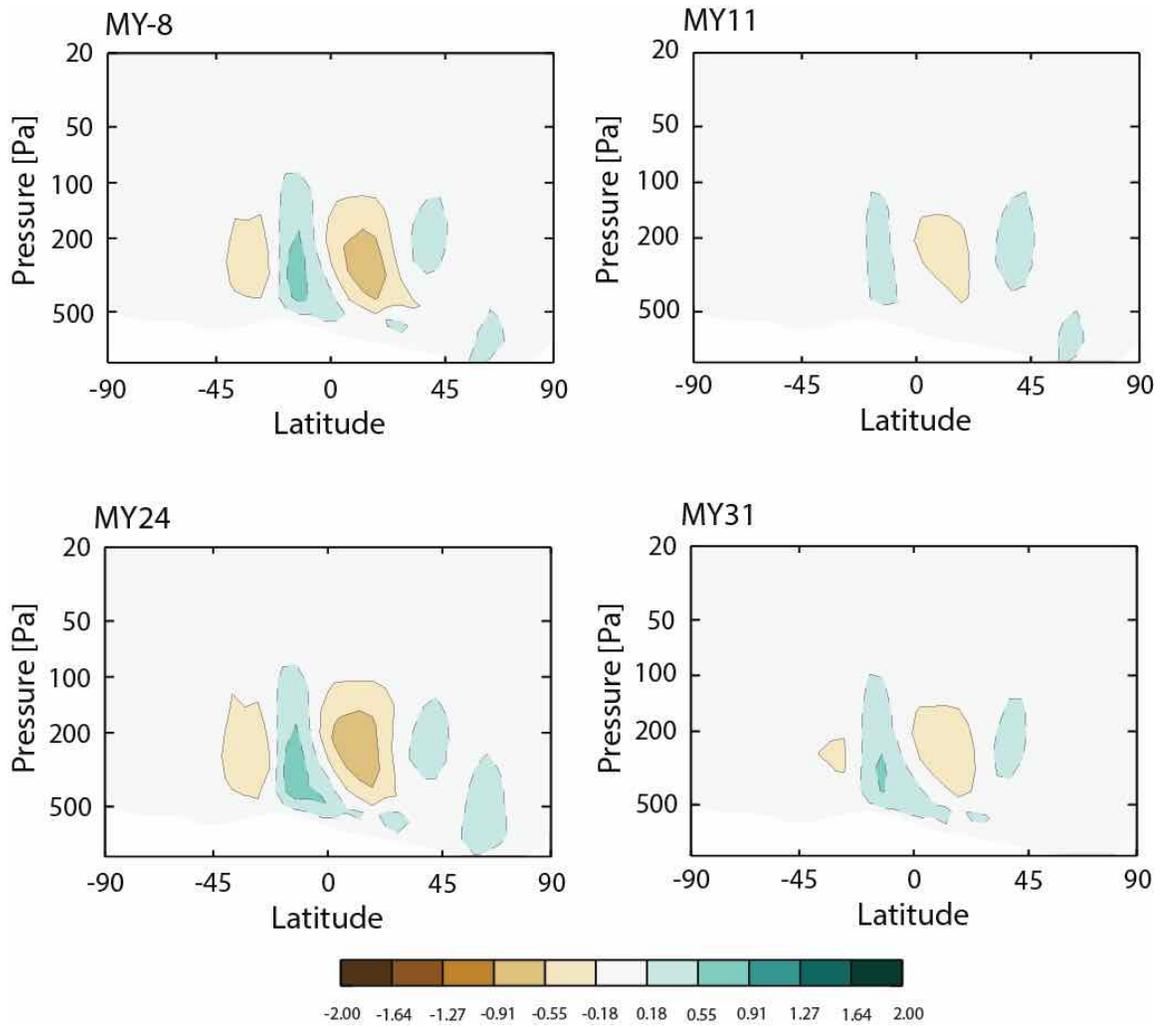

**Figure 18:** Streamfunction differences (forced minus unforced) for the negative polarity Mars years of Table 4 without GDS. In all cases, the seasonal interval modeled corresponds to the post-perihelion period ($L_s$=250-270°).

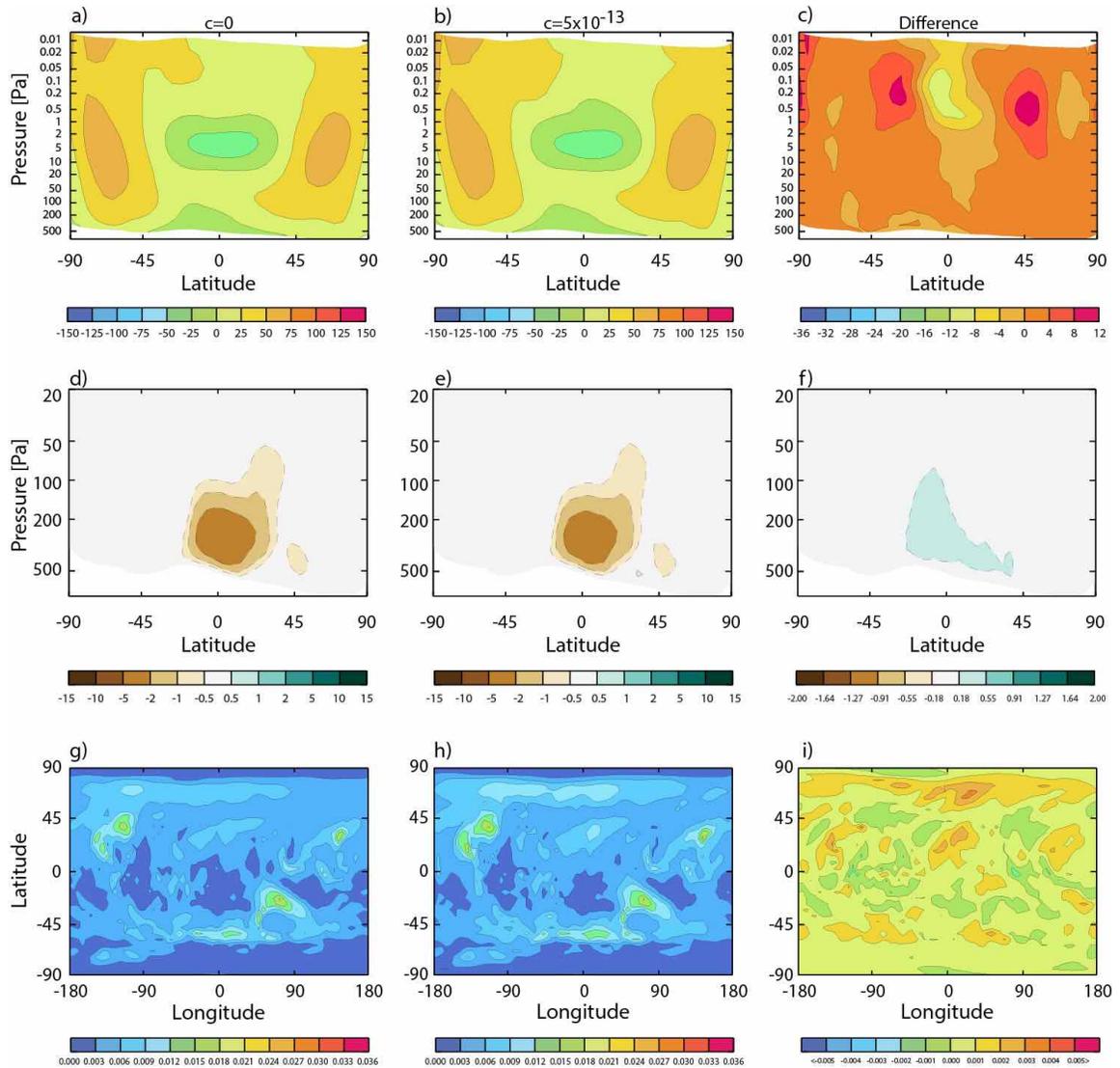

**Figure 19:** Same as Figs. 14 and 17, but for MY 25, a negative polarity year with equinox season GDS. Wind fields for the 'baseline' case (left column) and CTA case (middle column) along with differences between the two (right column) for (top) zonal-mean zonal wind [m/s], (middle) meridional streamfunction [$10^9$ kg/s], and (bottom) daytime-averaged surface stress [N m$^{-2}$] are shown. Note the difference in the vertical (pressure) scales of the top two rows. Values averaged over the season spanning $L_s$=165°-185°.

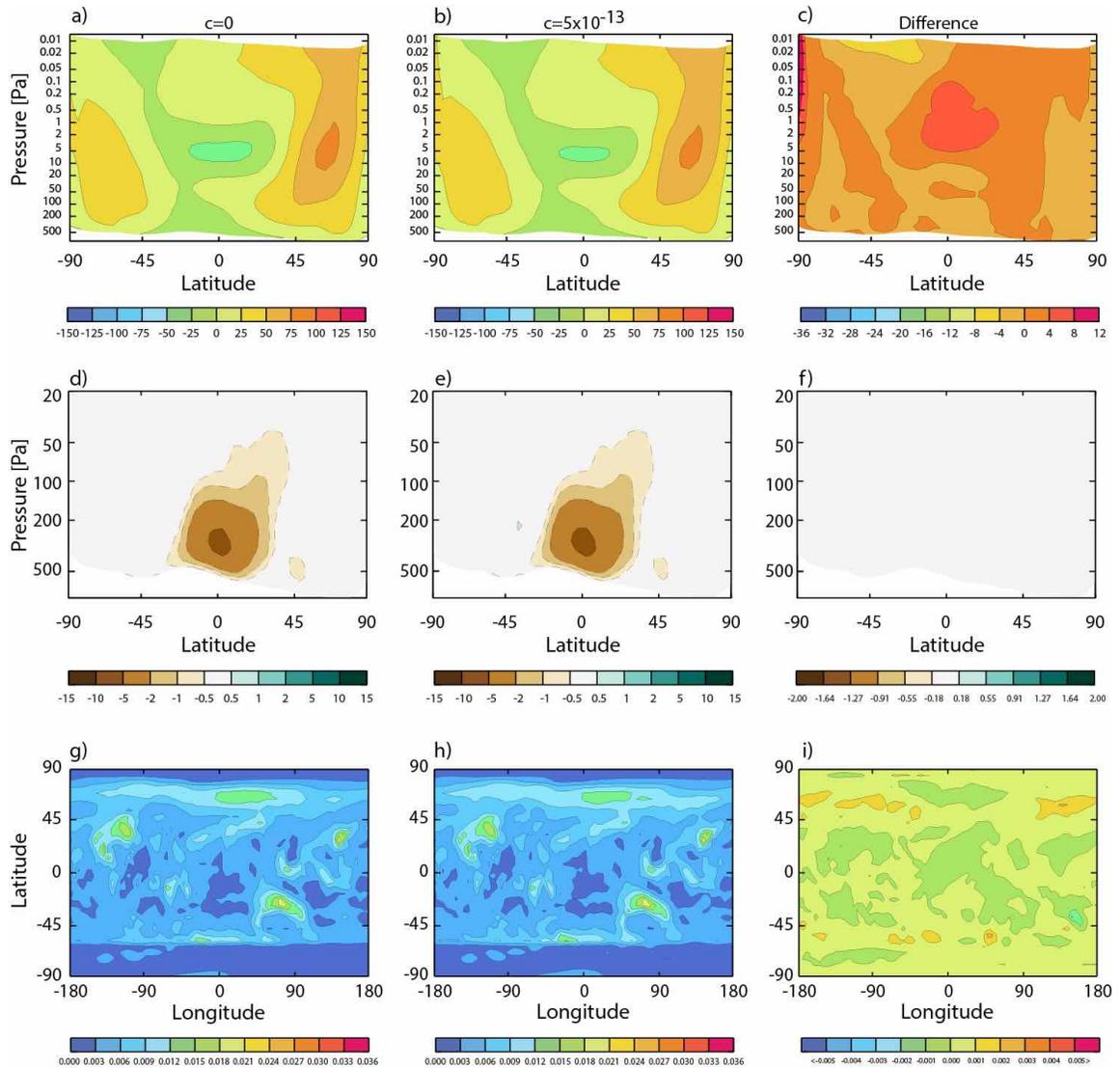

**Figure 20:** Same as Figs. 14, 17 and 19, but for MY 12, a negative polarity year with equinox season GDS. Wind fields for the 'baseline' case (left column) and CTA case (middle column) along with differences between the two (right column) for (top) zonal-mean zonal wind [m/s], (middle) meridional streamfunction [$10^9$ kg/s], and (bottom) daytime-averaged surface stress [N m$^{-2}$] are shown. Note the difference in the vertical (pressure) scales of the top two rows. Values averaged over the season spanning $L_s$=180°-200°.

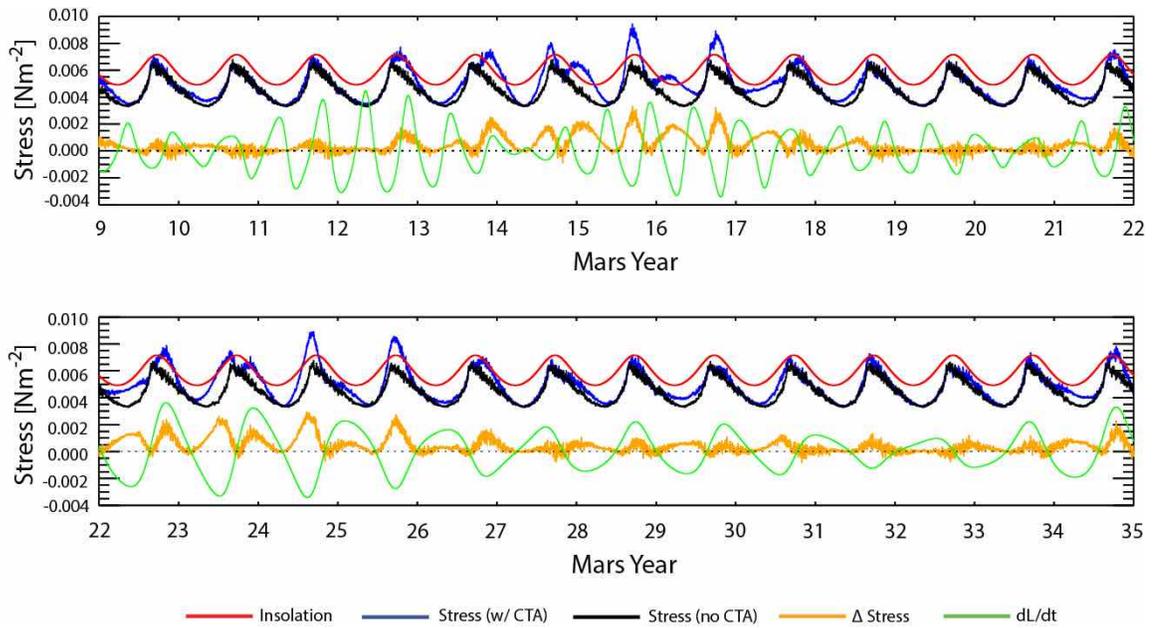

**Figure 21**: Global mean daytime surface wind stresses (N m$^{-2}$) with coupling term accelerations (blue) compared with the baseline case (in black) for MY 9-34. The surface wind stress differences between the two simulations are shown in gold. The $\dot{L}$ waveform is shown in green (arbitrary units, but scaled to the zero line), to illustrate the phasing of the putative forcing function, and the seasonal insolation cycle (also in arbitrary units) is shown in red. (As in Fig. 6, but expanded to encompass MY 9-34).

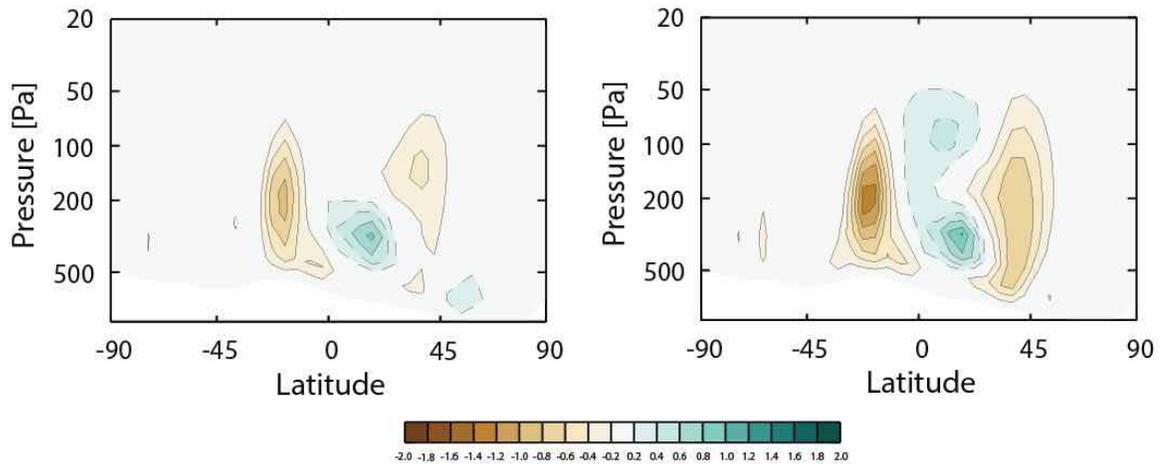

**Figure 22**: Meridional streamfunction differences (forced minus unforced) at perihelion ($L_s$=250-270°) for MY 33 (left) and MY 34 (right) as a CTA model prediction.

| Mars Year | $\varphi_{dL/dt}$ | Polarity (aphelion) |
|---|---|---|
| 25 | 122 | + |
| 26 | 52 | + |
| 27 | 355 | 0 |
| 28 | 299 | - |
| 29 | 234 | - |
| 30 | 187 | 0 |

**Table 1:** Polarity and phase assignments (with respect to aphelion) for the six Mars years illustrated in Fig. 4 (see discussion in text). The zero symbol signifies transitional interval.

| Mars Year | $dL/dt$ (aphelion) | Polarity | Control model Wind Stress at Aphelion | CTA model Wind Stress at Aphelion | Difference | Ratio |
|---|---|---|---|---|---|---|
| 25 | 122 | + | 0.00383 | 0.00441 | 0.00058 | 1.151 |
| 26 | 52 | + | 0.00383 | 0.00411 | 0.00027 | 1.072 |
| 27 | 355 | 0 | 0.00383 | 0.00389 | 0.00006 | 1.016 |
| 28 | 299 | - | 0.00383 | 0.00455 | 0.00072 | 1.187 |
| 29 | 234 | - | 0.00383 | 0.00471 | 0.00088 | 1.229 |
| 30 | 187 | 0 | 0.00383 | 0.00399 | 0.00016 | 1.042 |

**Table 2:** Comparison of modeled global mean daytime surface wind stresses, in N m$^{-2}$, with and without the orbit-spin coupling (CTA) accelerations, for MY 25-30, as illustrated in Figs. 5 and 6.

|           | Baseline  | MY 28     | MY 25     |
|-----------|-----------|-----------|-----------|
| Daytime   | 0.003832  | 0.004548  | 0.004410  |
| Nighttime | 0.002753  | 0.002842  | 0.003248  |
| Full Day  | 0.003459  | 0.003869  | 0.004000  |

**Table 3**: Global average surface wind stress (in N m$^{-2}$) for our baseline case, and for MY 25 and MY 28, partitioned as 'daytime', 'nighttime' and 'full day'. In all cases, the daytime surface stresses exceed the nighttime surface stresses—a result of the generally stronger winds during daytime hours on Mars (Section 2.3).

| Mars Year | Year (perihelion) | GDS inception (Ls) | $\varphi_{dL/dt}$ (perihelion) | Polarity | Pre-GDS Ls range |
|---|---|---|---|---|---|
| -16 | 1924 | 310 | 92.6 | + | 290-310 |
| -8 | 1939 |  | 309.1 | - |  |
| 1 | 1956 | 249 | 143.7 | + | 230-250 |
| 9 | 1971 | 260 | 92.5 | + | 240-260 |
| 10 | 1973 | 300 | 44.4 | + | 280-300 |
| 11 | 1975 |  | 302.5 | - |  |
| 12 | 1977 | 204, 268 | 232.7 | - | 180-200 |
| 15 | 1982 | 208 | 98.7 | + | 190-210 |
| 17 | 1986 |  | 38.6 | 0 |  |
| 18 | 1988 |  | 1.5 | 0 |  |
| 21 | 1994 | 254 | 70.3 | + | 230-250 |
| 23 | 1998 |  | 1.6 | 0 |  |
| 24 | 1999 |  | 313.7 | - |  |
| 25 | 2001 | 185 | 272.4 | - | 165-185 |
| 26 | 2003 |  | 213 | 0 |  |
| 27 | 2005 |  | 134.5 | + |  |
| 28 | 2007 | 262 | 82.4 | + | 240-260 |
| 29 | 2009 |  | 37.8 | 0 |  |
| 30 | 2011 |  | 342.9 | 0 |  |
| 31 | 2013 |  | 272.1 | - |  |
| 32 | 2014 |  | 174.6 | 0 |  |

**Table 4**: List of Mars years with and without global-scale dust storms (GDS), after *Shirley and Mischna* (2016). For years with GDS, the L$_s$ of storm initiation is indicated. Blue shading indicates equinox-season (or "early season") storms, while orange shading indicates years with perihelion-season GDS, as defined in *Shirley* (2015) and *Shirley and Mischna* (2016). Phase (φ) and polarity values were obtained as described in Section 3.3. Comparison of the polarity values of Table 4 for Mars years 25-30 with those of Table 1 reveals that the polarities are reversed from the earlier table. In Section 3 we were focusing on the aphelion time period (i.e., in Table 1 and Fig. 4), while here we focus instead on the perihelion season. The polarity of the $\dot{L}$ waveform for MY 28 is negative at aphelion, but is positive during the subsequent perihelion season, one-half of a Mars year later.

| Mars Year | $\varphi dL/dt$ (perihelion) | Polarity | CTA model Wind Stress at Perihelion | Pre-GDS $L_s$ range employed | CTA Model Wind stress prior to GDS | Baseline Model Wind stress prior to GDS |
|---|---|---|---|---|---|---|
| -16 | 92.6 | + | 0.00831316 | 290-310 | 0.00682366 | 0.00596749 |
| -8 | 309.1 | - | 0.0072988 | | | |
| 1 | 143.7 | + | 0.00684334 | 230-250 | 0.00836876 | 0.00672248 |
| 9 | 92.5 | + | 0.00733521 | 240-260 | 0.0076916 | 0.00643261 |
| 10 | 44.4 | + | 0.00662246 | 280-300 | 0.00659548 | 0.00602629 |
| 11 | 302.5 | - | 0.00664439 | | | |
| 12 | 232.7 | - | 0.00718216 | 180-200 | 0.00557475 | 0.00540606 |
| 15 | 98.7 | + | 0.0103328 | 190-210 | 0.00813745 | 0.00583158 |
| 17 | 38.6 | 0 | 0.00742453 | | | |
| 18 | 1.5 | 0 | 0.00629366 | | | |
| 21 | 70.3 | + | 0.008581 | 230-250 | 0.00848288 | 0.00672248 |
| 23 | 1.6 | 0 | 0.006349 | | | |
| 24 | 313.7 | - | 0.00757812 | | | |
| 25 | 272.4 | - | 0.00855013 | 165-185 | 0.00539678 | 0.00485226 |
| 26 | 213 | 0 | 0.0067593 | | | |
| 27 | 134.5 | + | 0.00649666 | | | |
| 28 | 82.4 | + | 0.00740818 | 240-260 | 0.00771413 | 0.00643261 |
| 29 | 37.8 | 0 | 0.00675689 | | | |
| 30 | 342.9 | 0 | 0.00636981 | | | |
| 31 | 272.1 | - | 0.00671778 | | | |
| 32 | 174.6 | 0 | 0.00620588 | | | |

**Table 5:** Model-derived global mean daytime surface wind stress values following perihelion for all Mars years of Table 4, in N m$^{-2}$. The interval for the perihelion period calculations is from $L_s$=250°-270°. Forced and unforced global mean daytime surface wind stress values are also calculated for pre-storm–inception intervals in years with global-scale dust storms (see discussion in text).

| Population 1 | $n_1$ | Population 2 | $n_2$ | Seasonal Interval | z | p |
|---|---|---|---|---|---|---|
| Years without global storms | 12 | Years with global storms | 9 | Perihelion season (Ls 250-270) | 2.59 | 0.005 |
| Transitional polarity years | 7 | Positive polarity years | 8 | Perihelion season (Ls 250-270) | 2.14 | 0.02 |
| Transitional polarity years | 7 | Negative polarity years | 6 | Perihelion season (Ls 250-270) | 1.79 | 0.04 |
| Transitional polarity years | 7 | (+) and (-) polarity years | 14 | Perihelion season (Ls 250-270) | 2.35 | 0.01 |
| Positive polarity years | 8 | Negative polarity years | 6 | Perihelion season (Ls 250-270) | 0.33 | 0.37 |
| Storms baseline model stresses | 9 | Storms forced model stresses | 9 | Pre-storm intervals (Table 5) | 2.03 | 0.02 |

**Table 6:** Results of the application of the Mann-Whitney Test for the equality of sample means as applied to the global mean daytime surface wind stress data of Table 5. The test *z* statistic and the associated probability level *p* are provided in the last two columns.

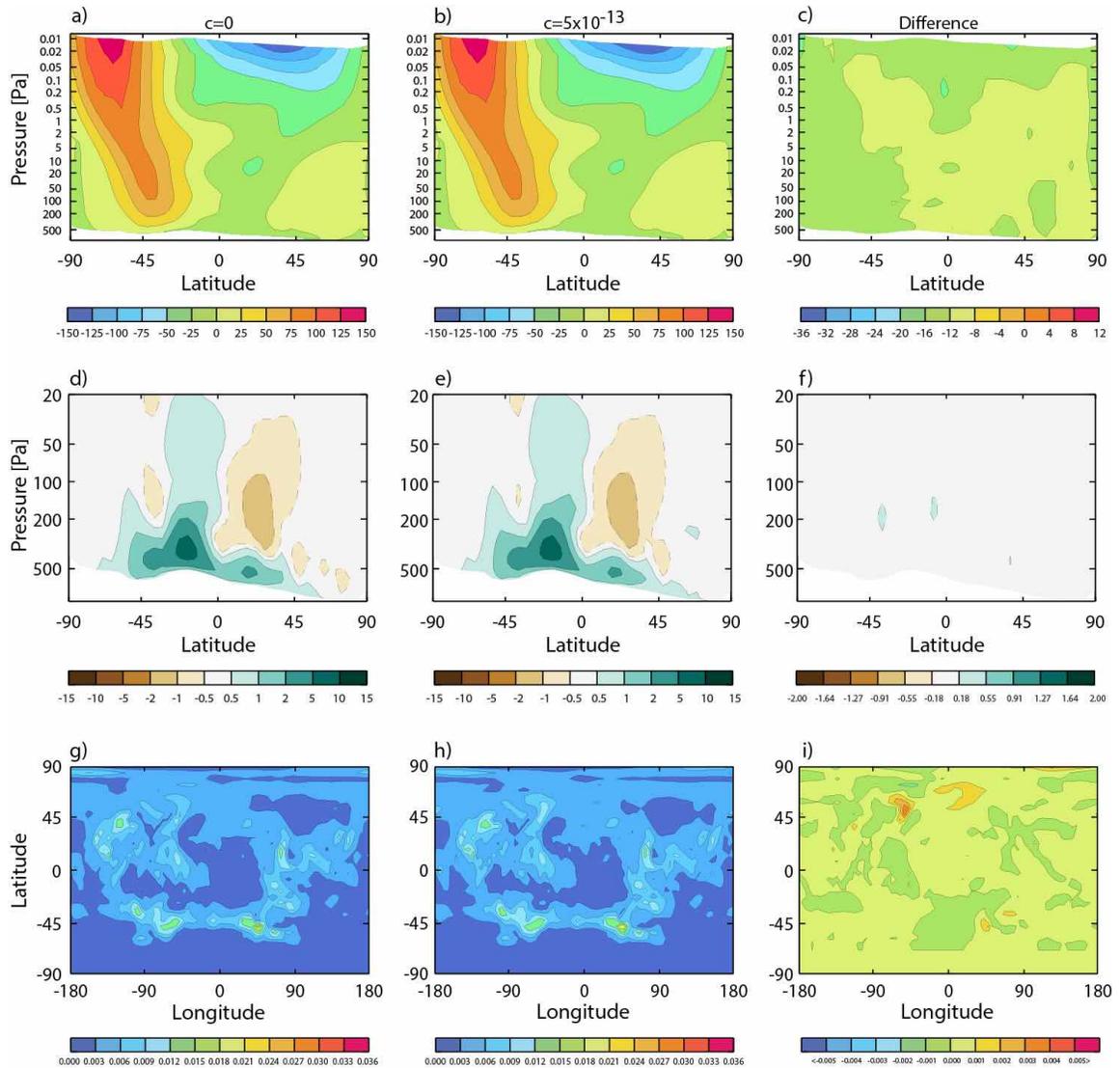

**Figure S1**: Mars Year 30 aphelion wind fields for the 'baseline' case (left column) and CTA case (middle column) along with differences between the two (right column) for (top) vertical slice of zonal-mean zonal wind [m/s], (middle) vertical slice of meridional streamfunction [$10^9$ kg/s], and (bottom) map view of daytime-averaged surface stress [N m$^{-2}$]. Note the difference in the vertical (pressure) scales of the top two rows.

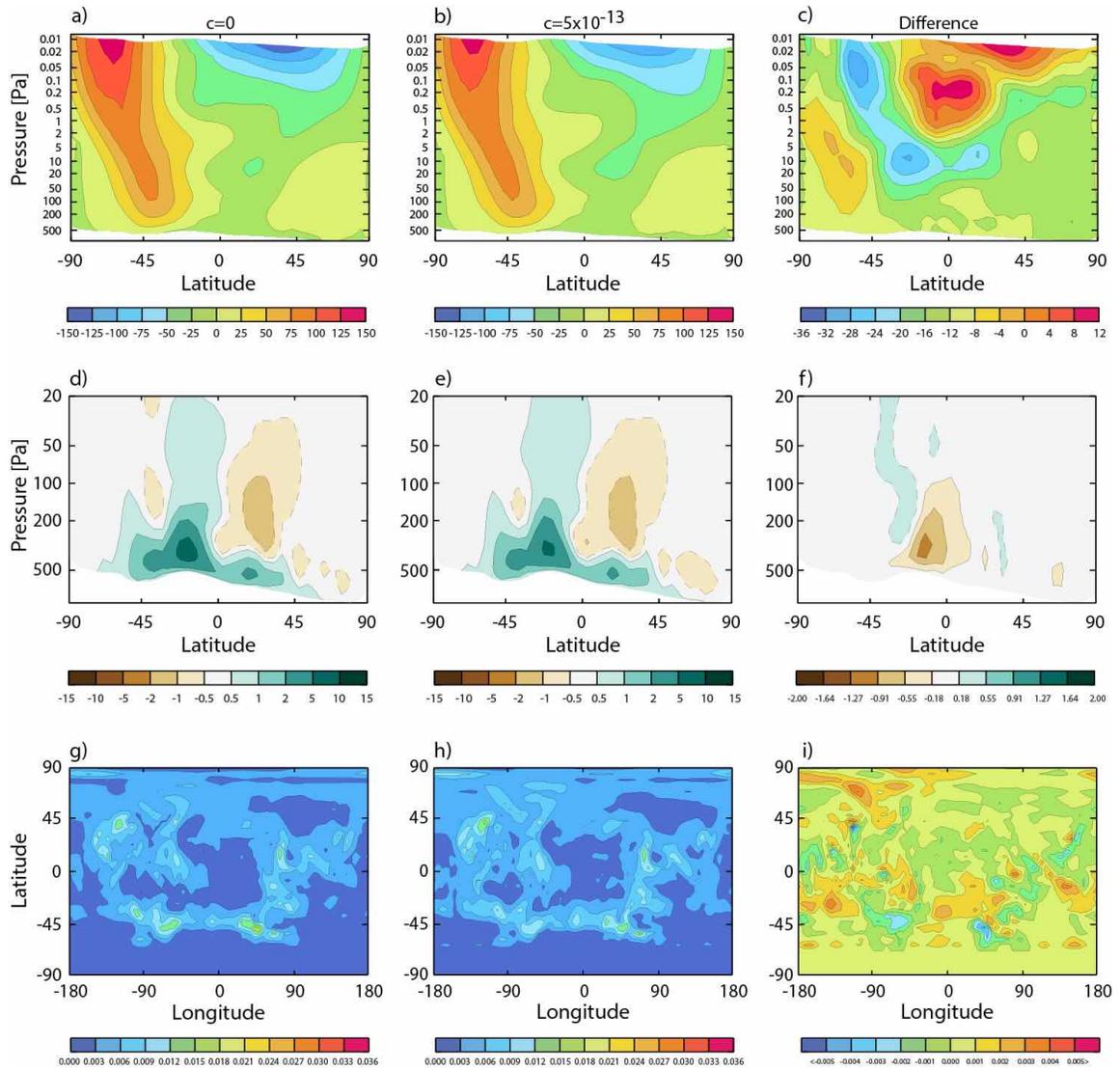

**Figure S2**: Mars Year 26 aphelion wind fields for the 'baseline' case (left column) and CTA case (middle column) along with differences between the two (right column) for (top) vertical slice of zonal-mean zonal wind [m/s], (middle) vertical slice of meridional streamfunction [$10^9$ kg/s], and (bottom) map view of daytime-averaged surface stress [N m$^{-2}$]. Note the difference in the vertical (pressure) scales of the top two rows.

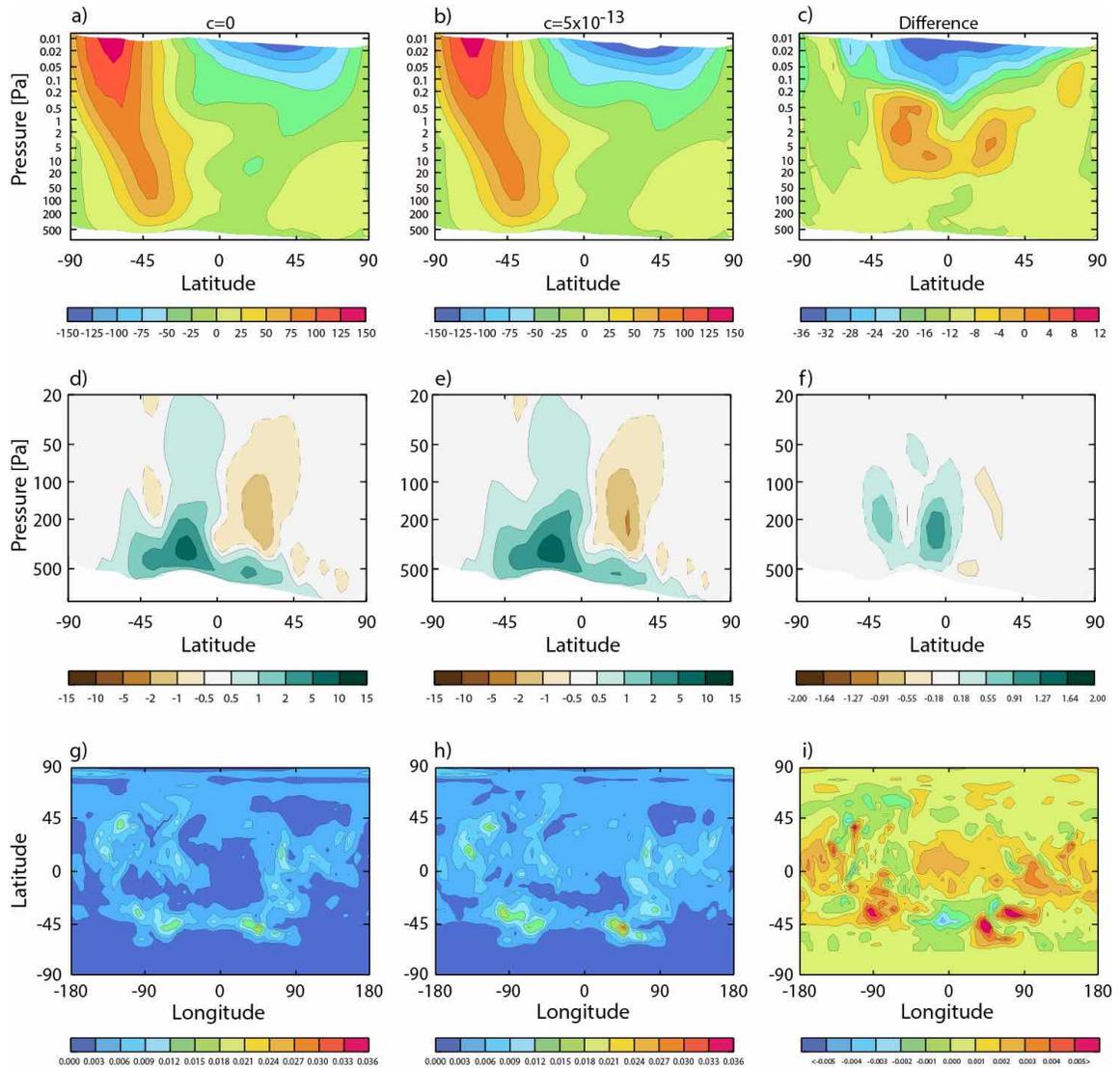

**Figure S3**: Mars Year 29 aphelion wind fields for the 'baseline' case (left column) and CTA case (middle column) along with differences between the two (right column) for (top) vertical slice of zonal-mean zonal wind [m/s], (middle) vertical slice of meridional streamfunction [$10^9$ kg/s], and (bottom) map view of daytime-averaged surface stress [N m$^{-2}$]. Note the difference in the vertical (pressure) scales of the top two rows.

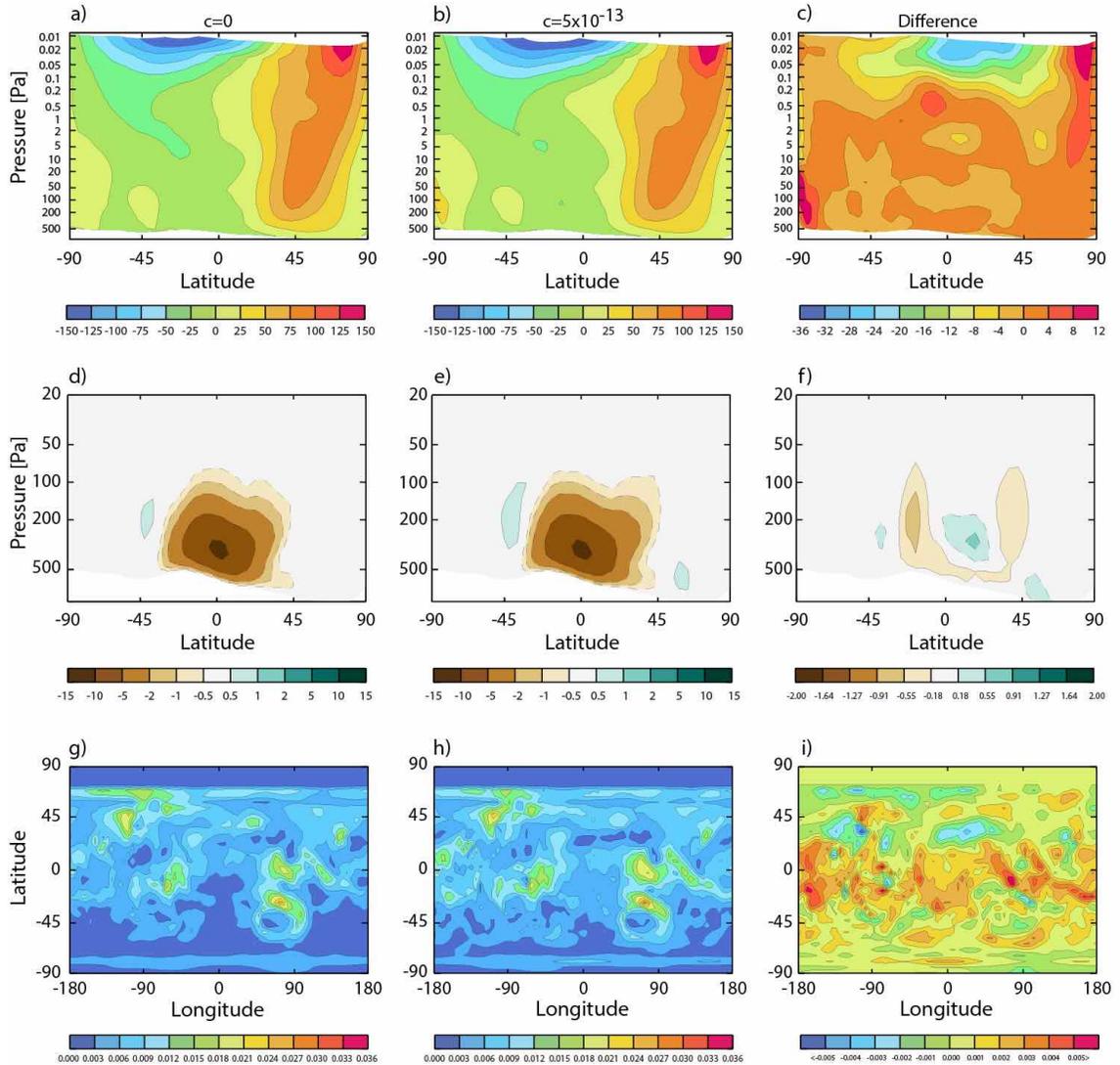

**Figure S4**: Mars Year -16 wind fields for the 'baseline' case (left column) and CTA case (middle column) along with differences between the two (right column) for (top) vertical slice of zonal-mean zonal wind [m/s], (middle) vertical slice of meridional streamfunction [$10^9$ kg/s], and (bottom) map view of daytime-averaged surface stress [N m$^{-2}$]. Timeframe spans $L_s$=290-310°. Note the difference in the vertical (pressure) scales of the top two rows.

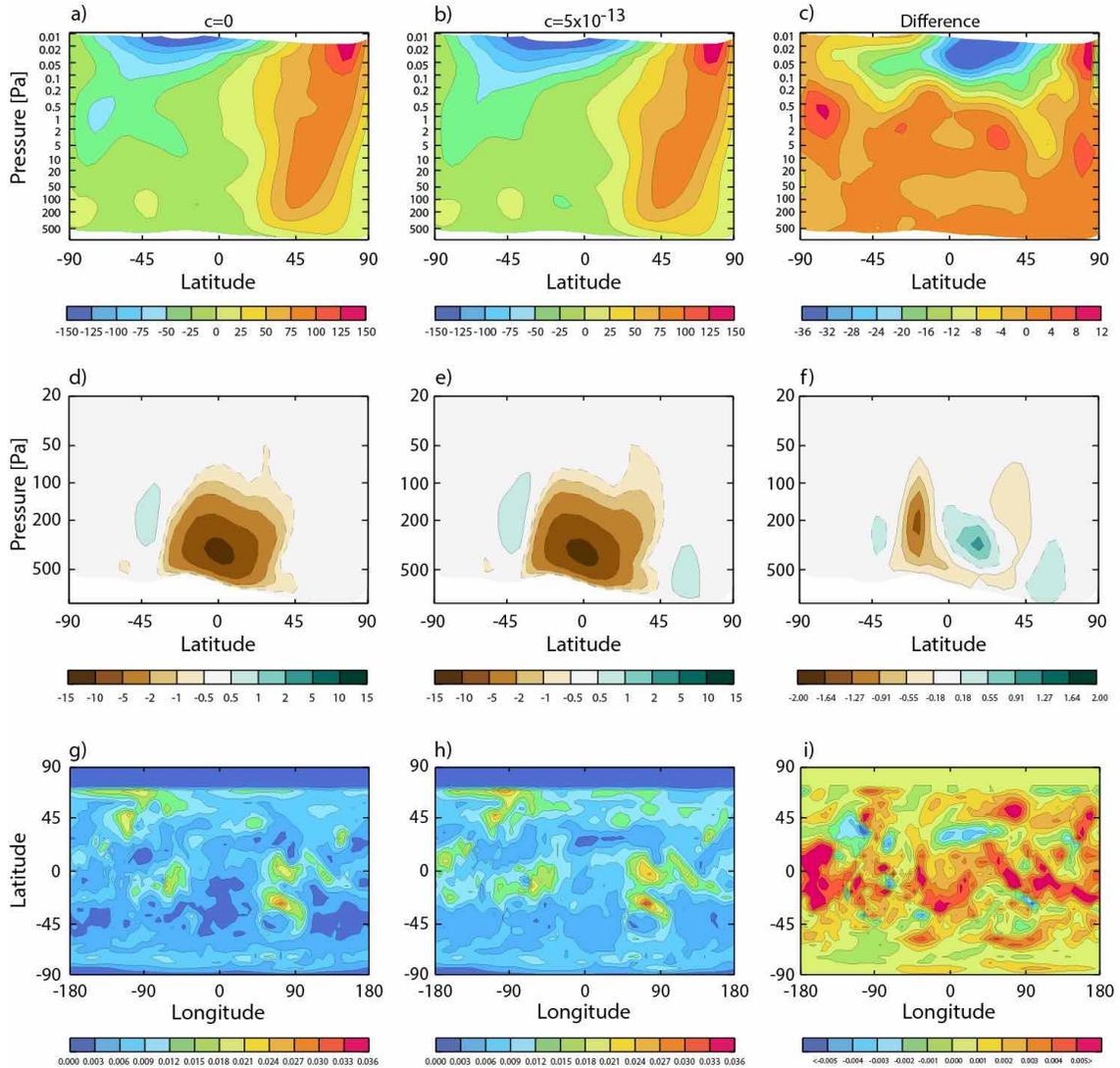

**Figure S5**: Mars Year 1 wind fields for the 'baseline' case (left column) and CTA case (middle column) along with differences between the two (right column) for (top) vertical slice of zonal-mean zonal wind [m/s], (middle) vertical slice of meridional streamfunction [$10^9$ kg/s], and (bottom) map view of daytime-averaged surface stress [N m$^{-2}$]. Timeframe spans $L_s$=230-250°. Note the difference in the vertical (pressure) scales of the top two rows.

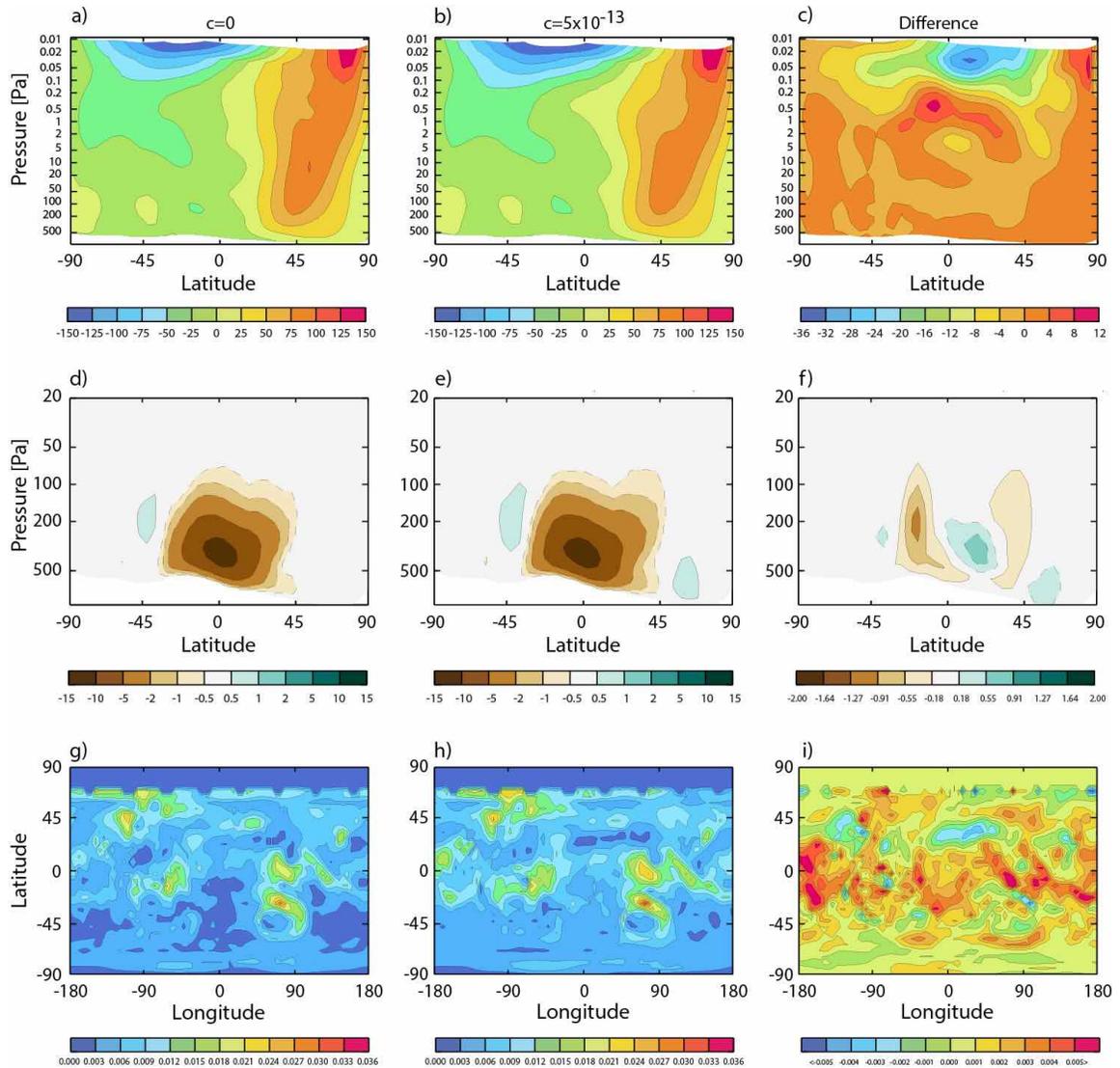

**Figure S6**: Mars Year 9 perihelion wind fields for the 'baseline' case (left column) and CTA case (middle column) along with differences between the two (right column) for (top) vertical slice of zonal-mean zonal wind [m/s], (middle) vertical slice of meridional streamfunction [$10^9$ kg/s], and (bottom) map view of daytime-averaged surface stress [N m$^{-2}$]. Note the difference in the vertical (pressure) scales of the top two rows.

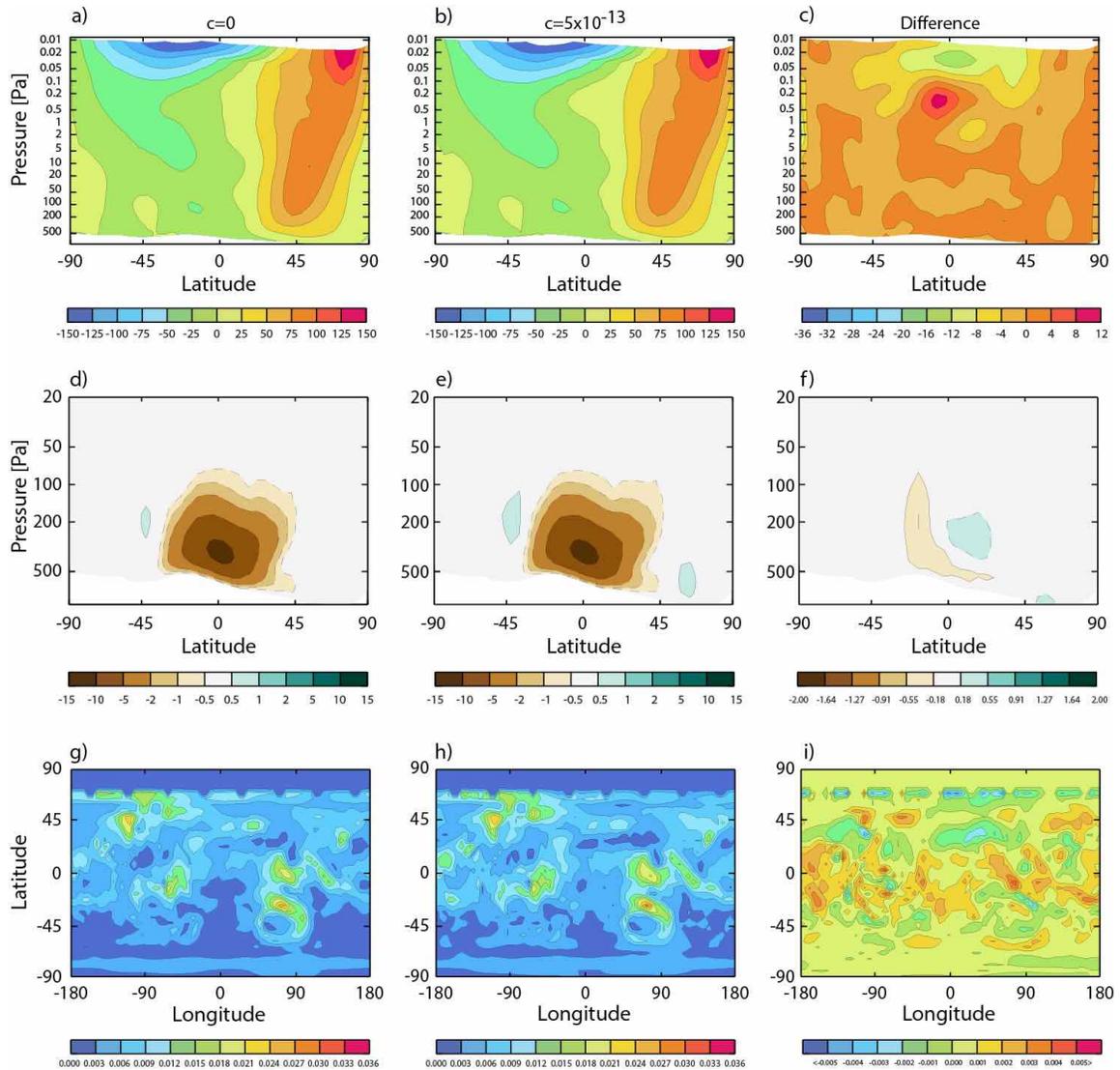

**Figure S7**: Mars Year 10 wind fields for the 'baseline' case (left column) and CTA case (middle column) along with differences between the two (right column) for (top) vertical slice of zonal-mean zonal wind [m/s], (middle) vertical slice of meridional streamfunction [$10^9$ kg/s], and (bottom) map view of daytime-averaged surface stress [N m$^{-2}$]. Timeframe spans $L_s$=280-300°. Note the difference in the vertical (pressure) scales of the top two rows.

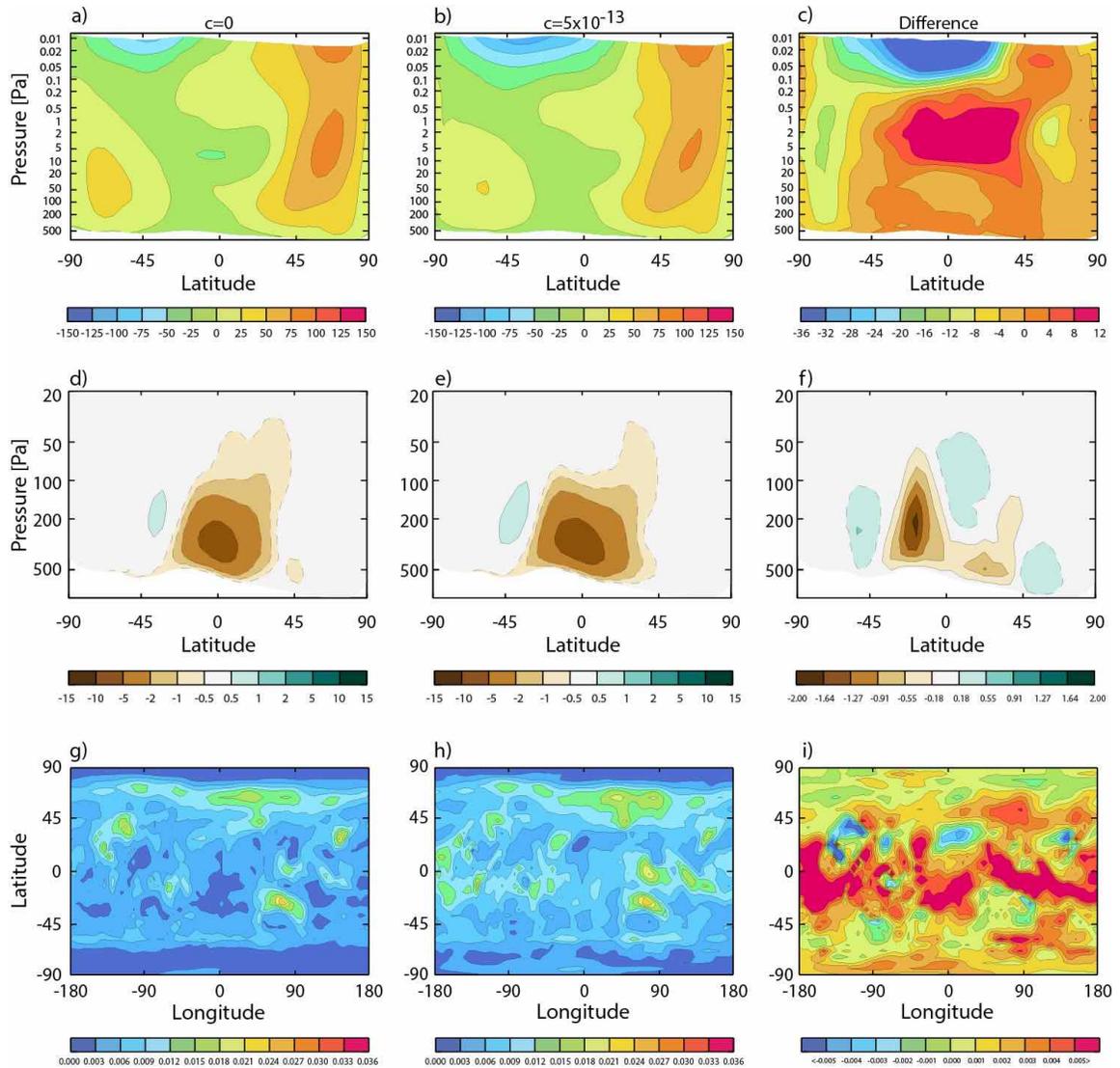

**Figure S8**: Mars Year 15 wind fields for the 'baseline' case (left column) and CTA case (middle column) along with differences between the two (right column) for (top) vertical slice of zonal-mean zonal wind [m/s], (middle) vertical slice of meridional streamfunction [$10^9$ kg/s], and (bottom) map view of daytime-averaged surface stress [N m$^{-2}$]. Timeframe spans $L_s$=190-210°. Note the difference in the vertical (pressure) scales of the top two rows.

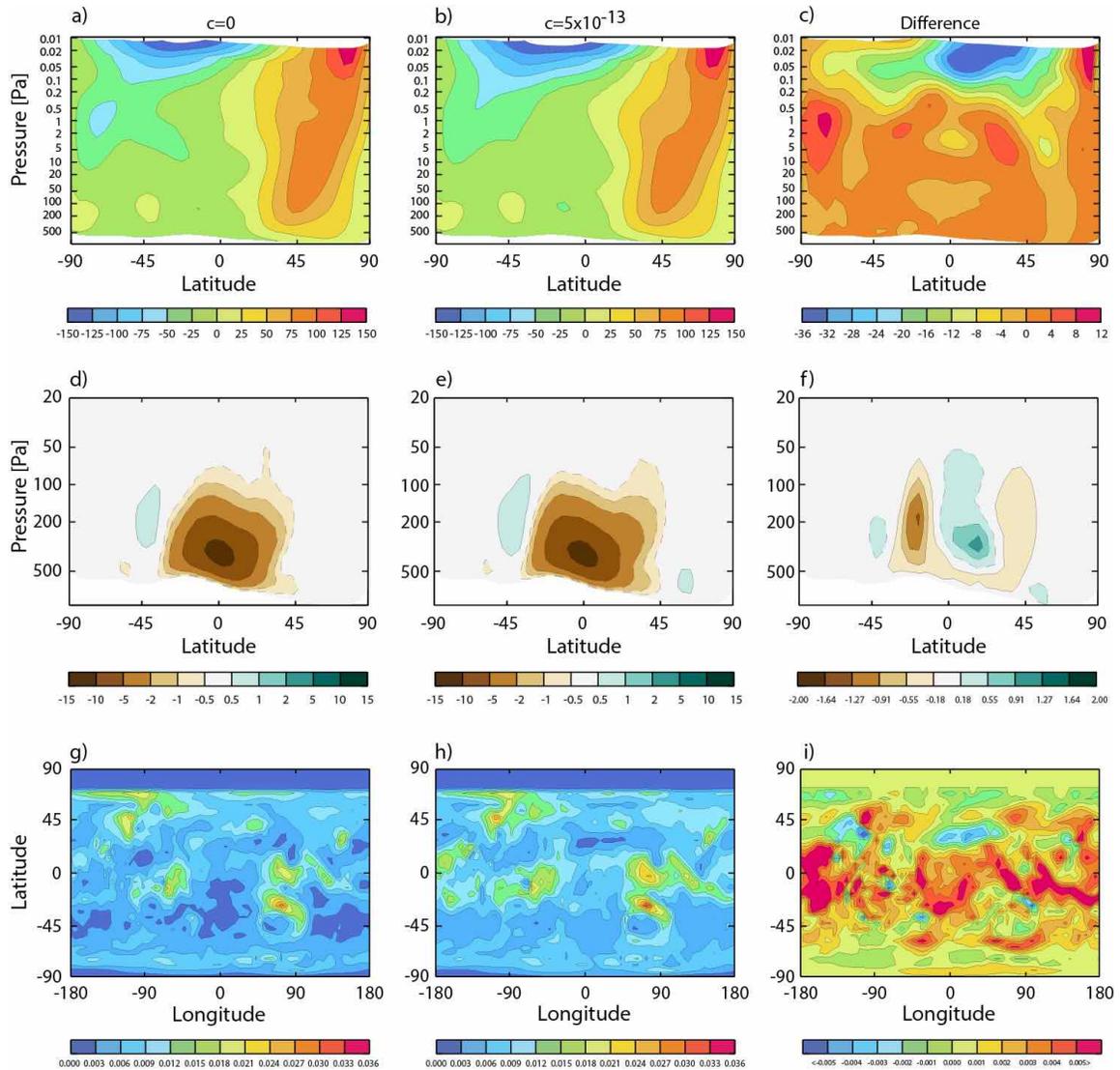

**Figure S9**: Mars Year 21 wind fields for the 'baseline' case (left column) and CTA case (middle column) along with differences between the two (right column) for (top) vertical slice of zonal-mean zonal wind [m/s], (middle) vertical slice of meridional streamfunction [$10^9$ kg/s], and (bottom) map view of daytime-averaged surface stress [N m$^{-2}$]. Timeframe spans $L_s$=230-250°. Note the difference in the vertical (pressure) scales of the top two rows.

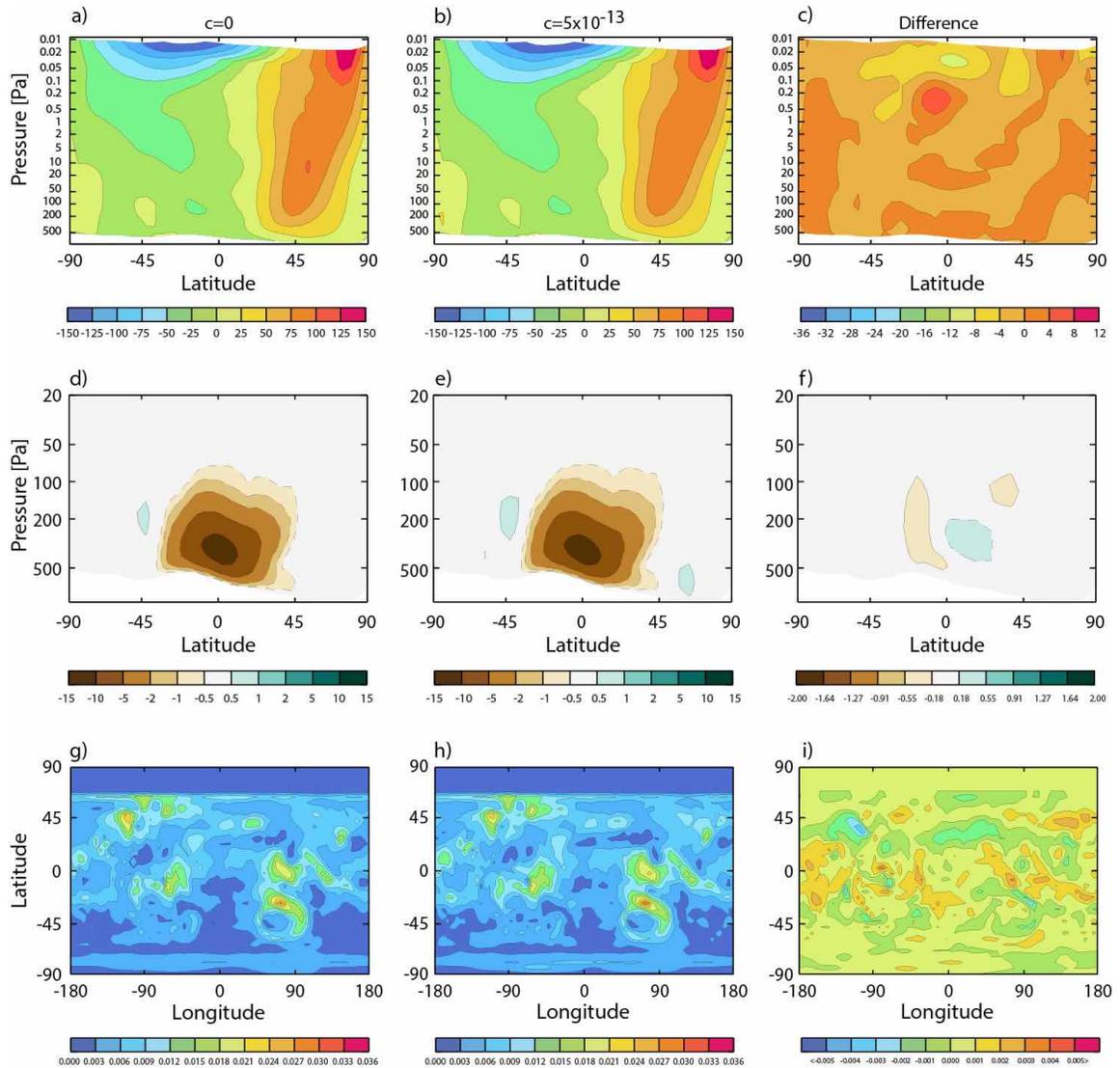

**Figure S10**: Mars Year 27 perihelion wind fields for the 'baseline' case (left column) and CTA case (middle column) along with differences between the two (right column) for (top) vertical slice of zonal-mean zonal wind [m/s], (middle) vertical slice of meridional streamfunction [$10^9$ kg/s], and (bottom) map view of daytime-averaged surface stress [N m$^{-2}$]. Note the difference in the vertical (pressure) scales of the top two rows.

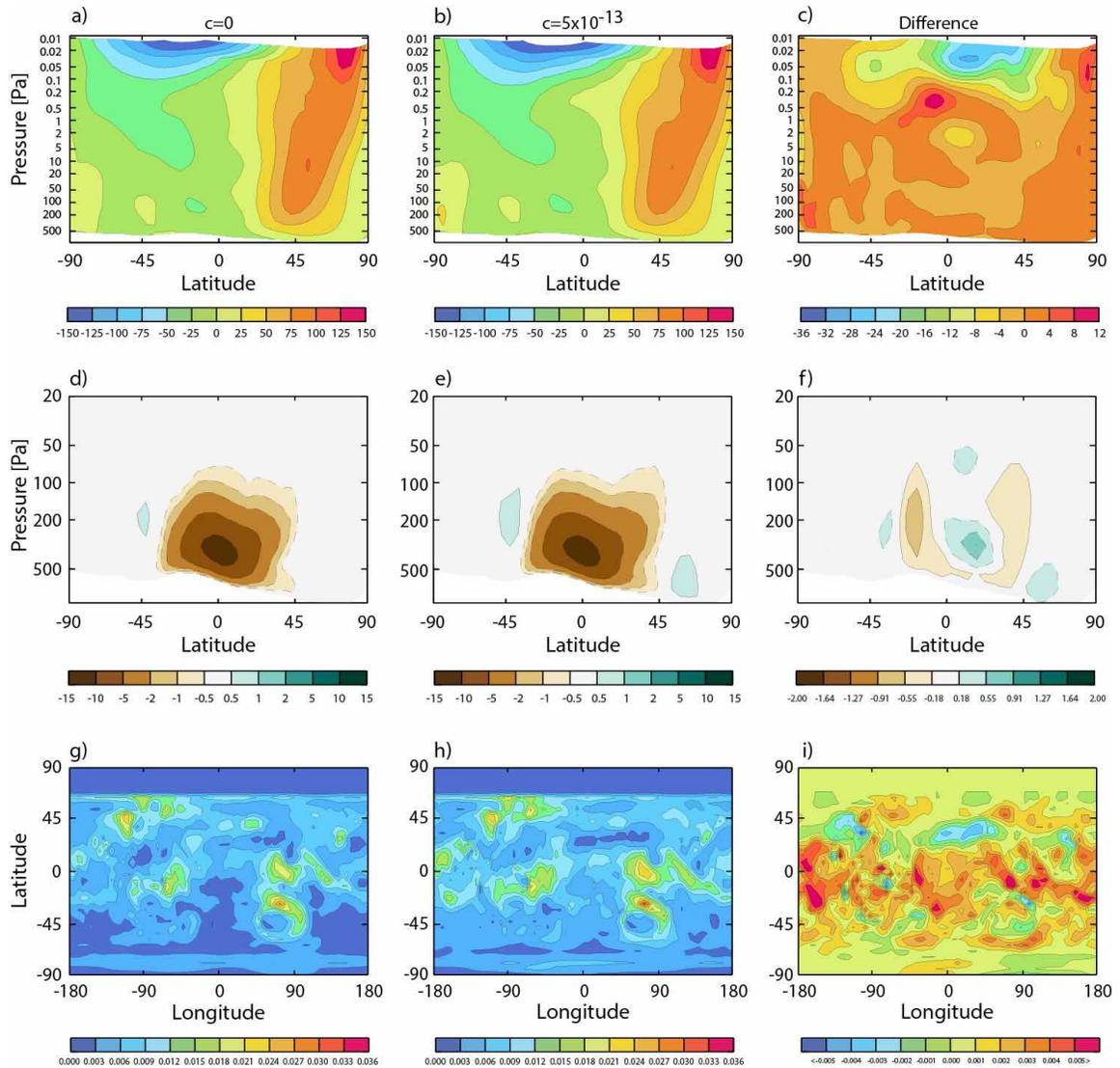

**Figure S11**: Mars Year 17 perihelion wind fields for the 'baseline' case (left column) and CTA case (middle column) along with differences between the two (right column) for (top) vertical slice of zonal-mean zonal wind [m/s], (middle) vertical slice of meridional streamfunction [$10^9$ kg/s], and (bottom) map view of daytime-averaged surface stress [N m$^{-2}$]. Note the difference in the vertical (pressure) scales of the top two rows.

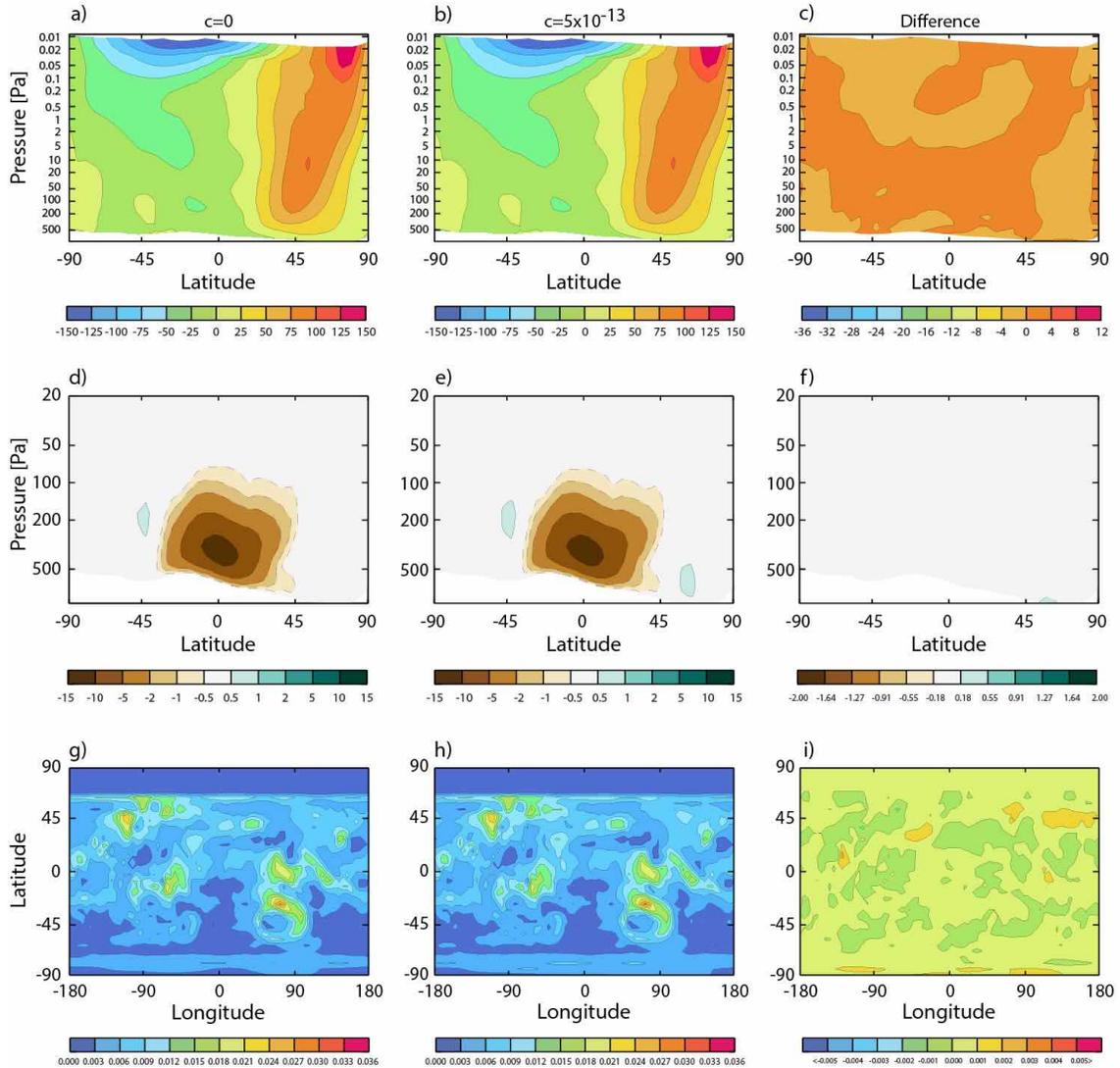

**Figure S12**: Mars Year 18 perihelion wind fields for the 'baseline' case (left column) and CTA case (middle column) along with differences between the two (right column) for (top) vertical slice of zonal-mean zonal wind [m/s], (middle) vertical slice of meridional streamfunction [$10^9$ kg/s], and (bottom) map view of daytime-averaged surface stress [N m$^{-2}$]. Note the difference in the vertical (pressure) scales of the top two rows.

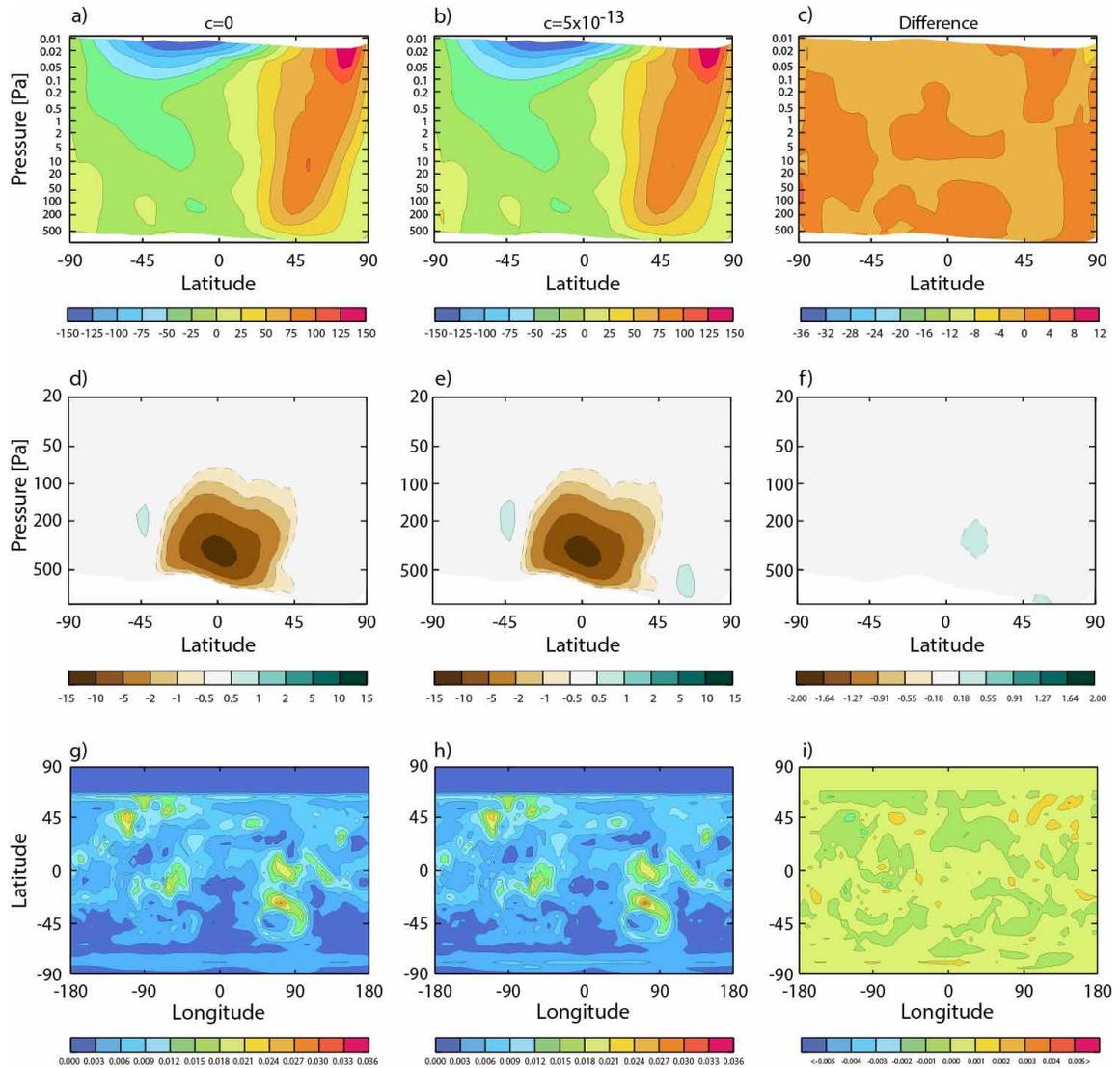

**Figure S13**: Mars Year 23 perihelion wind fields for the 'baseline' case (left column) and CTA case (middle column) along with differences between the two (right column) for (top) vertical slice of zonal-mean zonal wind [m/s], (middle) vertical slice of meridional streamfunction [$10^9$ kg/s], and (bottom) map view of daytime-averaged surface stress [N m$^{-2}$]. Note the difference in the vertical (pressure) scales of the top two rows.

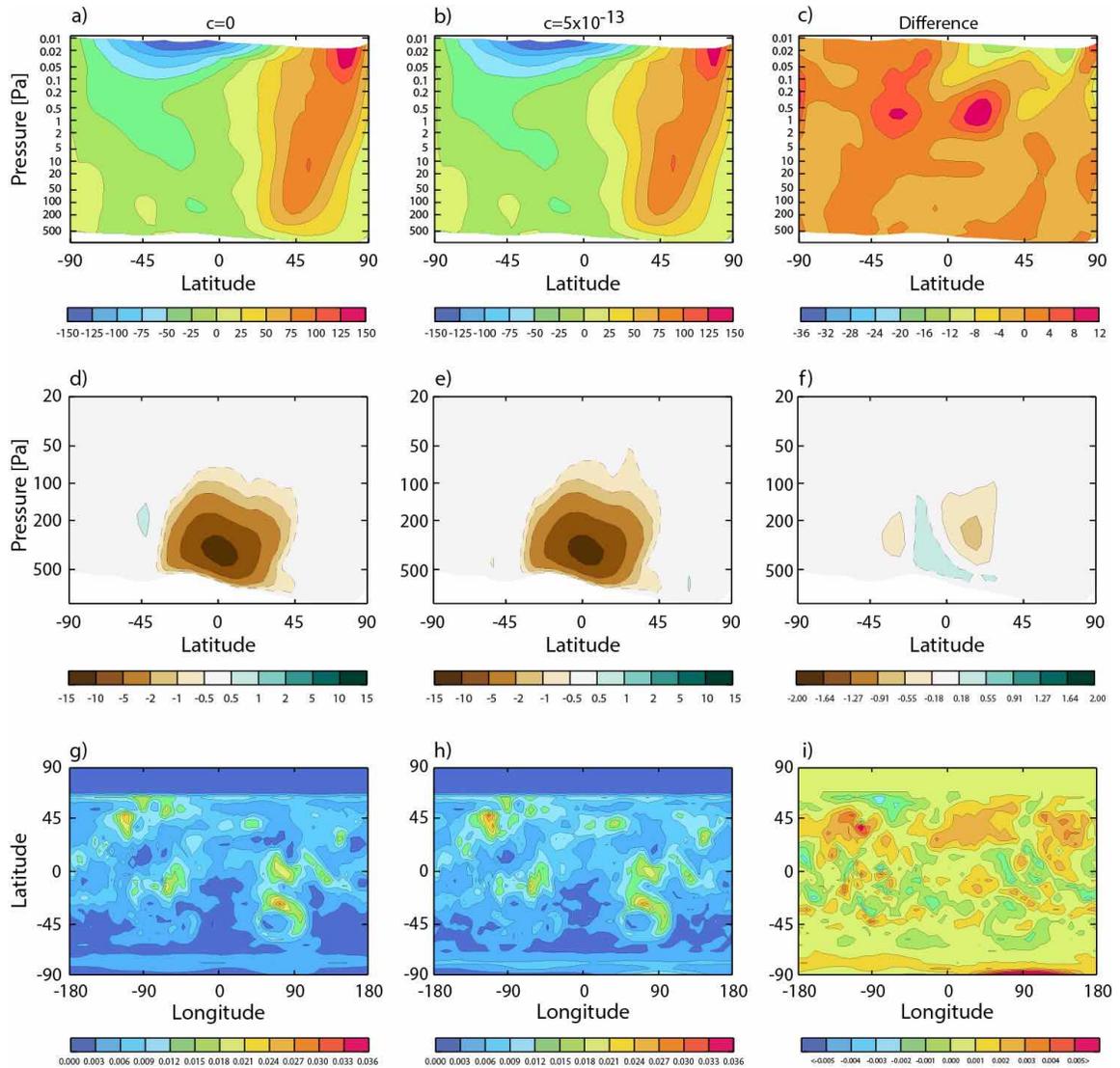

**Figure S14**: Mars Year 26 perihelion wind fields for the 'baseline' case (left column) and CTA case (middle column) along with differences between the two (right column) for (top) vertical slice of zonal-mean zonal wind [m/s], (middle) vertical slice of meridional streamfunction [$10^9$ kg/s], and (bottom) map view of daytime-averaged surface stress [N m$^{-2}$]. Note the difference in the vertical (pressure) scales of the top two rows.

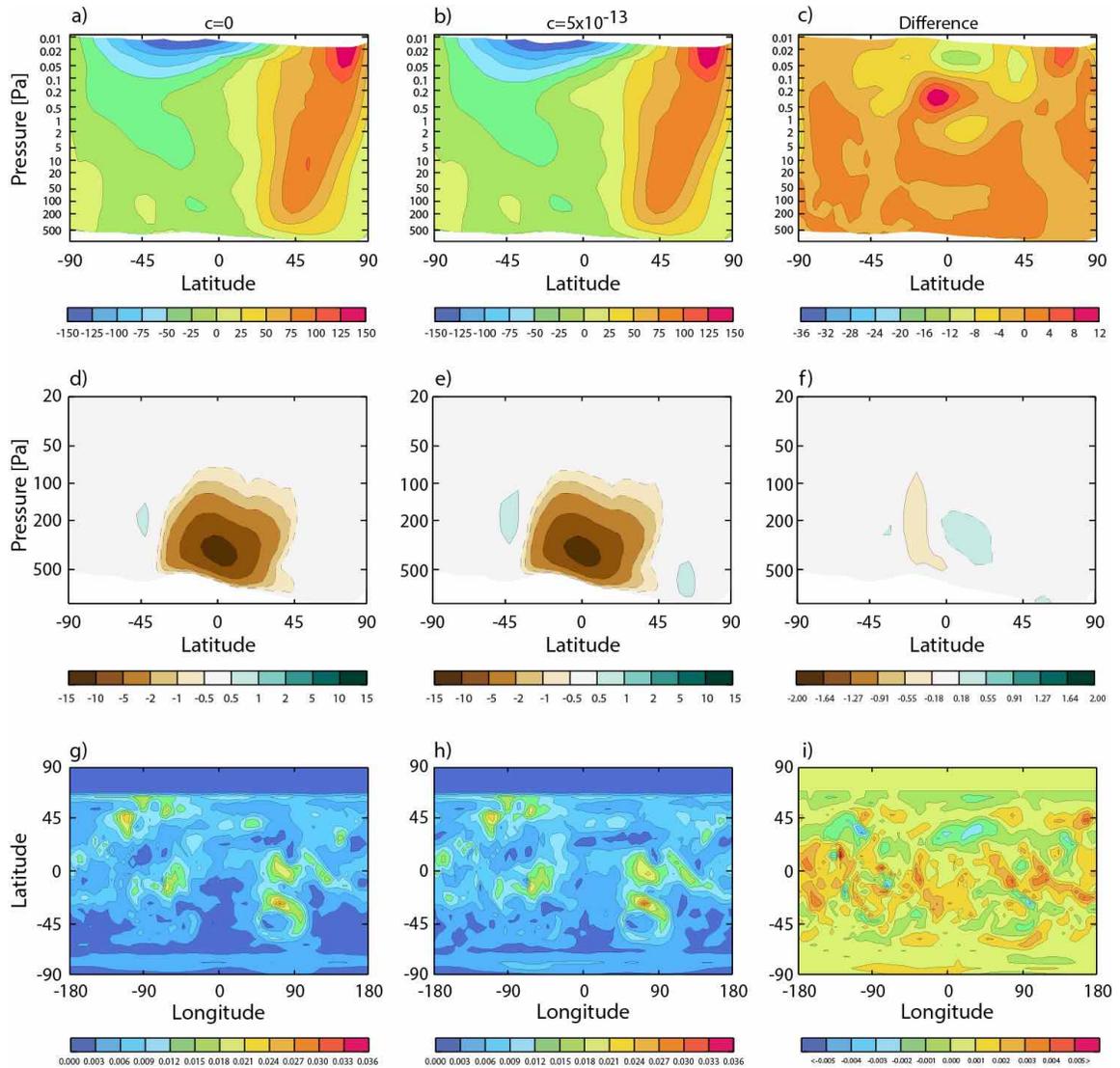

**Figure S15**: Mars Year 29 perihelion wind fields for the 'baseline' case (left column) and CTA case (middle column) along with differences between the two (right column) for (top) vertical slice of zonal-mean zonal wind [m/s], (middle) vertical slice of meridional streamfunction [$10^9$ kg/s], and (bottom) map view of daytime-averaged surface stress [N m$^{-2}$]. Note the difference in the vertical (pressure) scales of the top two rows.

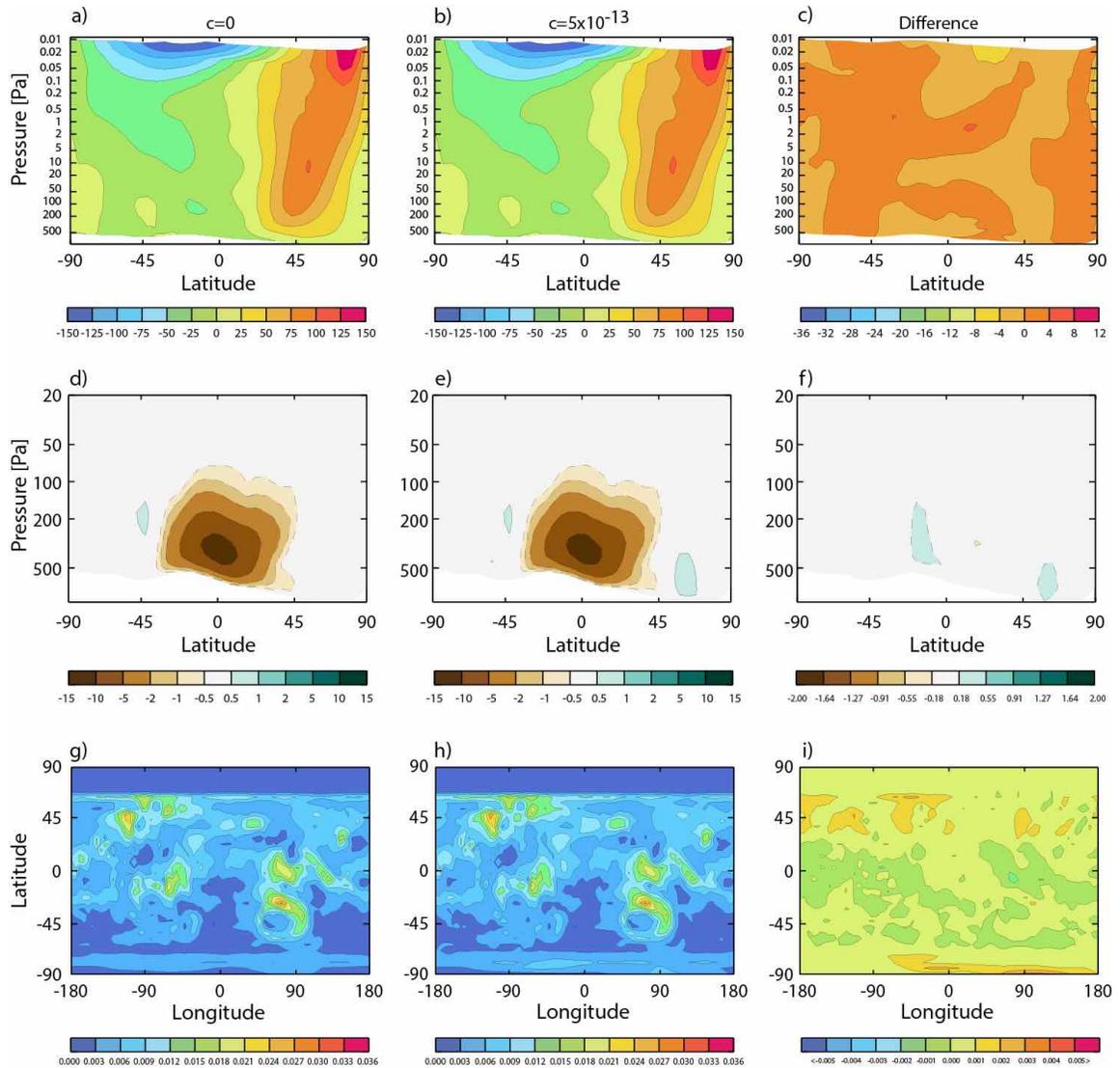

**Figure S16**: Mars Year 30 perihelion wind fields for the 'baseline' case (left column) and CTA case (middle column) along with differences between the two (right column) for (top) vertical slice of zonal-mean zonal wind [m/s], (middle) vertical slice of meridional streamfunction [$10^9$ kg/s], and (bottom) map view of daytime-averaged surface stress [N m$^{-2}$]. Note the difference in the vertical (pressure) scales of the top two rows.

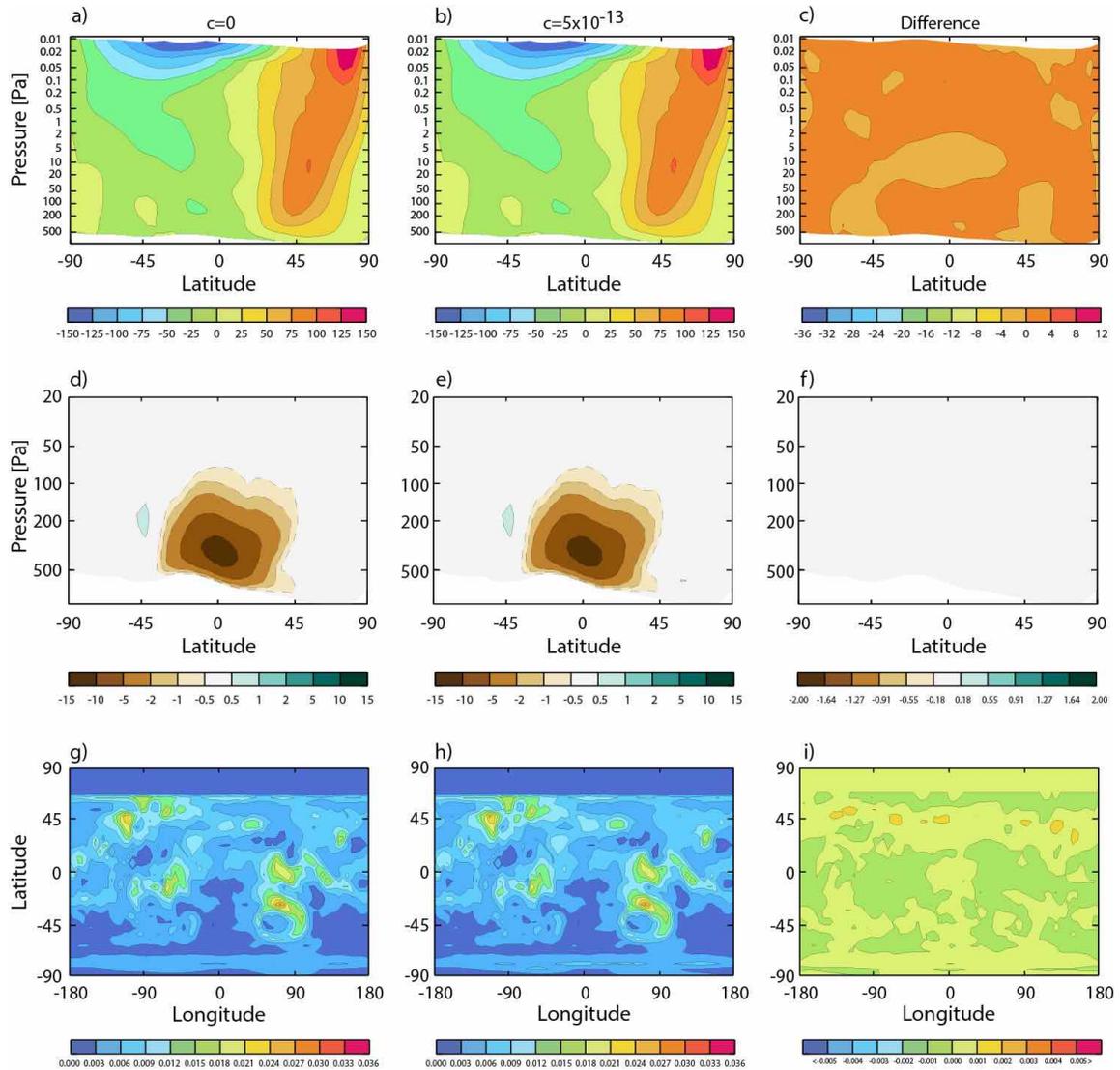

**Figure S17**: Mars Year 32 perihelion wind fields for the 'baseline' case (left column) and CTA case (middle column) along with differences between the two (right column) for (top) vertical slice of zonal-mean zonal wind [m/s], (middle) vertical slice of meridional streamfunction [$10^9$ kg/s], and (bottom) map view of daytime-averaged surface stress [N m$^{-2}$]. Note the difference in the vertical (pressure) scales of the top two rows.

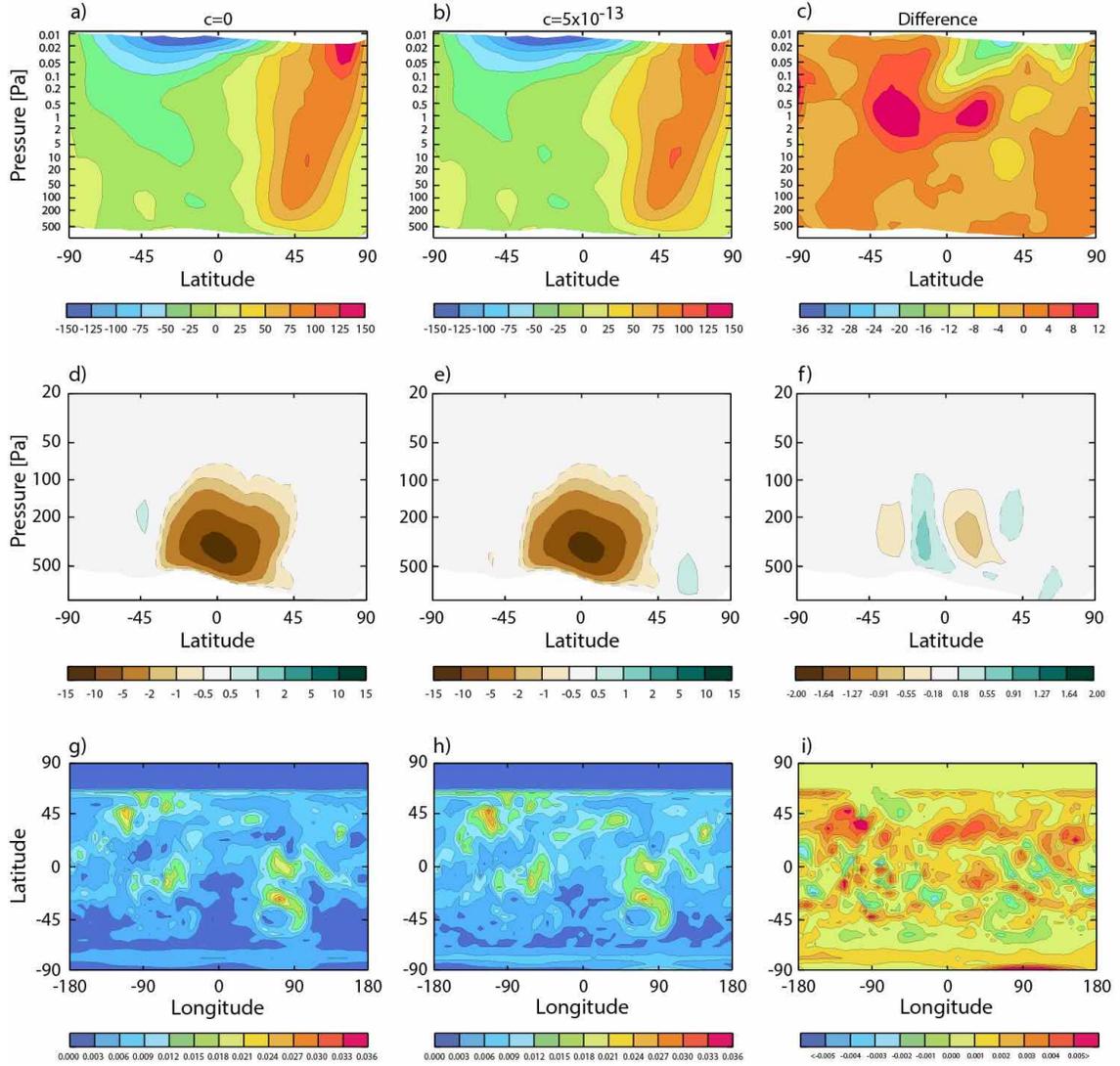

**Figure S18**: Mars Year -8 perihelion wind fields for the 'baseline' case (left column) and CTA case (middle column) along with differences between the two (right column) for (top) vertical slice of zonal-mean zonal wind [m/s], (middle) vertical slice of meridional streamfunction [$10^9$ kg/s], and (bottom) map view of daytime-averaged surface stress [N m$^{-2}$]. Note the difference in the vertical (pressure) scales of the top two rows.

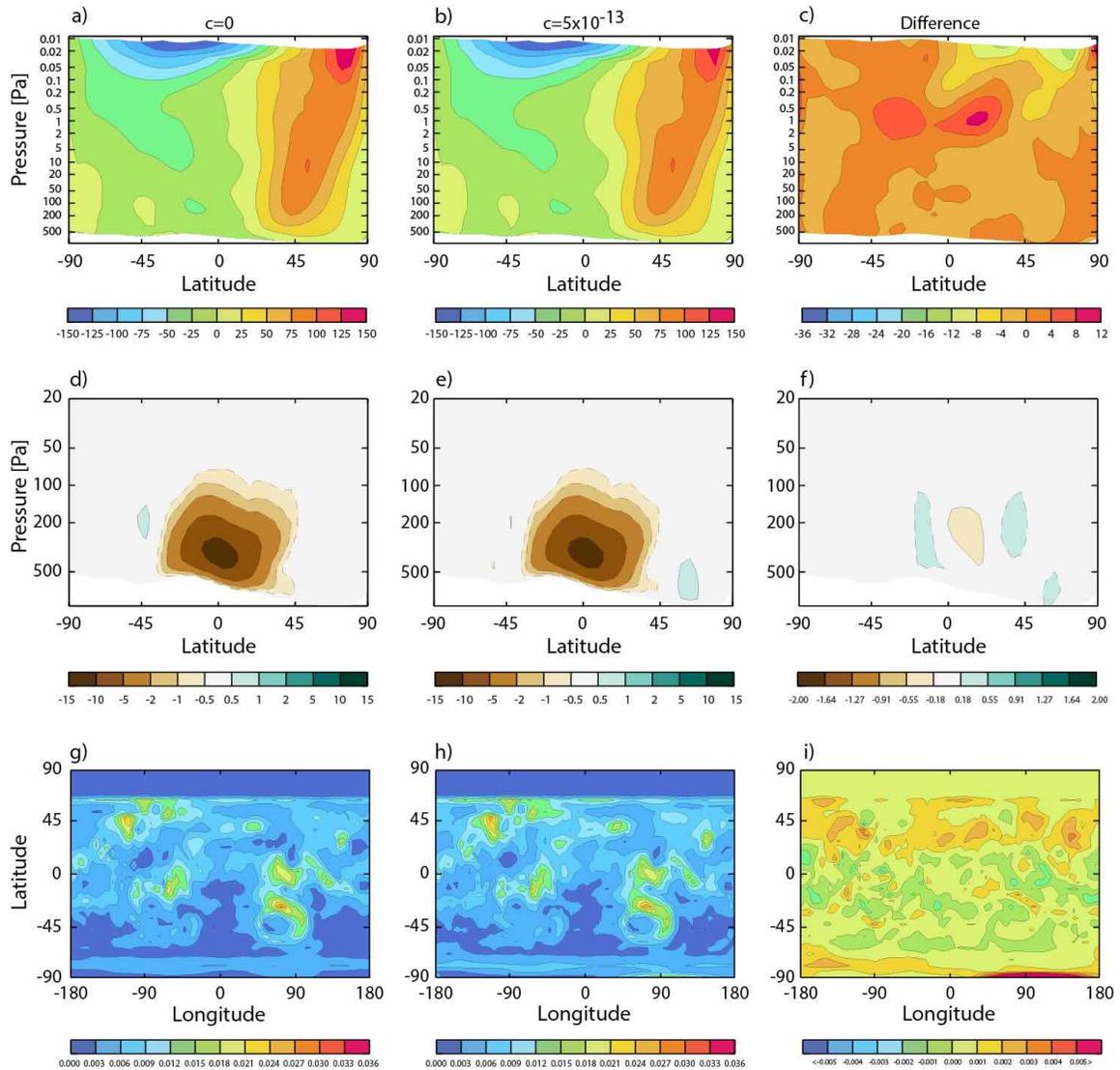

**Figure S19**: Mars Year 11 perihelion wind fields for the 'baseline' case (left column) and CTA case (middle column) along with differences between the two (right column) for (top) vertical slice of zonal-mean zonal wind [m/s], (middle) vertical slice of meridional streamfunction [$10^9$ kg/s], and (bottom) map view of daytime-averaged surface stress [N m$^{-2}$]. Note the difference in the vertical (pressure) scales of the top two rows.

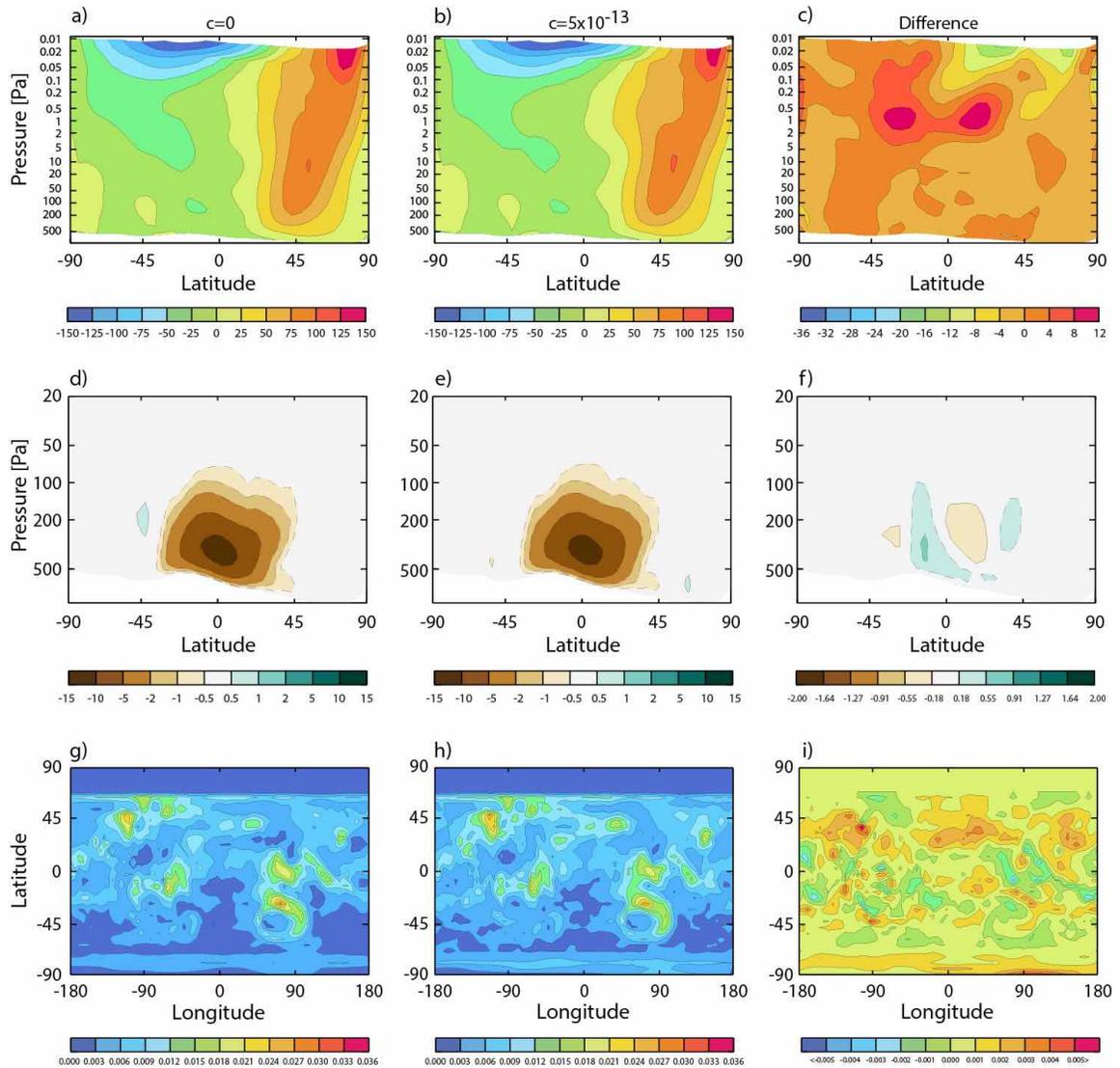

**Figure S20**: Mars Year 31 perihelion wind fields for the 'baseline' case (left column) and CTA case (middle column) along with differences between the two (right column) for (top) vertical slice of zonal-mean zonal wind [m/s], (middle) vertical slice of meridional streamfunction [$10^9$ kg/s], and (bottom) map view of daytime-averaged surface stress [N m$^{-2}$]. Note the difference in the vertical (pressure) scales of the top two rows.

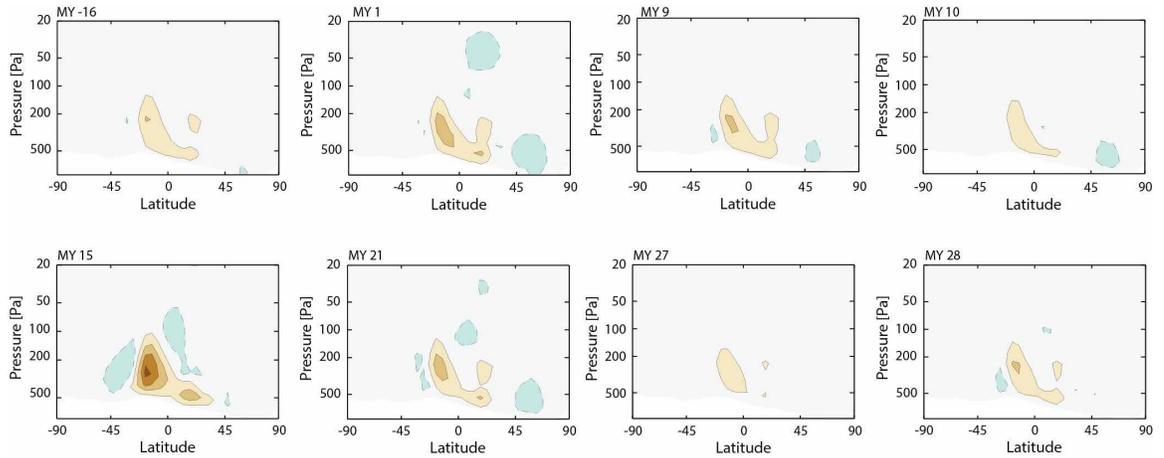

**Figure S21**: Streamfunction differences (forced minus unforced) for the positive polarity Mars years of Table 4 with radiatively active dust included in the simulation (compare to Figure 15). For years with GDS, the illustrated season corresponds to the period just prior to GDS initiation (Table 4). For the one non-GDS year (MY 27), the illustrated seasonal interval corresponds to the post-perihelion period ($L_s$=250-270°).

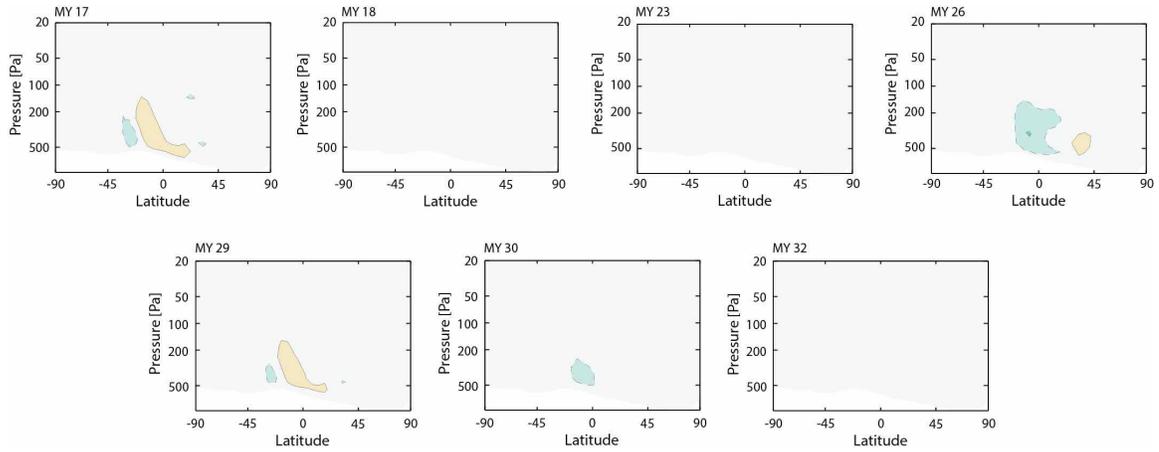

**Figure S22**: Streamfunction differences (forced minus unforced) for the transitional polarity (zero-crossing) Mars years of Table 4 with radiatively active dust included in the simulation (compare to Figure 16). In all cases, the seasonal interval modeled corresponds to the post-perihelion period ($L_s$=250-270°).

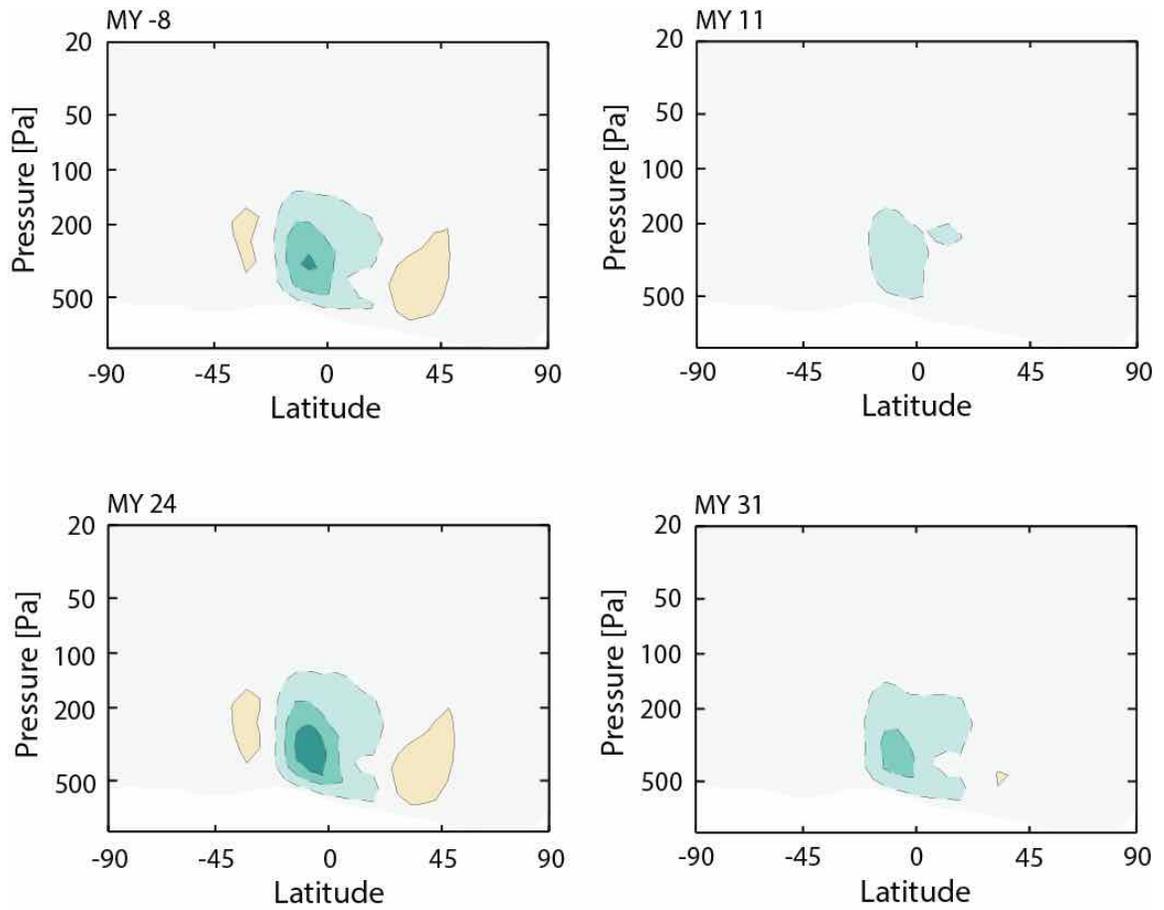

**Figure S23**: Streamfunction differences (forced minus unforced) for the negative polarity Mars years of Table 4 without GDS, but with radiatively active dust included in the simulation (compare to Figure 18). In all cases, the seasonal interval modeled corresponds to the post-perihelion period ($L_s$=250-270°).